\documentclass[preprint]{ptephy}
\preprintnumber{} 
\usepackage{array}
\usepackage{graphicx}
\usepackage{epsfig}
\usepackage{subfigure}
\usepackage{xspace}
\usepackage{dcolumn}
\usepackage{multirow}
\usepackage{longtable}
\usepackage{mathtools}
\usepackage{xcolor}
\usepackage{verbatim}
\usepackage{bm}
\usepackage{collectbox}
\usepackage{float}
\usepackage{amsmath}
\usepackage{lineno}

\DeclareMathOperator*{\argmax}{argmax} 

\makeatletter
\newcommand{\mybox}{    \collectbox{        \setlength{\fboxsep}{1pt}        \fbox{\BOXCONTENT}    }}
\makeatother

\newcolumntype{P}[1]{>{\centering\arraybackslash}p{#1}}

\newcommand{\val}[2]{\ensuremath{#1 \; \mathrm{#2}\xspace}}

\newcommand{\UP}{UP-$\mu$\xspace}

\newcommand{\numu}{\ensuremath{\nu_{\mu}}\xspace}
\newcommand{\nubar}{\ensuremath{\bar{\nu}_{\mu}}\xspace}
\newcommand{\numubar}{\nubar}
\newcommand{\nue}{\ensuremath{\nu_{e}}\xspace}
\newcommand{\nuebar}{\ensuremath{\bar{\nu}_{e}}\xspace}
\newcommand{\nutau}{\ensuremath{\nu_{\tau}}\xspace}

\newcommand{\pizero}{\ensuremath{\pi^0}\xspace}

\newcommand{\dm}{\ensuremath{\Delta m^{2}}\xspace}
\newcommand{\dmsq}[1]{\ensuremath{\dm_{#1}}\xspace}

\newcommand{\sn}[1]{\ensuremath{ \sin^{2}(\theta_{#1}) }\xspace }

\newcolumntype{d}[1]{D{.}{\cdot}{#1}}
\newcolumntype{.}{D{.}{.}{-1}}
\newcolumntype{,}{D{,}{,}{2}}

\begin{document}

\title{Atmospheric Neutrino Oscillation Analysis With Improved Event Reconstruction in Super-Kamiokande IV}

\newcommand{\AFFtokyo}{35}
\newcommand{\AFFregina}{29}
\newcommand{\AFFfukuoka}{11}
\newcommand{\AFFqmul}{28}
\newcommand{\AFFduke}{10}
\newcommand{\AFFkobe}{17}
\newcommand{\AFFokayama}{25}
\newcommand{\AFFbu}{6}
\newcommand{\AFFroma}{47}
\newcommand{\AFFtokai}{41}
\newcommand{\AFFtit}{37}
\newcommand{\AFFnagoya}{21}
\newcommand{\AFFshizuokasc}{32}
\newcommand{\AFFstfc}{33}
\newcommand{\AFFynu}{43}
\newcommand{\AFFosaka}{26}
\newcommand{\AFFtsinghua}{42}
\newcommand{\AFFcnm}{9}
\newcommand{\AFFpadova}{48}
\newcommand{\AFFcsu}{8}
\newcommand{\AFFtus}{38}
\newcommand{\AFFuci}{7}
\newcommand{\AFFtriumf}{40}
\newcommand{\AFFpol}{23}
\newcommand{\AFFskk}{34}
\newcommand{\AFFuh}{14}
\newcommand{\AFFgifu}{12}
\newcommand{\AFFmad}{4}
\newcommand{\AFFgist}{13}
\newcommand{\AFFox}{27}
\newcommand{\AFFsheff}{31}
\newcommand{\AFFliv}{19}
\newcommand{\AFFkashiwa}{2}
\newcommand{\AFFsuny}{24}
\newcommand{\AFFkek}{16}
\newcommand{\AFFicl}{15}
\newcommand{\AFFkyoto}{18}
\newcommand{\AFFtoronto}{39}
\newcommand{\AFFkmi}{22}
\newcommand{\AFFnapoli}{46}
\newcommand{\AFFicrr}{1}
\newcommand{\AFFbari}{45}
\newcommand{\AFFmiyagi}{20}
\newcommand{\AFFseoul}{30}
\newcommand{\AFFllr}{44}
\newcommand{\AFFipmu}{3}
\newcommand{\AFFubc}{5}
\newcommand{\AFFtodai}{36}
\author{
\name{M.~Jiang}{\AFFkyoto},
\name{K.~Abe}{\AFFicrr},
\name{C.~Bronner}{\AFFicrr},
\name{Y.~Hayato}{\AFFicrr},
\name{M.~Ikeda}{\AFFicrr},
\name{K.~Iyogi}{\AFFicrr },
\name{J.~Kameda}{\AFFicrr},
\name{Y.~Kato}{\AFFicrr},
\name{Y.~Kishimoto}{\AFFicrr},
\name{Ll.~Marti}{\AFFicrr}
% ...
\\
\name{(The Super-Kamiokande Collaboration)}{}}
\address{
\affil{\AFFicrr}{{Kamioka Observatory, Institute for Cosmic Ray Research, University of Tokyo, Kamioka, Gifu 506-1205, Japan}}
\affil{\AFFkashiwa}{{Research Center for Cosmic Neutrinos, Institute for Cosmic Ray Research, University of Tokyo, Kashiwa, Chiba 277-8582, Japan}}
\affil{\AFFipmu}{{Kavli Institute for the Physics andMathematics of the Universe (WPI), The University of Tokyo Institutes for Advanced Study,University of Tokyo, Kashiwa, Chiba 277-8583, Japan }}
\affil{\AFFmad}{{Department of Theoretical Physics, University Autonoma Madrid, 28049 Madrid, Spain}}
\affil{\AFFubc}{{Department of Physics and Astronomy, University of British Columbia, Vancouver, BC, V6T1Z4, Canada}}
\affil{\AFFbu}{{Department of Physics, Boston University, Boston, MA 02215, USA}}
\affil{\AFFuci}{{Department of Physics and Astronomy, University of California, Irvine, Irvine, CA 92697-4575, USA }}
\affil{\AFFcsu}{{Department of Physics, California State University, Dominguez Hills, Carson, CA 90747, USA}}
\affil{\AFFcnm}{{Department of Physics, Chonnam National University, Kwangju 500-757, Korea}}
\affil{\AFFduke}{{Department of Physics, Duke University, Durham NC 27708, USA}}
\affil{\AFFfukuoka}{{Junior College, Fukuoka Institute of Technology, Fukuoka, Fukuoka 811-0295, Japan}}
\affil{\AFFgifu}{{Department of Physics, Gifu University, Gifu, Gifu 501-1193, Japan}}
\affil{\AFFgist}{{GIST College, Gwangju Institute of Science and Technology, Gwangju 500-712, Korea}}
\affil{\AFFuh}{{Department of Physics and Astronomy, University of Hawaii, Honolulu, HI 96822, USA}}
\affil{\AFFicl}{{Department of Physics, Imperial College London , London, SW7 2AZ, United Kingdom }}
\affil{\AFFkek}{{High Energy Accelerator Research Organization (KEK), Tsukuba, Ibaraki 305-0801, Japan }}
\affil{\AFFkobe}{{Department of Physics, Kobe University, Kobe, Hyogo 657-8501, Japan}}
\affil{\AFFkyoto}{{Department of Physics, Kyoto University, Kyoto, Kyoto 606-8502, Japan}}
\affil{\AFFliv}{{Department of Physics, University of Liverpool, Liverpool, L69 7ZE, United Kingdom}}
\affil{\AFFmiyagi}{{Department of Physics, Miyagi University of Education, Sendai, Miyagi 980-0845, Japan}}
\affil{\AFFnagoya}{{Institute for Space-Earth Environmental Research, Nagoya University, Nagoya, Aichi 464-8602, Japan}}
\affil{\AFFkmi}{{Kobayashi-Maskawa Institute for the Origin of Particles and the Universe, Nagoya University, Nagoya, Aichi 464-8602, Japan}}
\affil{\AFFpol}{{National Centre For Nuclear Research, 00-681 Warsaw, Poland}}
\affil{\AFFsuny}{{Department of Physics and Astronomy, State University of New York at Stony Brook, NY 11794-3800, USA}}
\affil{\AFFokayama}{{Department of Physics, Okayama University, Okayama, Okayama 700-8530, Japan }}
\affil{\AFFosaka}{{Department of Physics, Osaka University, Toyonaka, Osaka 560-0043, Japan}}
\affil{\AFFox}{{Department of Physics, Oxford University, Oxford, OX1 3PU, United Kingdom}}
\affil{\AFFqmul}{{School of Physics and Astronomy, Queen Mary University of London, London, E1 4NS, United Kingdom}}
\affil{\AFFregina}{{Department of Physics, University of Regina, 3737 Wascana Parkway, Regina, SK, S4SOA2, Canada}}
\affil{\AFFseoul}{{Department of Physics, Seoul National University, Seoul 151-742, Korea}}
\affil{\AFFsheff}{{Department of Physics and Astronomy, University of Sheffield, S3 7RH, Sheffield, United Kingdom}}
\affil{\AFFshizuokasc}{{Department of Informatics inSocial Welfare, Shizuoka University of Welfare, Yaizu, Shizuoka, 425-8611, Japan}}
\affil{\AFFstfc}{{STFC, Rutherford Appleton Laboratory, Harwell Oxford, and Daresbury Laboratory, Warrington, OX11 0QX, United Kingdom}}
\affil{\AFFskk}{{Department of Physics, Sungkyunkwan University, Suwon 440-746, Korea}}
\affil{\AFFtokyo}{{The University of Tokyo, Bunkyo, Tokyo 113-0033, Japan }}
\affil{\AFFtodai}{{Department of Physics, University of Tokyo, Bunkyo, Tokyo 113-0033, Japan }}
\affil{\AFFtit}{{Department of Physics,Tokyo Institute of Technology, Meguro, Tokyo 152-8551, Japan }}
\affil{\AFFtus}{{Department of Physics, Faculty of Science and Technology, Tokyo University of Science, Noda, Chiba 278-8510, Japan }}
\affil{\AFFtoronto}{{Department of Physics, University of Toronto, ON, M5S 1A7, Canada }}
\affil{\AFFtriumf}{{TRIUMF, 4004 Wesbrook Mall, Vancouver, BC, V6T2A3, Canada }}
\affil{\AFFtokai}{{Department of Physics, Tokai University, Hiratsuka, Kanagawa 259-1292, Japan}}
\affil{\AFFtsinghua}{{Department of Engineering Physics, Tsinghua University, Beijing, 100084, China}}
\affil{\AFFynu}{{Faculty of Engineering, Yokohama National University, Yokohama, 240-8501, Japan}}
\affil{\AFFllr}{{Ecole Polytechnique, IN2P3-CNRS, Laboratoire Leprince-Ringuet, F-91120 Palaiseau, France }}
\affil{\AFFbari}{{ Dipartimento Interuniversitario di Fisica, INFN Sezione di Bari and Universit\`a e Politecnico di Bari, I-70125, Bari, Italy}}
\affil{\AFFnapoli}{{Dipartimento di Fisica, INFN Sezione di Napoli and Universit\`a di Napoli, I-80126, Napoli, Italy}}
\affil{\AFFroma}{{INFN Sezione di Roma and Universit\`a di Roma ``La Sapienza'', I-00185, Roma, Italy}}
\affil{\AFFpadova}{{Dipartimento di Fisica, INFN Sezione di Padova and Universit\`a di Padova, I-35131, Padova, Italy}}
\email{mjiang@scphys.kyoto-u.ac.jp}
}
 \date{\today}

\begin{abstract}

A new event reconstruction algorithm based on a maximum likelihood method has been developed for Super-Kamiokande.
Its improved kinematic and particle identification capabilities enable the analysis of atmospheric neutrino data 
in a detector volume 32\% larger than previous analyses and increases sensitivity to the neutrino mass hierarchy.
Analysis of a 253.9~kton-year exposure of the Super-Kamiokande IV atmospheric neutrino data has yielded 
a weak preference for the normal hierarchy, disfavoring the inverted hierarchy at 74\% assuming oscillations at 
the best fit of the analysis.
\end{abstract}

\maketitle

%\linenumbers

\section{Introduction}
\label{sec:intro}

Neutrino oscillations have been confirmed by a variety of experiments using both natural 
and artificial sources.  
At present the data are well described assuming mixing among all three active 
neutrinos with the Pontecorvo-Maki-Nakagawa-Sakata (PMNS) formalism~\cite{Pontecorvo:1967fh,Maki:1962mu}.
Though most of its parameters have been experimentally measured\cite{Olive:2016xmw}, 
the ordering of the mass states with the largest splitting (known as the mass hierarchy), 
the octant of the atmospheric mixing angle $\theta_{23}$, and the value of its CP-violating phase are unknown.
These unresolved issues have been at the forefront of results from the T2K~\cite{Abe:2017uxa}, NOvA~\cite{Adamson:2017gxd} 
and Super-Kamiokande~\cite{Abe:2017osc} (Super-K, SK) experiments and are the focus of next-generation experiments planned in  
the U.S.~\cite{Acciarri:2016crz}, China~\cite{An:2015jdp}, and Japan~\cite{Abe:2018uyc}. 

The atmospheric neutrino flux provides a source of both neutrinos and antineutrinos with a wide variety of 
energies and path lengths suitable to probe each of these open questions. 
Matter effects influence the oscillations of neutrinos passing through the Earth and 
produce a resonance enhancement of the oscillation probability that depends on the 
value of $\theta_{13}$, $\theta_{23}$, and $\delta_{CP}$. 
Importantly this resonance occurs only for neutrinos if the hierarchy is normal (largest splitting is between the two heaviest states)
and only for antineutrinos if it is inverted.
The hierarchy signal manifests most strongly as an 
upward-going excess of electron neutrino or antineutrino events with multi-GeV energies,
a region where interactions frequently have multi-particle final states that complicate 
particle identification and where the flux is orders of magnitude lower than its peak. 
Accordingly, recent Super-Kamiokande results~\cite{Abe:2017osc} have been limited by 
both a lack of statistics and event miscategorization in the signal region. 
 
A new event reconstruction algorithm has been developed 
based on a maximum likelihood method designed to extract specific event topologies 
and determine the best set of kinematic parameters.
The new algorithm demonstrates improved reconstruction performance across a variety of 
metrics including vertex resolution, particle identification probability, and 
momentum resolution and does so across a larger volume of the Super-K detector than 
its predecessor.  
As a result, it is expected to improve Super-K's sensitivity to 
the remaining parameters of the PMNS mixing paradigm.

This paper presents an analysis of 3118.5 days of SK-IV atmospheric data for a 253.9 kiloton$\cdot$year exposure
with a 32\% larger fiducial volume (FV) that previous Super-K analyses. 
Section~\ref{sec:oscillations} briefly reviews atmospheric neutrino oscillations
before the detector is introduced in Section~\ref{sec:detector}.
In Section~\ref{sec:reduc} the atmospheric neutrino data reduction and event categorization are discussed.
The new reconstruction algorithm and its performance in comparison to the conventional reconstruction 
are detailed in Section~\ref{sec:fQRec}.
Based on this performance improvement the expansion of the FV and 
corresponding systematic errors are presented in Sections~\ref{sec:fvexp} and ~\ref{sec:syst}, respectively.
An analysis of the atmospheric neutrino data by themselves is presented in
Section~\ref{sec:atm_only} and is followed by an analysis in which $\theta_{13}$ is constrained by the world data in Section~\ref{sec:gaibu}, before concluding in Section~\ref{sec:conclusion}.
 \section{Neutrino Oscillations}
\label{sec:oscillations}
Neutrino oscillations are a result of the neutrino mass eigenstates differing from 
their weak-interaction (flavor) eigenstates.
For neutrinos in vacuum, their oscillation probability in the standard three-flavor paradigm can be described by six parameters, 
nearly all of which have been confirmed to have non-zero values based on the 
results of reactor, atmospheric, solar, and long-baseline neutrino experiments~\cite{Olive:2016xmw}.
For brevity the details of the PMNS oscillation formalism will be omitted here, more details about the treatment of neutrino oscillation in Super-K are at~\cite{Abe:2017osc}.

In vacuum the leading terms in the oscillations of atmospheric $\nu_e$ and $\nu_{\mu}$ in can be written as:
\begin{eqnarray}
P(\nu_e \rightarrow \nu_e) &\cong& 1-\sin^2 2\theta_{13} \sin^2\left( \frac{1.27 \Delta m_{32}^2 L}{E} \right) \nonumber \\
P(\nu_{\mu} \rightarrow \nu_{\mu}) &\cong& 1-4\cos^2 \theta_{13} \sin^2 \theta_{23} (1-\cos^2 \theta_{13}\sin^2\theta_{23}) \nonumber \\
& & \times \sin^2 \left(\frac{1.27 \Delta m_{32}^2 L}{E} \right) \nonumber \\
P(\nu_{\mu} \leftrightarrow \nu_e) &\cong& \sin^2 \theta_{23} \sin^2 2\theta_{13} \sin^2\left(\frac{1.27 \Delta m_{32}^2 L}{E} \right)
\label{eqn:oscprob_atm_vac}
\end{eqnarray}
\noindent using the assumption that $\Delta m_{31}^2 \approx \Delta m_{32}^2$ since $\Delta m_{21}^2$ is known to be small in comparison.
In these equations the neutrino travel length is represented by $L$ (km) and its energy by $E$ (GeV). The unit of $\Delta m^2$ is eV$^2$.
Atmospheric neutrinos are sensitive to $\sin^2 2\theta_{13}$ and the mass splitting $\Delta m_{32}^2$ via $\nu_{\mu} \leftrightarrow \nu_e$ oscillations, 
which depend on $\sin^2\theta_{23}$.

When neutrinos propagate in matter, the oscillation probabilities are modified due to interactions  
with the electrons in the earth.
Indeed the forward scattering amplitude of $\nu_{e}$ differs from those of the other flavors, producing an effective potential 
felt only by this species. 
This phenomenon is known as the matter effect or the MSW effect of neutrino oscillations~\cite{Wolfenstein:1978msw,Mikheev:1985msw}. 
When neutrinos travel through homogeneous matter, the oscillation parameters $\sin^2 \theta_{13}$ and $\Delta m_{32}^2$ 
in Equation \ref{eqn:oscprob_atm_vac} are replaced by their matter-equivalents~\cite{Giunti:1997MSW},
\begin{equation}
\sin^2 2\theta_{13,M} = \frac{\sin^{2}2\theta_{13}}{(\cos2\theta_{13}-A_{CC}/\Delta m^2_{32})^2+\sin^2 2\theta_{13}} 
\label{eqn:oscprob_inmatter}
\end{equation}
\begin{equation}
\Delta m_{32,M}^2 = \Delta m_{32}^2 \sqrt{(\cos2\theta_{13}-A_{CC}/\Delta m^2_{32})^2+\sin^2 2\theta_{13}} 
\end{equation}
\noindent where $A_{CC}=\pm2\sqrt{2}G_FN_eE_\nu$ and the electron density $N_e$ is assumed to be constant. 
Here $G_F$ represents the Fermi constant.

Resonant enhancement of these matter variables occurs when $A_{CC}/\Delta m^2_{32} = \cos 2\theta_{13}$
and is dependent upon the sign of $\Delta m_{32}^2$ and whether the neutrino is a particle ($A_{CC} > 0$) or an antiparticle ($A_{CC} < 0$).
Note that the enhancement only occurs for neutrinos if the hierarchy is normal ($\Delta m_{32}^2 > 0$) whereas it only happens for antineutrinos if it is inverted ($\Delta m_{32}^2 < 0$).
Though atmospheric neutrinos generally experience a varying matter profile, and hence electron density $N_e$ changes as they travel through the earth, this phenomena is nonetheless present. The calculation of oscillation probability in this analysis takes such variation on matter density into consideration, with a simplified version of the preliminary reference Earth model (PREM) (c.f.~\cite{Abe:2017osc}).
Typically neutrinos with a few to ten GeV of energy that travel through the Earth's core experience matter effects the most strongly. 

 \section{The Super-Kamiokande Detector}
\label{sec:detector}

Super-Kamiokande is a cylindrical water Cherenkov detector with a total volume of 50~kilotons and 
located inside the Kamioka mine in Gifu Prefecture, Japan. 
It is optically separated into two regions, an inner detector (ID), which forms the primary neutrino target 
and has a total volume of 32~kilotons and an outer detector (OD), which is a two-meter cylindrical shell 
surrounding the ID and used primarily as a veto. 
The ID is instrumented with more than 11,000 inward-facing 20-inch photomultiplier tubes (PMTs),
representing 40\% photocathode coverage of the target volume. 
A total of 1,885 8-inch PMTs coupled to wavelength-shifting plates are mounted on the inner surface 
of the OD, while its outer surface is covered with reflective sheets to increase light collection.

Since the start of operations in 1996, Super-Kamiokande has gone through four data-taking periods, SK-I, -II, -III, and -IV.
The present work focuses on the latter, which started in September 2008 and ended on May 31$^{\text{st}}$ 2018 when the 
detector began upgrade work for its next phase.
Though the topology of the detector in the SK-III and -IV periods are the same, 
at the start of SK-IV the front end electronics were upgraded to 
a system based on ASIC that uses a high-speed charge-to-time converter~\cite{Nishino:2009zu}.
After this upgrade Super-Kamiokande is able to collect all PMT hits above threshold 
without incurring any dead time.
More detailed descriptions of the detector, its electronics are presented in~\cite{Fukuda:2002uc,Nishino:2009zu,Abe:2013gga}.

Atmospheric neutrino interactions in the detector are simulated 
using the Honda et. al flux calculation~\cite{Honda:2011nf} and the NEUT~\cite{Hayato:2002sd} 
neutrino interaction software (version 5.4.0). 
Particles emerging from the interactions are tracked through a simulation of 
the detector based on GEANT3~\cite{Brun:1994aa}.
The present work uses an updated version of NEUT relative to the previous 
atmospheric neutrino analysis (c.f.~\cite{Abe:2017osc}).
In particular, charged-current quasi-elastic (CCQE) interactions in the new simulation are 
on the local Fermi-Gas model of Nieves \cite{Nieves:2013wx, Nieves:2011wx}, assuming 
an axial mass $M_{A} = 1.05~\mbox{GeV/c}^{2}$ and using the random phase approximation correction.
Further, deep inelastic scattering (DIS) is modeled using 
the GRV98 parton distribution function~\cite{Gluck:1998xa} and utilizing CKM matrix elements for the calculation of structure functions.
Corrections for low $q^{2}$ scattering have been updated to those of Bodek and Yang~\cite{Bodek:2005de}, where $q^{2}$ denotes the square of the transferred four-momentum.

 \section{Event sample and reduction}
\label{sec:reduc}

The current analysis utilizes atmospheric neutrino data collected during the SK-IV period with a total livetime of 3118.45 days. 
As in previous Super-K analyses, the atmospheric neutrino data are separated into three general categories, fully contained (FC), 
partially contained (PC), and upward-going muons (Up-$\mu$).

For events classified as FC and PC, the neutrino interacts within the fiducial volume, defined as the region located more than 2~m from the ID wall in previous analyses. 
Events with no activity in the outer detector are classified as FC.
If energy deposition in the OD is observed, generally from a high energy muon exiting the ID, the event is classified as PC. 
Muons created by neutrino interactions in the rock around SK or in the OD water and traveling upward through the detector 
form the Up-$\mu$ sample.

At trigger level the Super-Kamiokande event sample consists mainly of downward-going cosmic ray muons and low energy radioactivity from contaminants in the water such as radon. 
To remove these backgrounds, reduction processes specific to the three main sample categories are applied to the data, whose detail can be found in~\cite{Ashie:2005osc}.

As an example, five steps of data reduction criteria are used for FC sample. 
In the first and second reduction steps, criteria on the total charge collected by ID and the number of hits on OD are applied to remove most cosmic ray muons. 
In the third reduction step, a dedicated cosmic ray muon reconstruction algorithm is used to apply a cut on OD hits near the entrance or exit point of a reconstructed track. The fourth step is designed to remove so-called ``flasher'' events caused by internal corona discharge of a PMT.
Failing tubes often produce multiple events with similar topologies as these discharges repeat.
Typical flasher events have a broad distribution in time among hit PMTs and spatially similar hit distributions.
Cuts based on the hit timing distribution and the charge pattern correlation with other events that passed the third step reduction are applied to remove flasher events.
The fifth reduction rejects the remaining cosmic ray muons, flasher events and electronic noise further using a more precise fitter.
Finally, the reconstructed vertex is required to be within a specified fiducial volume to further reduce non-neutrino background events 
and to prevent the reconstruction performance from deteriorating near the detector wall. 
These steps are particularly relevant to the new reconstruction algorithm considered here and will be discussed in Section \ref{sec:fvexp} in detail.
An event's visible energy ($E_{\mbox{vis}}$), which is defined as the energy of an electromagnetic shower producing the same amount of Cherenkov light as observed in the event,
is also required to be larger than 30~MeV to remove low energy backgrounds.
All events passing the reduction are scanned by eye to determine the level of background contamination in the final analysis sample and to estimate the uncertainty inherent to the reduction process.

After reduction, the FC data are sub-divided based upon the number of observed Cherenkov rings, the particle ID (PID) of the most energetic ring, and the number of observed electrons from muon decays into combinations of single- or multi-ring, electron-like ($e$-like) or muon-like ($\mu$-like), and 
sub-GeV ($E_{\mbox{vis}} < 1330.0$ MeV) or multi-GeV ($E_{\mbox{vis}} > 1330.0$ MeV) categories, as well as the number of decay electrons (0, 1, or 2 or more).
In the present study, all FC events have been reconstructed using the new algorithm, which will be introduced in the next section.

The PC and Up-$\mu$ samples also have their own reduction processes, which are optimized for the topology of OD activity of neutrino events.
These two samples are reconstructed using the pre-existing  algorithm and are divided into ``stopping'' and ``through-going'' subsamples based on the estimated muon stopping point for PC and Up-$\mu$ events.
 The ``through-going'' events in the Up-$\mu$ sample are further divided in to ``showering'' and ``non-showering'' based on whether the 
event induces an electromagnetic shower while traversing the ID.
Although Super-K also works as the far detector for T2K experiment~\cite{Abe:2013nka} and detects accelerator-generated neutrinos,
they are excluded based on event time information in this analysis.
After all selections there are a total of 13 FC analysis samples, two PC samples and three Up-$\mu$ samples.
 \section{FC Event Reconstruction}
\label{sec:fQRec}

The Super-K event reconstruction algorithms determine an event's physical properties such as the interaction vertex, number of particles, the particle types and momenta based on PMT hit information.

The conventional event reconstruction algorithm, APFit, was introduced at the beginning of Super-K, contributing to the discovery of atmospheric neutrino oscillation and has been used in both the K2K and T2K experiments. 
APFit is a single iteration fitter based on the time and charge information of hit PMTs~\cite{Shiozawa:1999ap}. 
The interaction vertex is reconstructed based on hit timing information after accounting for the photon time of flight. 
Then the direction of the first found ring, usually also the most energetic ring, is determined based on the observed charge distribution with respect to the interaction vertex. 
Additional ring candidates are found using a Hough-transform based method and selected by a likelihood function optimized to reject spurious ring candidates. 
After determining the number of Cherenkov rings in the event, 
the particle type of each ring is determined based upon the Cherenkov ring pattern and opening angle.
Rings from electrons tend to have rough edges produced by the light from their electromagnetic showers, 
while rings from muons or charged pions predominately produce crisp edges. 
In the last step of the fit, the momentum of each ring is evaluated based on the observed charge inside cone with a half angle of $70^\circ$ drawn along 
the line connecting the interaction vertex to each ring's center. 
Corrections and adjustments are made to account for charge sharing between overlapping rings.
It is worth noting that the hit time information is only used during the first step to find the initial vertex candidate in APFit.

The new algorithm, named fiTQun, employs a maximum likelihood method to reconstruct particle types and determine kinematics in the detector simultaneously. The algorithm is based on methods developed for the MiniBooNE experiment\cite{Pat:2009mini}, 
but has been developed from scratch for Super-K with additional features such as multi-ring reconstruction for events with multiple final-state particles. 
Compared to APFit, fiTQun uses more information, including information from PMT hits outside of the expected Cherenkov cone 
and hit timing information, during the fitting procedure. 
For a given event fiTQun's fit procedure will run multiple times to determine the best kinematic parameters for each possible 
particle configuration hypothesis, while APFit only fits those parameters once. 
The remarkable evolution of computing power since the start of Super-Kamiokande has enabled fiTQun to achieve 
higher reconstruction precision on similar time scales as APFit used to be. 
FiTQun has already been used in the T2K analyses using an expanded fiducial volume due its improved resolution of reconstructed quantities and particle identification~\cite{Abe:2018wpn}.

\subsection{Likelihood Function}

An event topology hypothesis $\Gamma$ (e.g. single-ring $e$-like) together with its associated kinematic parameters $\theta$, which include the vertex position, particle creation times, the azimuthal and zenith angles of the particle directions, as well as their momenta are considered in the likelihood function during a fit. 
Based on the observed charge and hit time of each PMT, fiTQun constructs the following likelihood function for a given the hypothesis to estimate the kinematic variables:

\begin{eqnarray}
L(\Gamma, \theta)&=& \prod^{\text{unhit}}_{j} P_j(\text{unhit}|\Gamma, \theta) \prod^{\text{hit}}_{i}\{ 1-P_i(\text{unhit}|\Gamma, \theta)\} \nonumber \\
&& \times f_q(q_i|\Gamma, \theta)f_t(t_i|\Gamma, \theta).
\label{eq:lklhd}
\end{eqnarray}

\noindent In this equation, the index $j$ runs over all PMTs which did not register a hit (``unhit'' PMTs) and for each of these 
the probability that it does not register a hit given the fitting hypothesis ($\Gamma, \theta$) is calculated 
as $P_j(\text{unhit}|\Gamma, \theta)$.
PMTs which did register a hit are indexed with $i$.
For such PMTs the likelihood density for observing a charge $q_i$ under the fitting hypothesis is represented by charge likelihood 
$f_q(q_i|\Gamma, \theta)$.
The likelihood density of producing a hit at the observed time $t_i$ is defined similarly as $f_t(t_i|\Gamma, \theta)$.

Since the processes of particle and optical photon propagation are decoupled from the response of the PMT and the electronics,
the charge likelihood can be rewritten in terms of the expected number of photoelectrons produced at the $i$-th PMT given the hypothesis (the predicted charge), 
$\mu_i(\Gamma, \theta)$, as 
\begin{eqnarray}
L(\Gamma, \theta)&=& \prod^{\text{unhit}}_{j} P_j(\text{unhit}|\mu_j) \prod^{\text{hit}}_{i}\{ 1-P_i(\text{unhit}|\mu_i)\} \nonumber \\
&& \times f_q(q_i|\mu_i)f_t(t_i|\Gamma, \theta)
\label{eq:lklhdmu}
\end{eqnarray}
$P_j(\text{unhit}|\mu_j)$ and $f_q(q_i|\mu_i)$ are properties of the PMT and the electronics, and therefore do not explicitly depend on the process of Cherenkov photon emission and propagation in water. 

In the calculation of the predicted charge, the contributions from direct light and light that has scattered or been reflected (indirect light)
are considered separately and summed to form the final $\mu_i$.
The predicted charge from direct light reaching a PMT 
is calculated by integrating the Cherenkov emission profile along the track 
while correcting for the distance from the track to the PMT, 
the light transmission in water, and the PMT angular acceptance. 
Charge produced by indirect light reaching the PMT is predicted by integrating the product of the direct light emission profile 
and a scattering 
function that has been generated in advance based on simulation and incorporates effects arising from the relative position of the PMT and light source. 
For events with multiple Cherenkov rings the predicted charge of each ring is first calculated separately and then the sum from all the rings is used to calculate the total expected $\mu_i$. 
The final charge likelihood $f_q(q_i|\mu_i)$ is obtained by comparing the observed charge in a PMT against the prediction assuming photoelectrons generated according to Poisson statistics.

The time likelihood term can be expressed as $f_t(t_i|t^\text{exp}_i, \Gamma, p, \mu_i)$, where $p$ is the momentum for the topology $\Gamma$ and 
 $t^\text{exp}_i$ is the expected hit time.
The latter is defined as the arrival time of unscattered photons emitted at the track midpoint and traveling directly to the PMT as
\begin{eqnarray}
t^\text{exp}_i=t+s_{\text{mid}}/c+|\bm{R}_i^{\text{PMT}}-\bm{x}-s_{\text{mid}}\bm{d}|/c_n ,
\label{eq:time}
\end{eqnarray}
\noindent where $\bm{x}$ and $t$ are the vertex and creation time of the particle, respectively, $\bm{d}$ is the particle direction, $\bm{R}_i^{\text{PMT}}$ is the position of the $i$-th PMT and $s_{\text{mid}}$ represents half of the track length. 
Here $c$ and $c_n$ represent the light velocity in vacuum and in water, respectively.
The time likelihood depends on the predicted charge since a hit is recorded by the first photon arriving at a PMT, which leads a narrower distribution of hit time for higher numbers of incident photons and hence, more predicted charge.
The track length of a particle $s$, which is determined by the topology $\Gamma$ and momentum $p$, also affects the shape of time likelihood since not all photons are generated at the track midpoint. 
Contributions to the likelihood from direct and indirect photon hits 
are calculated separately and then merged according to their relative intensities to obtain the final time likelihood. 
The time likelihoods are determined using particle gun simulations.
For multi-particle hypotheses, the time likelihood is calculated ring-by-ring and then merged to a final likelihood function assuming the photons from a particle with earlier $t^\text{exp}_i$ always arrive earlier than the photons from any other particles with later $t^\text{exp}_i$ values.

Once the likelihood function is defined, the best set of kinematic parameters $\hat\theta$ for a given event topology hypothesis 
is defined as that which maximizes $L(\Gamma, \theta)$. 
The best estimate for the particle content of a given event is determined by comparison of $L(\Gamma, \hat{\theta})$ among all hypotheses, $\Gamma$. 

\subsection{Fitting procedure}

The fiTQun reconstruction process can be divided into four steps. 
The first step, vertex pre-fitting, roughly estimates the interaction vertex based on the PMT timing information. 
During the second step, clusters of PMT hits in time are identified as candidate for particle activity.
Thereafter the single-ring reconstruction, which performs fits assuming event topologies with only a single light-producting particle, 
and multi-ring reconstruction, which fits using hypotheses with multiple particles, are performed in sequence. 
During these fits the negative log likelihood -ln $L(\Gamma, \theta)$ is minimized with respect to $\theta$, using the MINUIT\cite{James:1994vla} package.

\subsubsection{Vertex pre-fitting}
The vertex pre-fitter is a fast algorithm which uses only the hit time information from PMTs around the primary event trigger 
to estimate an initial vertex position 
assuming all observed light was emitted from a common point.
This is done by searching for the vertex position $\bm{x}$ and time $t$ which maximize the goodness function 

\begin{eqnarray}
G(\bm{x},t) &=& \sum^{\text{hit}}_{i} e^{(-(T_i^{\text{res}}/\sigma)^2/2)},
\label{fqgood}
\end{eqnarray}
where
\begin{eqnarray}
T_i^{\text{res}} &=& t_i - t - |\bm{R}_i^{\text{PMT}}-\bm{x}|/c_n
\end{eqnarray}

\noindent is the residual hit time calculated on the assumption of a point-like light source and subtracting the photon time of flight. The value of parameter $\sigma$ determines the precision of pre-fitting. 
Here $\bm{R}_i^{\text{PMT}}$ is the position of the $i^{th}$ PMT. 
When the vertex and time get close to their true values by doing a grid search, the $T_i^{\text{res}}$ distribute near zero, which results in a large value of the goodness.
The pre-fitter is executed several times with gradually shrinking the grid size and $\sigma$ to achieve high precision efficiently.
The fitted vertex from this step, called pre-fit vertex, is just a rough estimation and will be fitted again with higher precision
during minimization of $L(\Gamma, \theta)$.

\subsubsection{Hit clustering}
Events in Super-K are defined by detector activity in an $O(10~\mu$s) time window around an event trigger but may contain multiple subevents representing clusters of PMT hits separated in time from the primary trigger. 
A common example is muon decay in which the muon produces the primary event trigger and its decay produces additional delayed detector hits that form a subevent.
A hit clustering algorithm is used to search for activity around the primary trigger and locates any additional subevents 
for further fitting with the more precise reconstruction methods discussed below.

\begin{figure}[htbp]
\centering
    \includegraphics[width=0.5\textwidth]{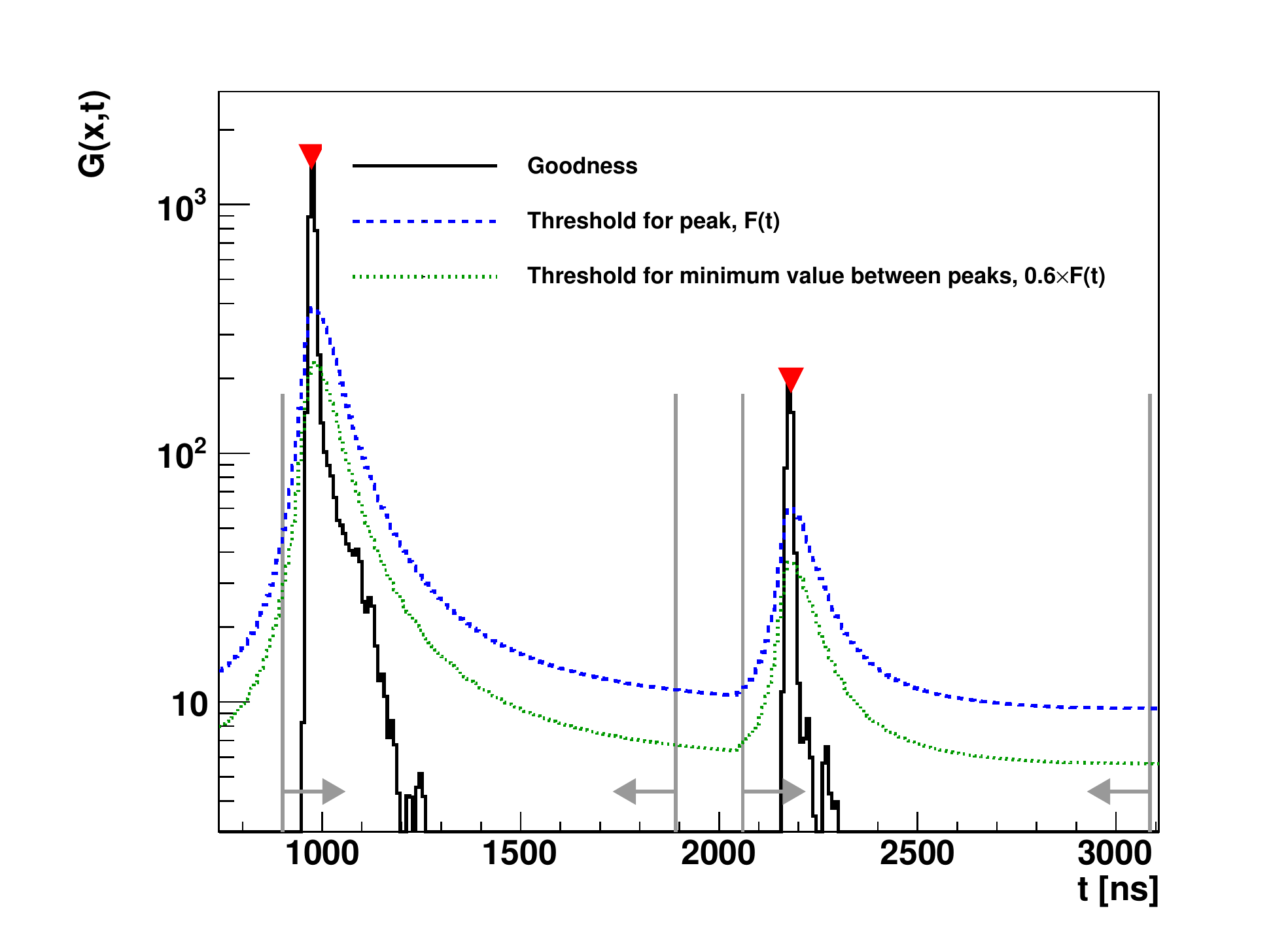}
  \caption{ Goodness as a function of hit time for an event with a muon and a Michel electron.
 The black line shows the goodness as a function of hit time with the vertex fixed to the pre-fit vertex. 
Blue (dashed) and green (dotted) curves denote thresholds for identifying candidate peaks. 
Red triangles denote the time of candidates and gray vertical lines indicate time windows around those candidates 
as determined by the algorithm. 
If such windows overlap, the candidates are merged into one.}
\label{fig:vertexfit}
\end{figure}

The hit clustering algorithm starts by searching in time for subevent activity around the event trigger using a peak-finding algorithm. 
The vertex goodness $G(\bm{x},t)$ from Equation \ref{fqgood} is scanned in $t$ with the vertex position fixed to the pre-fit vertex to 
search for additional peaks from particle activity in the detector.
An example of the goodness distribution from a muon decay event is shown in Figure \ref{fig:vertexfit}.
The two dominant peaks in the distribution correspond to the parent muon and its Michel electron, respectively.
To avoid counting peaks created by scattering or reflection processes within a cluster of hit activity, 
peaks are required to be above a minimum threshold $F(t)$, which is defined as:
\begin{equation}
 F(t):=0.25~{\argmax_{i\in M}}\{\frac{G(\bm{x},t_{i})}{(1+((t-t_{i})/\gamma)^2}\}+\eta
 \end{equation}
where $M$ represents all local maxima of goodness function $G(\bm{x},t)$. The value of time constant $\gamma$ is 25 ns when $t<t_{i}$ and 70 ns otherwise.  An offset $\eta$ = 9 is added to the threshold function to suppress the effect of dark hits. The function $F(t)$ is the blue curve in Figure~\ref{fig:vertexfit}.
The minimum goodness between any two peaks must be lower than a second threshold, $0.6\times F(t)$, which is shown as green dashed curve. 
Under these criteria only the peaks labelled with a triangle are identified as candidates for further fitting.

During this procedure all vertex positions are assumed to lie close to the pre-fit vertex when the peak-finding algorithm runs.
This assumption is broken when the primary particle travels a significant distance from the interaction vertex, as for high momenta muons.
Therefore, the vertex pre-fitting and peak-finding algorithm are rerun after masking the hits caused by the primary particle to improve decay electron reconstruction efficiency. 
The vertex position used in the goodness function $\bm{x}$ is then close to the vertex of the secondary particle.

A time window defined as -180 ns $< T_i^\text{res} <$ 800 ns around each found peak is defined to contain its associated hits.
Afterwards the vertex pre-fitter and peak finder will be run once again in each time window using only the hits within each window. 
Peaks remaining after this step become the final candidates for full event reconstruction.

\subsubsection{Single-ring reconstruction}

\begin{figure*}[htpb]
\subfigure[ Sub-GeV events]{
  \includegraphics[width=0.45\textwidth]{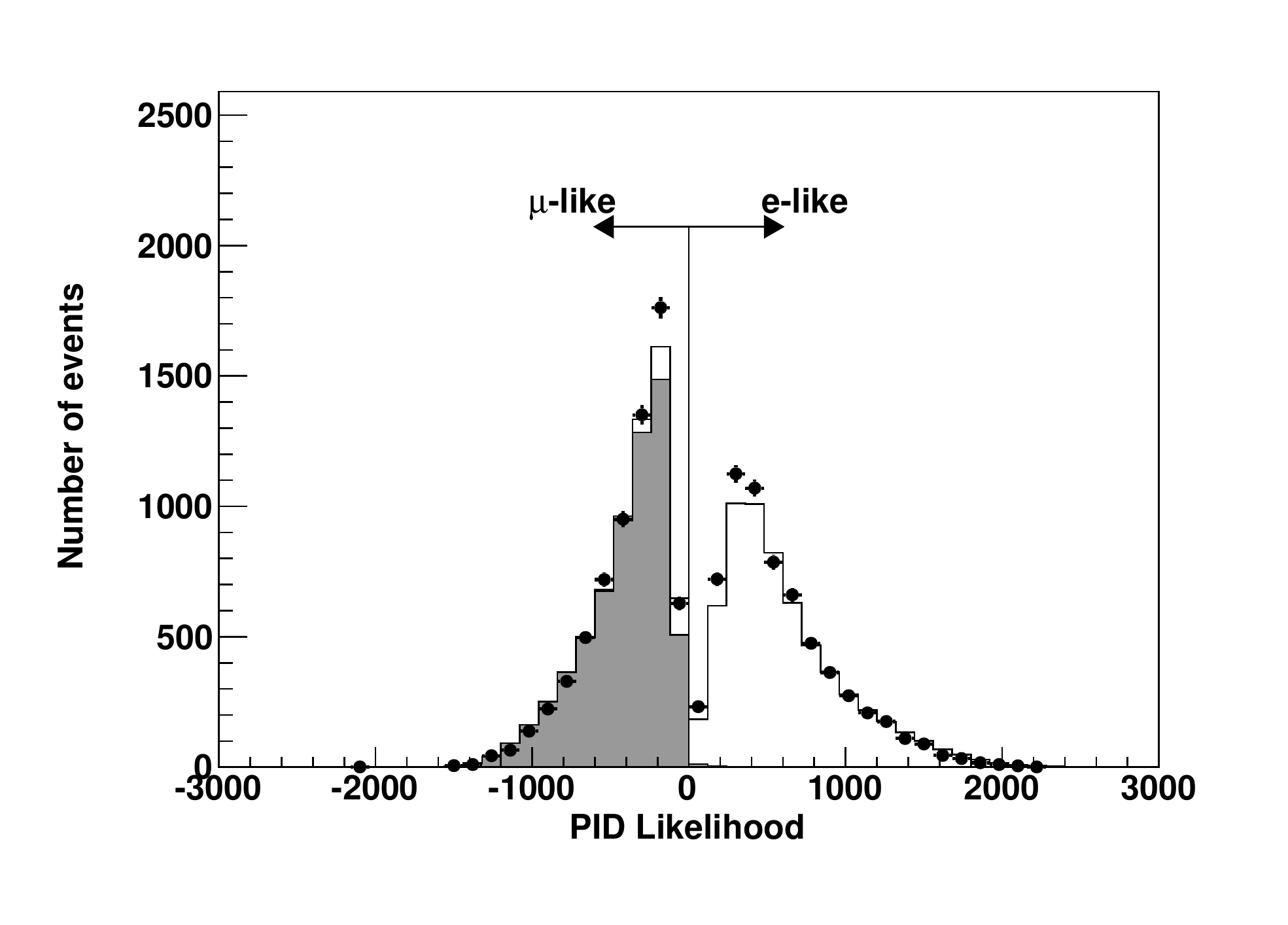}}
  \subfigure[ Multi-GeV events]{
  \includegraphics[width=0.45\textwidth]{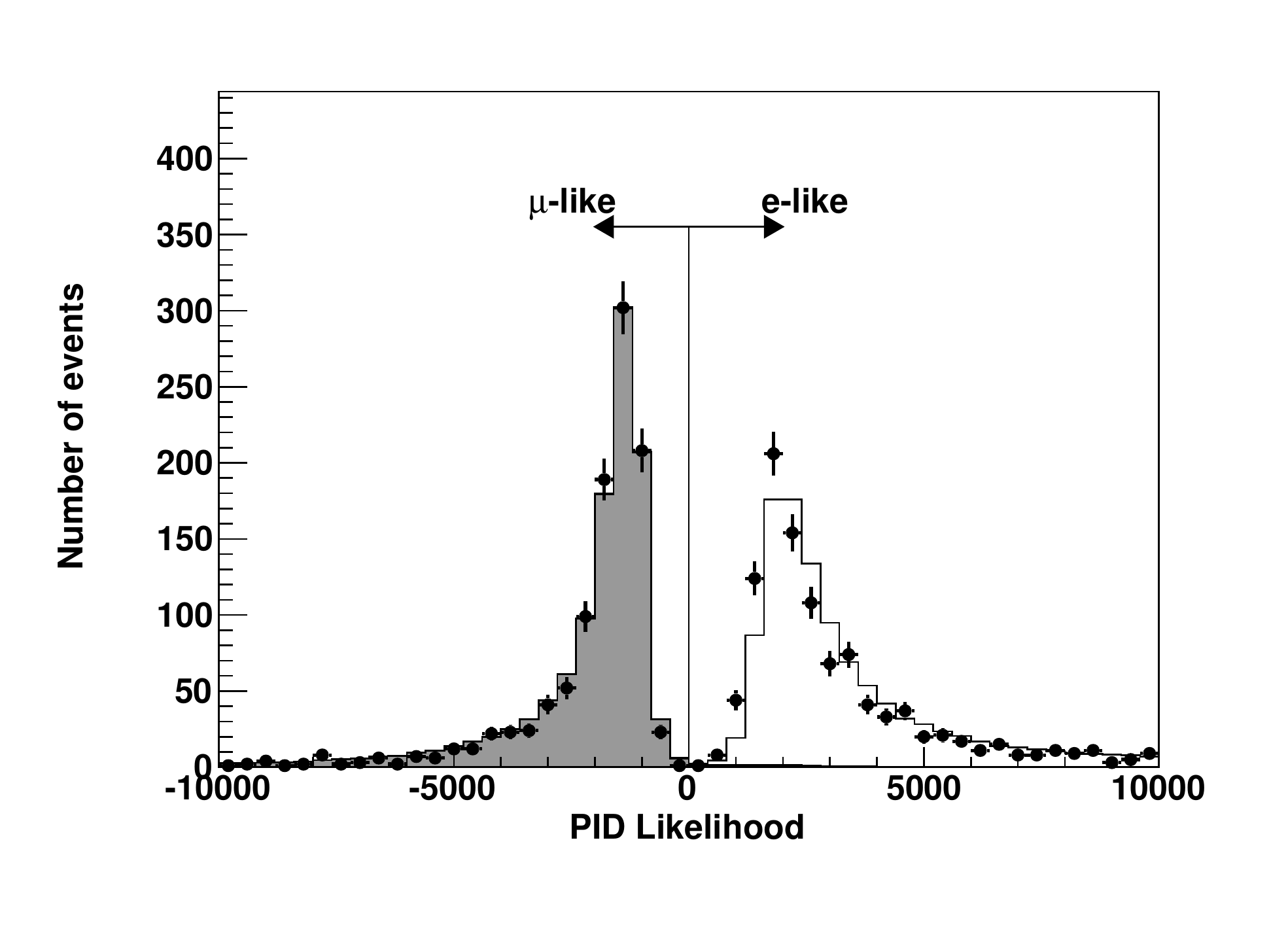}}
  \caption{ PID likelihood distribution for FC single-ring sub-GeV (left) and multi-GeV (right) comparing data (points) and atmospheric neutrino MC (histograms).
  Neutrino oscillations are taken into account with the normal hierarchy assumed and the oscillation parameters taken to be  
            $\Delta m_{23}^{2} = 2.4 \times 10^{-3} \mbox{eV}^{2}$, 
            $\mbox{sin}^{2} \theta_{23} = 0.5$, 
            $\mbox{sin}^{2} \theta_{13} = 0.0210$, and
            $\delta_{CP} = 0 $. The shaded histograms show charged-current $\nu_\mu$ interactions. 
Error bars show the statistical error.
In this figure the reconstructed event vertex is required to be at least 200~cm away from the ID wall.}
\label{fig:1rpid}
\end{figure*}

The most basic reconstruction which is applied to the time windows defined in the previous step is the single-ring fitter.
This fitter poses single-particle hypotheses for the likelihood function in Equation \ref{eq:lklhd}.
Three types of single-ring hypothesis are considered in fiTQun, that of an electron, a muon, and a charged-pion.
For each hypothesis the kinematic parameters of the event, including the interaction vertex, the particle momentum, and 
its direction are varied to maximize the likelihood function against the observation.
Particle identification (PID) is based on the best-fit likelihood values for each of these hypotheses.
Electrons and muons, for example, are separated by cutting on ln($L_e/L_\mu$), 
the logarithm of the likelihood ratio between the best-fit electron and muon hypotheses. 
The distribution of this variable is shown for FC atmospheric neutrino data and Monte Carlo (MC) simulation result with sub-GeV and multi-GeV energies in Figure \ref{fig:1rpid}. 
In both cases there is a clear separation of the likelihood variable between electron-like ($e$-like) and muon-like ($\mu$-like) events.

 \begin{figure*}[htpb]
\centering
\subfigure[ True electron event with single ring]{
  \includegraphics[width=0.45\textwidth]{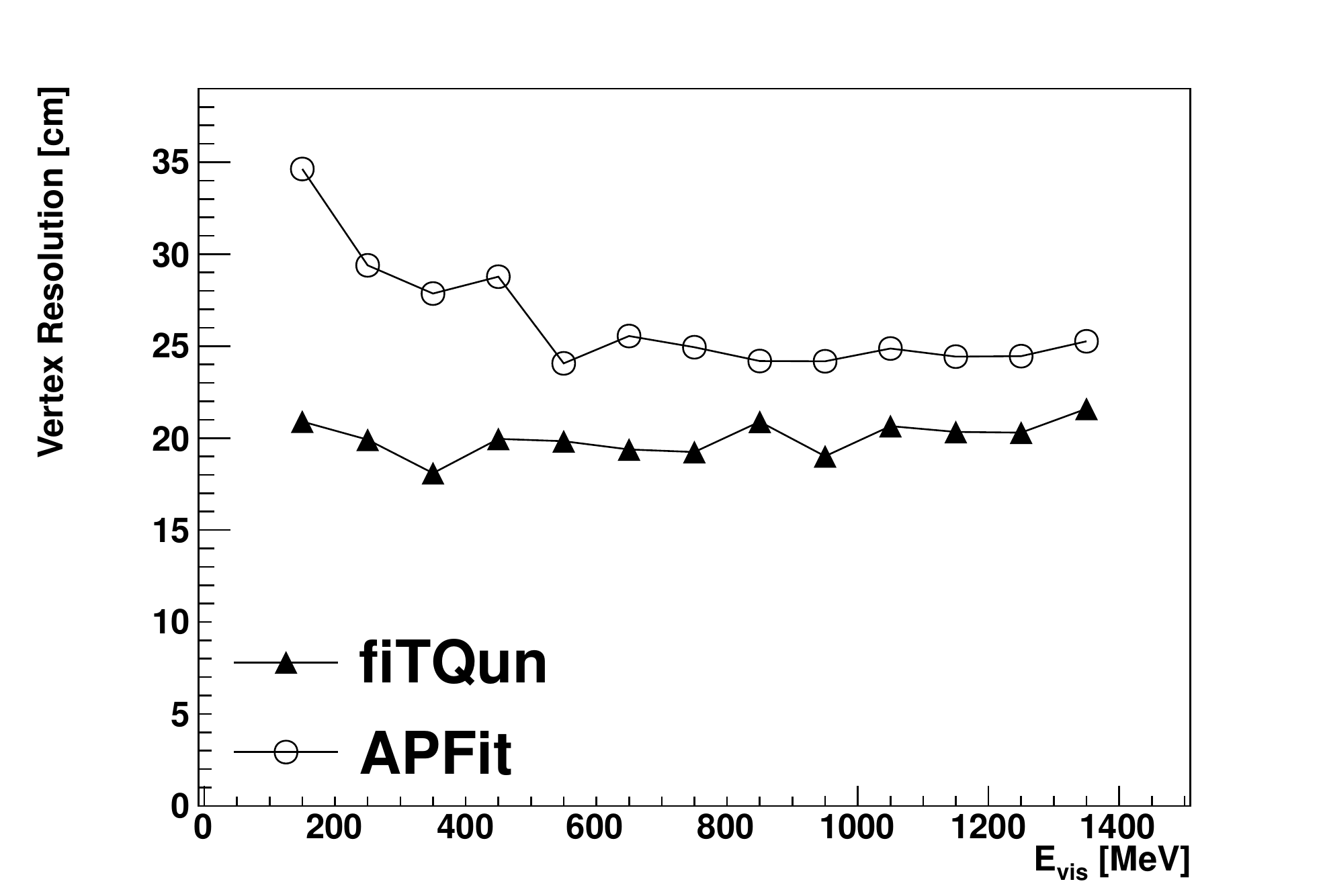}}
  \subfigure[ True muon event with single ring]{
  \includegraphics[width=0.45\textwidth]{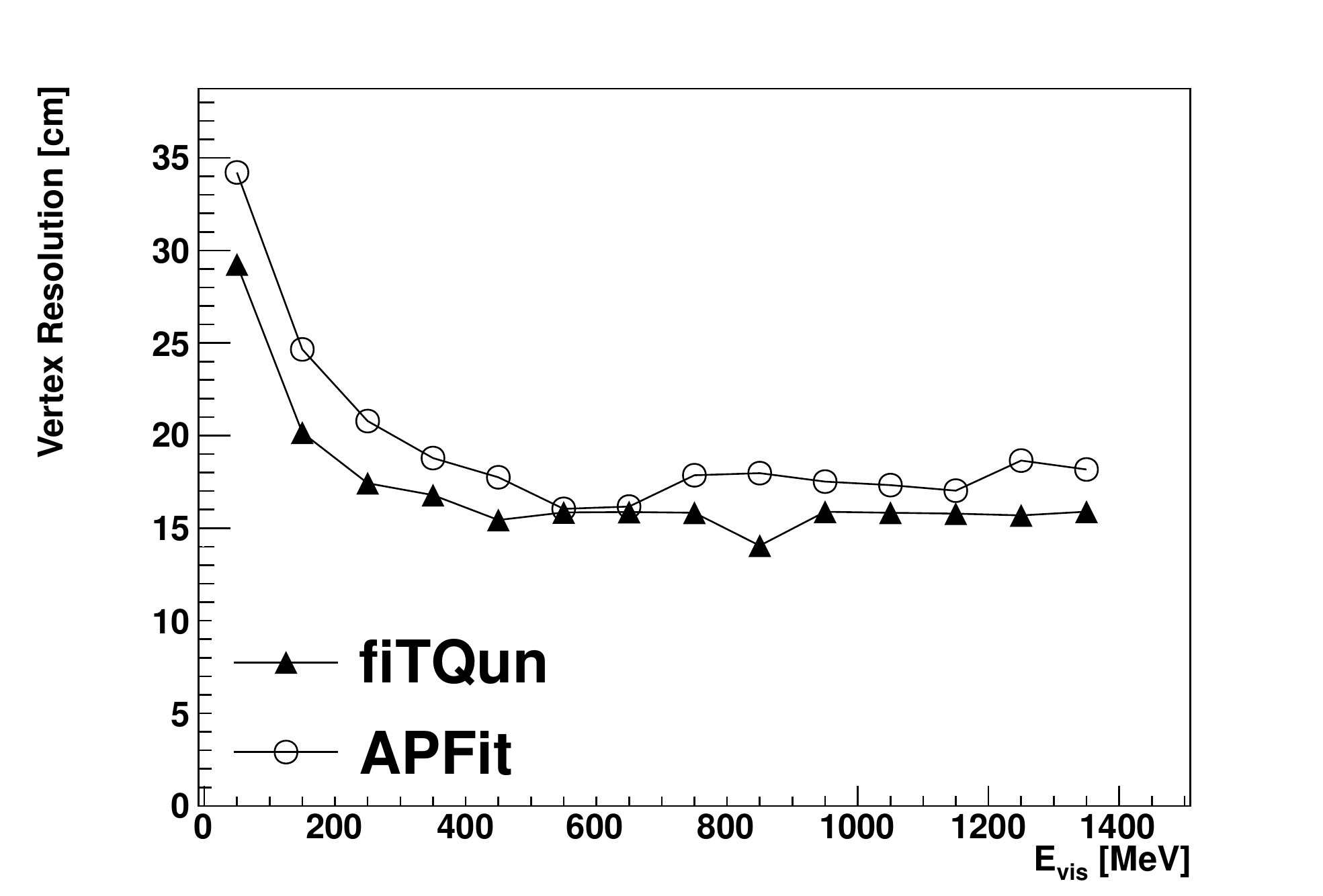}}
     \caption{ Vertex resolution of single-ring electron (left) and muon (right) events in the FC CCQE event sample in the atmospheric neutrino MC, plotted as a function of visible energy. The red full triangle indicate the performance of fiTQun and the black open circle are for APFit. The true event vertex is required to be at least 200~cm away from the ID wall.}
  \label{fig:vexevis}
\end{figure*}

\begin{figure*}[htpb]
\centering
\subfigure[ Momentum resolution of true single-electron events.]{
  \includegraphics[width=0.45\textwidth]{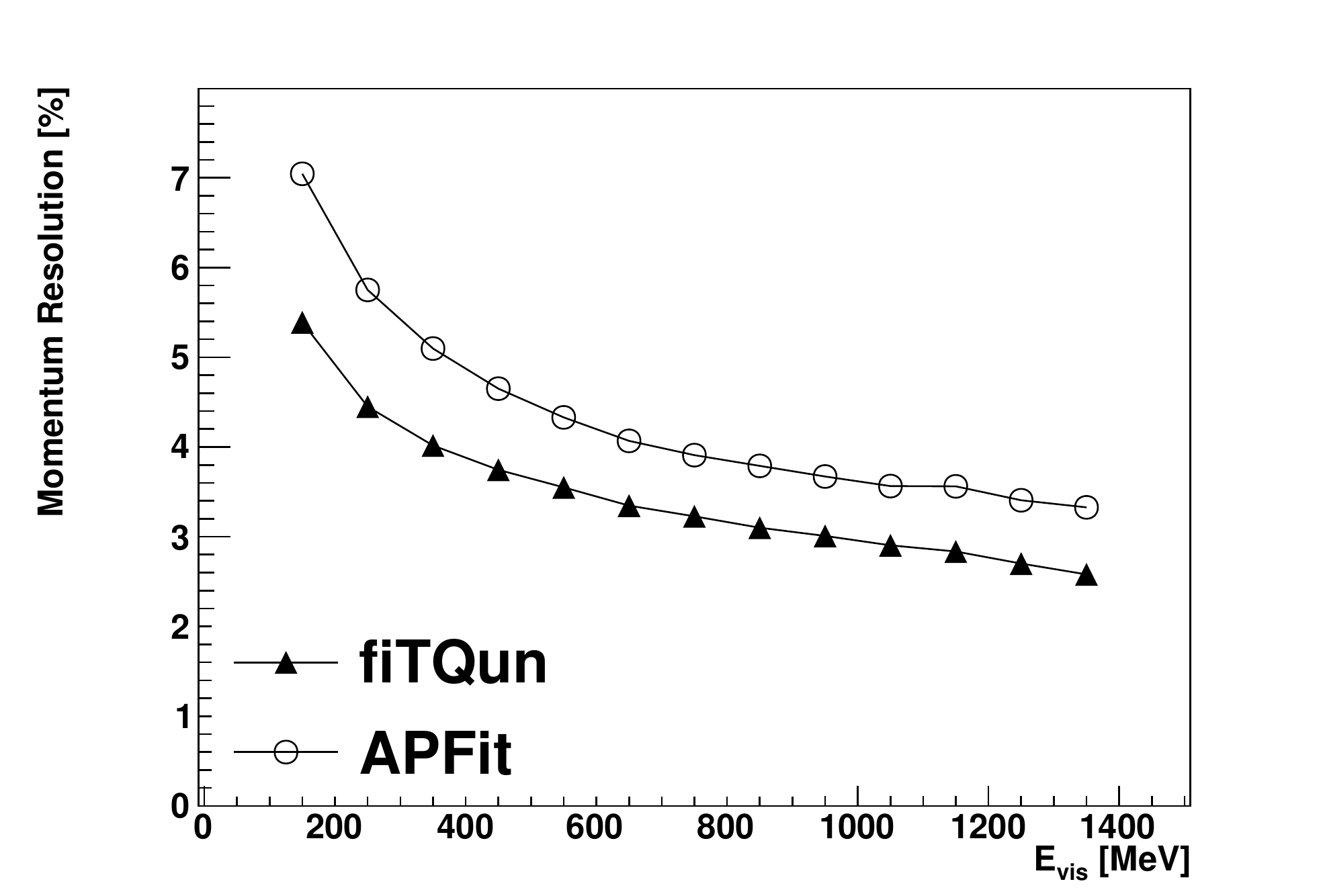}}
  \subfigure[Momentum resolution of true single-muon events.]{
  \includegraphics[width=0.45\textwidth]{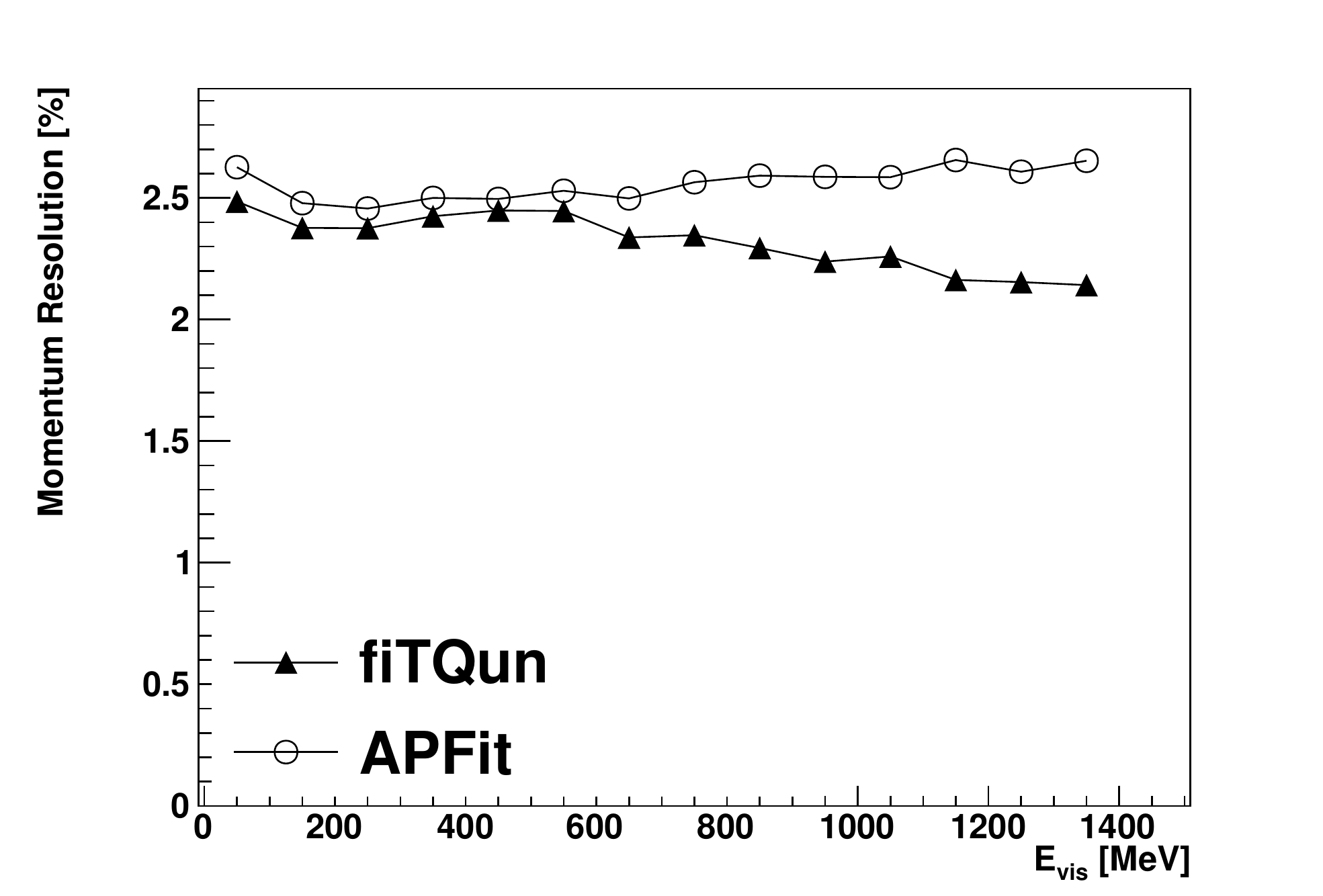}} \\
  \subfigure[ Momentum bias of true single-electron events.]{
  \includegraphics[width=0.45\textwidth]{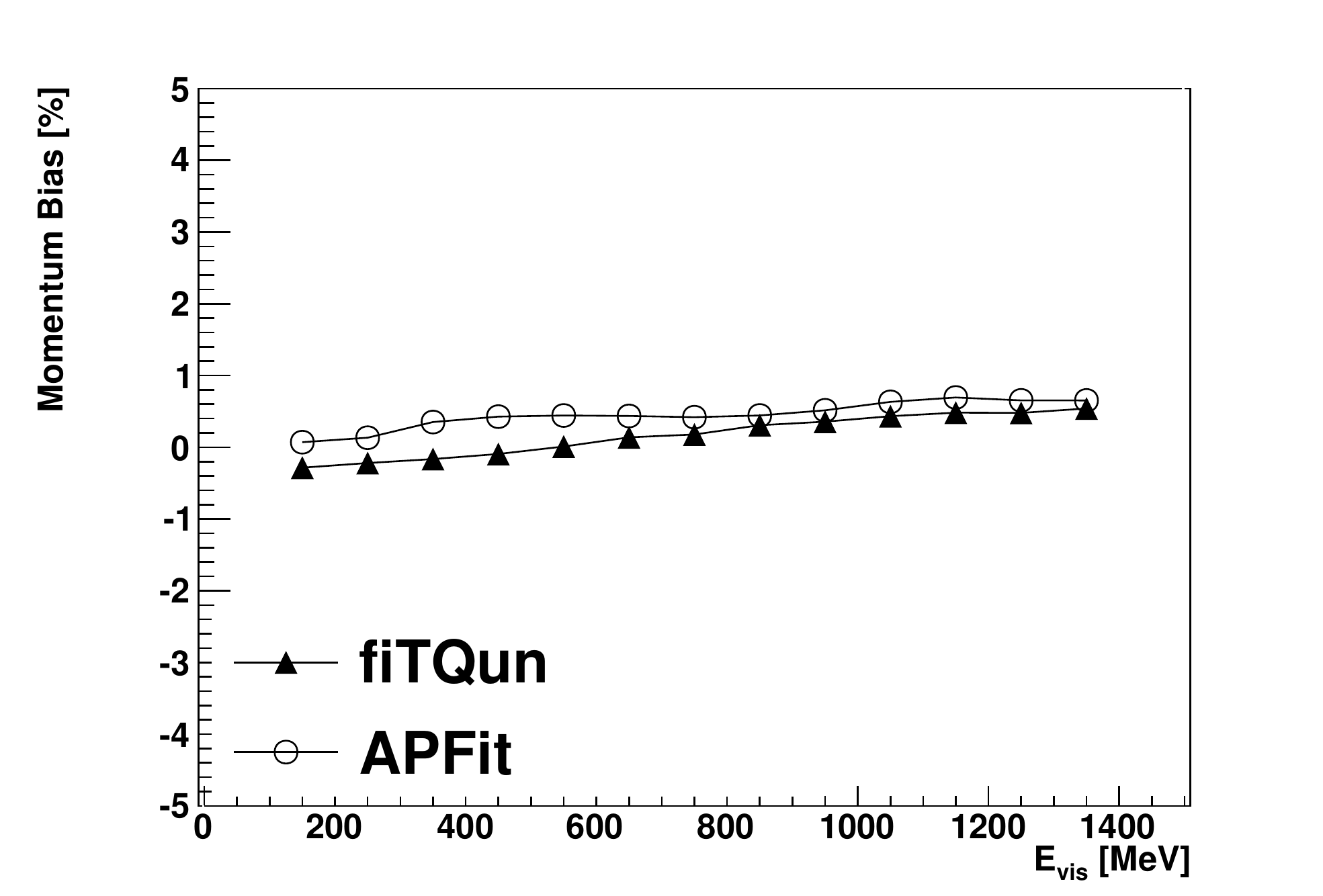}}
  \subfigure[ Momentum bias of true single-muon events]{
  \includegraphics[width=0.45\textwidth]{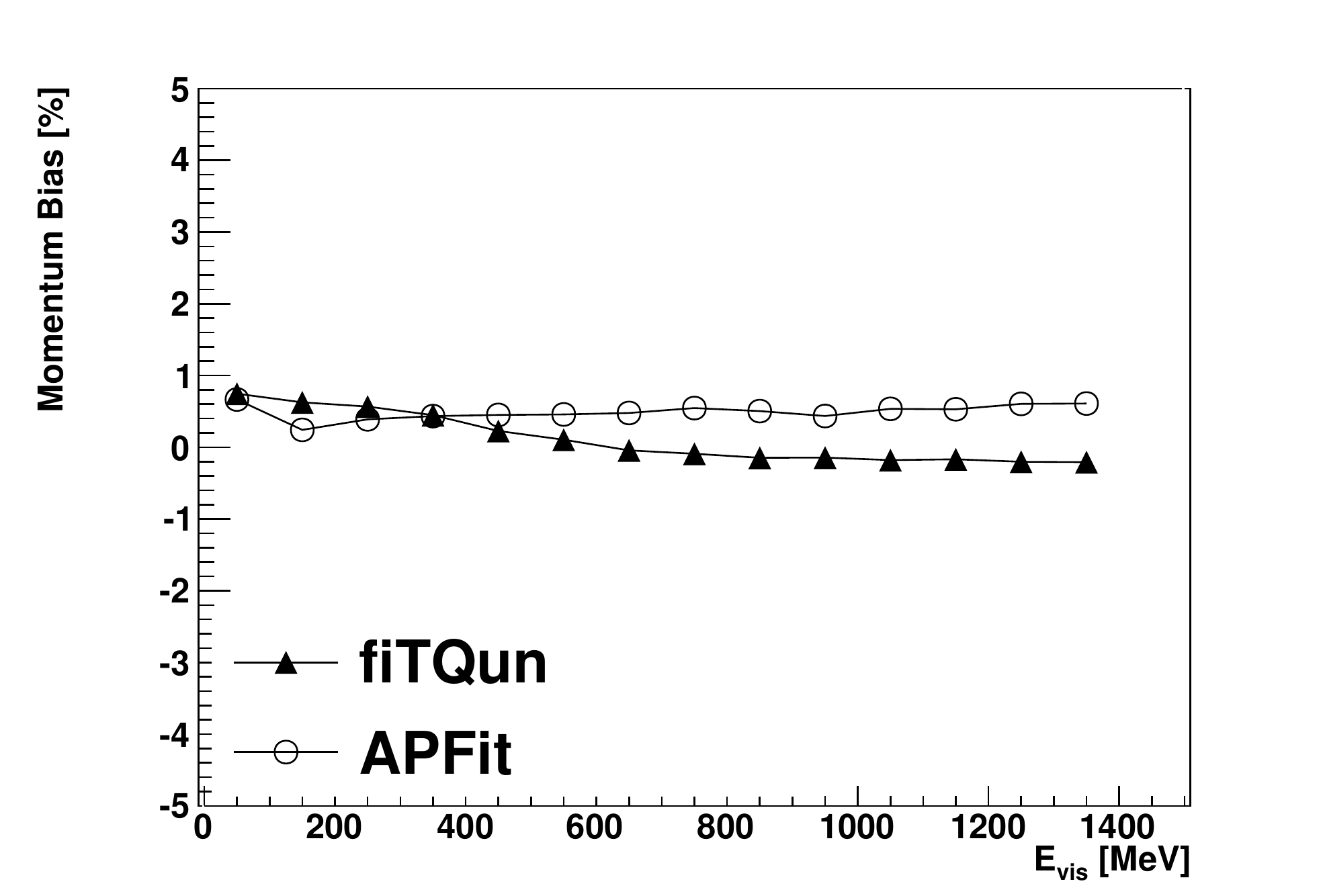}}
      \caption{ Momentum resolution (top) and bias (bottom) of single-ring electron (left) and muon (right) events in the FC CCQE event sample in the atmospheric neutrino MC, plotted as a function of visible energy. The red full triangle indicate the performance of fiTQun and the black open circle are for APFit. The reconstructed event vertex is required to be at least 200~cm away from the ID wall.}
  \label{fig:momevis}
\end{figure*}

\begin{figure*}[htpb]
  \subfigure[True single-electron events.]{
  \includegraphics[width=0.45\textwidth]{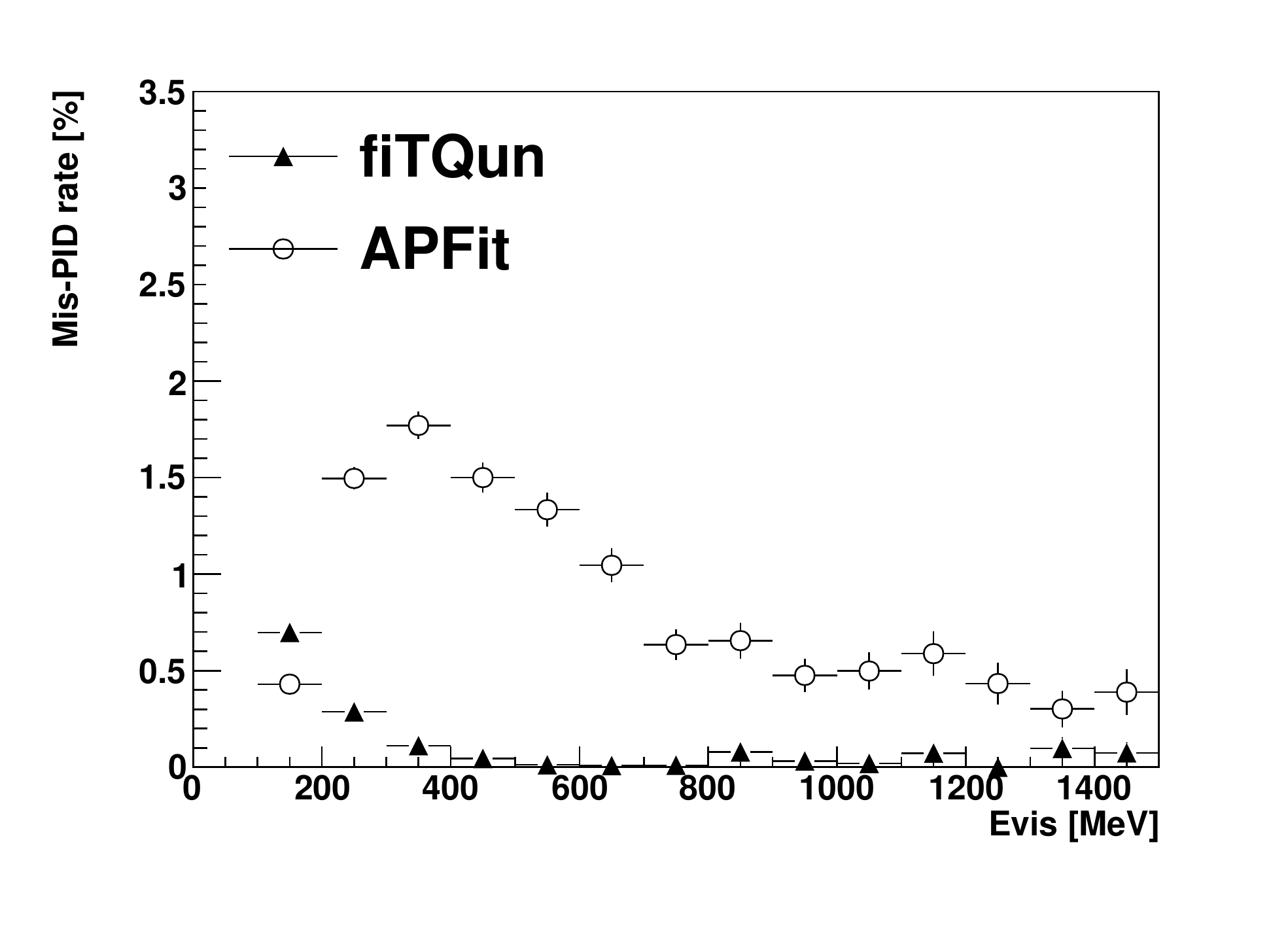}}
  \subfigure[ True single-muon events.]{
 \includegraphics[width=0.45\textwidth]{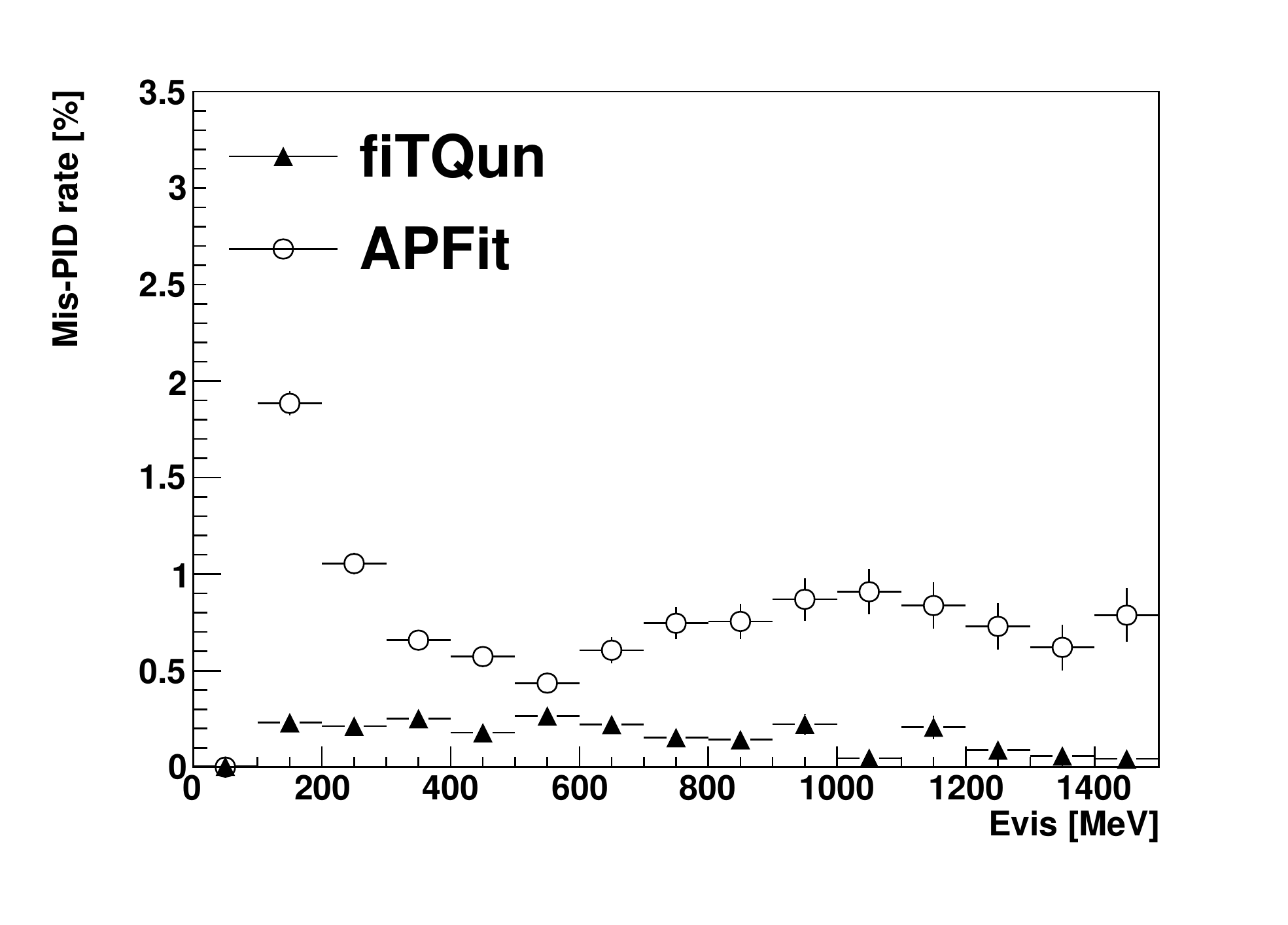}}
  \caption{ Mis-identification rate of single-ring electron (left) and muon (right) events from FC CCQE atmospheric neutrino MC events
plotted as a function of visible energy. 
The open circles indicate the performance of APFit and the triangles are for fiTQun. The reconstructed event vertex is required to be at least 200~cm away from the ID wall.}
  \label{fig:mispid}
\end{figure*}

Figure~\ref{fig:vexevis} shows the vertex resolution as the function of visible energy for the FC single-ring charged-current quasi-elastic (CCQE) event sample in the atmospheric neutrino MC whose true interaction vertex lies within the conventional fiducial volume,  the region located more than 2~m from the ID wall. 
As the visible energy increases from 100 MeV to 1330 MeV 
the vertex resolution of fiTQun for CCQE $\nue$ events is stable at 20.64~cm , 
while the resolution of APFit varies from 34.63~cm to 25.26~cm. For the CCQE $\numu$ events, the vertex resolution of fiTQun, which changes from 29.24~cm to 15.89~cm, is better than that of APFit, which varies from 34.21~cm to 18.16~cm.
In the same energy range, the momentum resolution of fiTQun for CCQE $\nue$ events improves from 5.39\% to 2.58\% as the visible energy increases, while the resolution of APFit changes from 7.04\% to 3.32\%, as shown in the top left plot of Figure~\ref{fig:momevis}. For CCQE numu events the momentum resolution is constant across the energy range, being lower than $2.5\%$ for fiTQun and slightly worse for APFit.
Similarly fiTQun shows an improved ability to discriminate between electrons and muons with less than a 1\% mis-identification rate for visible energies less than 1330 MeV (Figure~\ref{fig:mispid}). 
Other metrics used to measure fit quality are generally stable and their typical values can be found in Table~\ref{tbl:1rperf}. 
In general fiTQun performs as well or better than APFit.
\begin{table}
\begin{center}
\caption{ Summary of the basic performance of the APFit and fiTQun reconstruction algorithms on the fully contained CCQE single ring event sample with visible energy of 1 GeV. }
\label{tbl:1rperf}
\begin{tabular}{lcc}
\hline                                                          
\hline    
Reconstruction 			&  fiTQun  	&   APFit 	\\
\hline                                                             
\multicolumn{3}{l}{\textbf{True CCQE $\nue$ sample}} \\
\hspace*{4pt}Vertex Resolution     	& 20.64~cm  	& 24.86~cm 	\\  
\hspace*{4pt}Direction Resolution & 1.48$^\circ$ 	& 1.68$^\circ$ 	\\
\hspace*{4pt}Momentum Bias 		& 0.43\%  	& 0.63\% 	\\  
\hspace*{4pt}Momentum Resolution     	& 2.90\%  	& 3.56\% 	\\  
\hspace*{4pt}Mis-PID rate	& 0.02\% 	& 0.50\% 	\\
\multicolumn{3}{l}{\textbf{True CCQE $\numu$ sample}}       \\                                                     
\hspace*{4pt}Vertex Resolution 	& 15.83~cm  	& 17.32~cm 	\\ 
\hspace*{4pt}Direction Resolution & 1.00$^\circ$ 	& 1.28$^\circ$ 	\\
\hspace*{4pt}Momentum Bias 	& -0.18\%  	& 0.54\% 	\\  
\hspace*{4pt}Momentum Resolution   	& 2.26\%  	& 2.60\% 	\\  
\hspace*{4pt}Mis-PID rate 	& 0.05\% 	& 0.91\% 	\\
\hline 
\hline 
\end{tabular}
\end{center}
\end{table}

\subsubsection{Multi-ring reconstruction}

\begin{figure*}[htpb]
    \subfigure[ Sub-GeV events]{
  \includegraphics[width=0.45\textwidth]{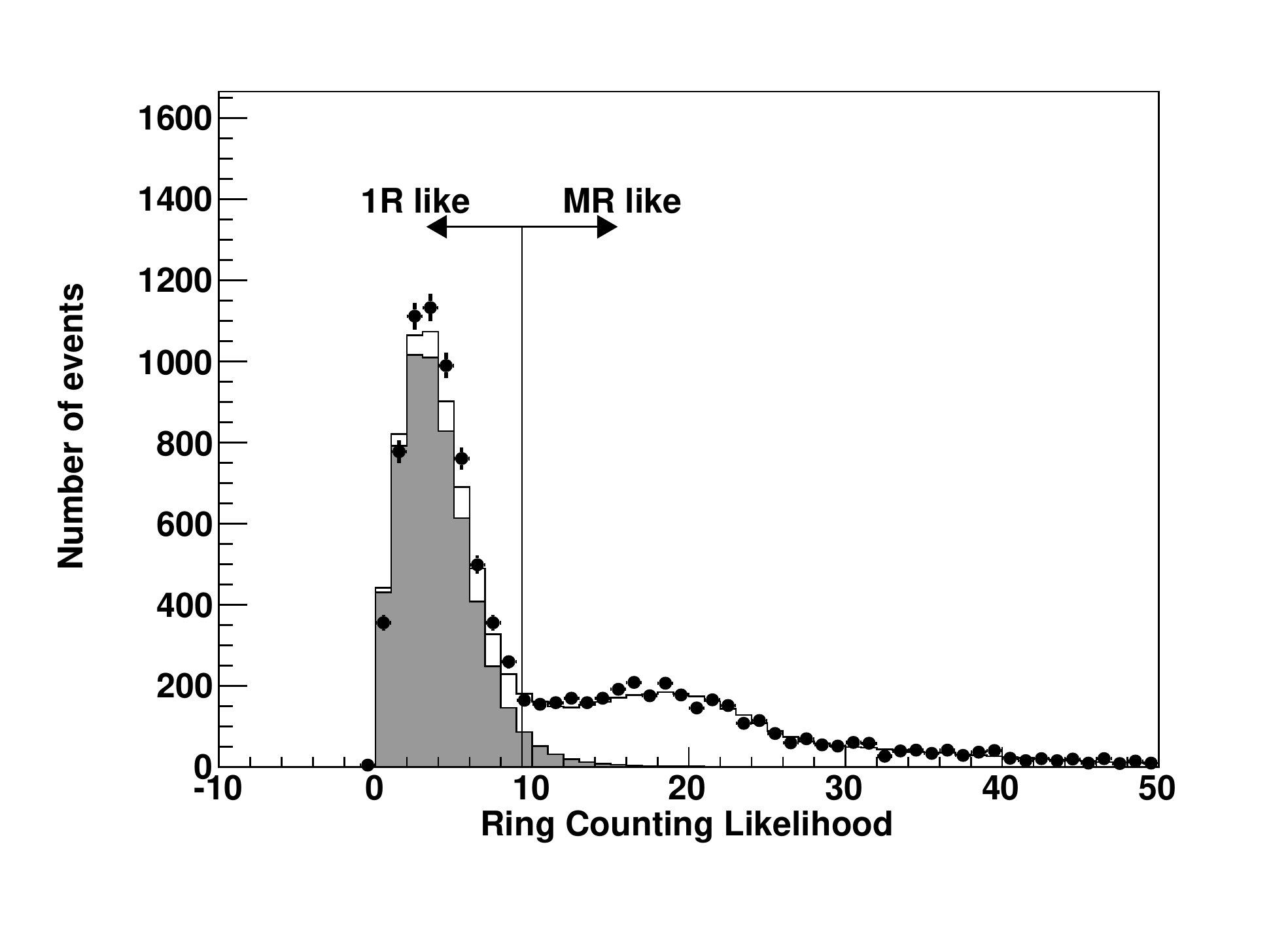}}
    \subfigure[ Multi-GeV events]{
  \includegraphics[width=0.45\textwidth]{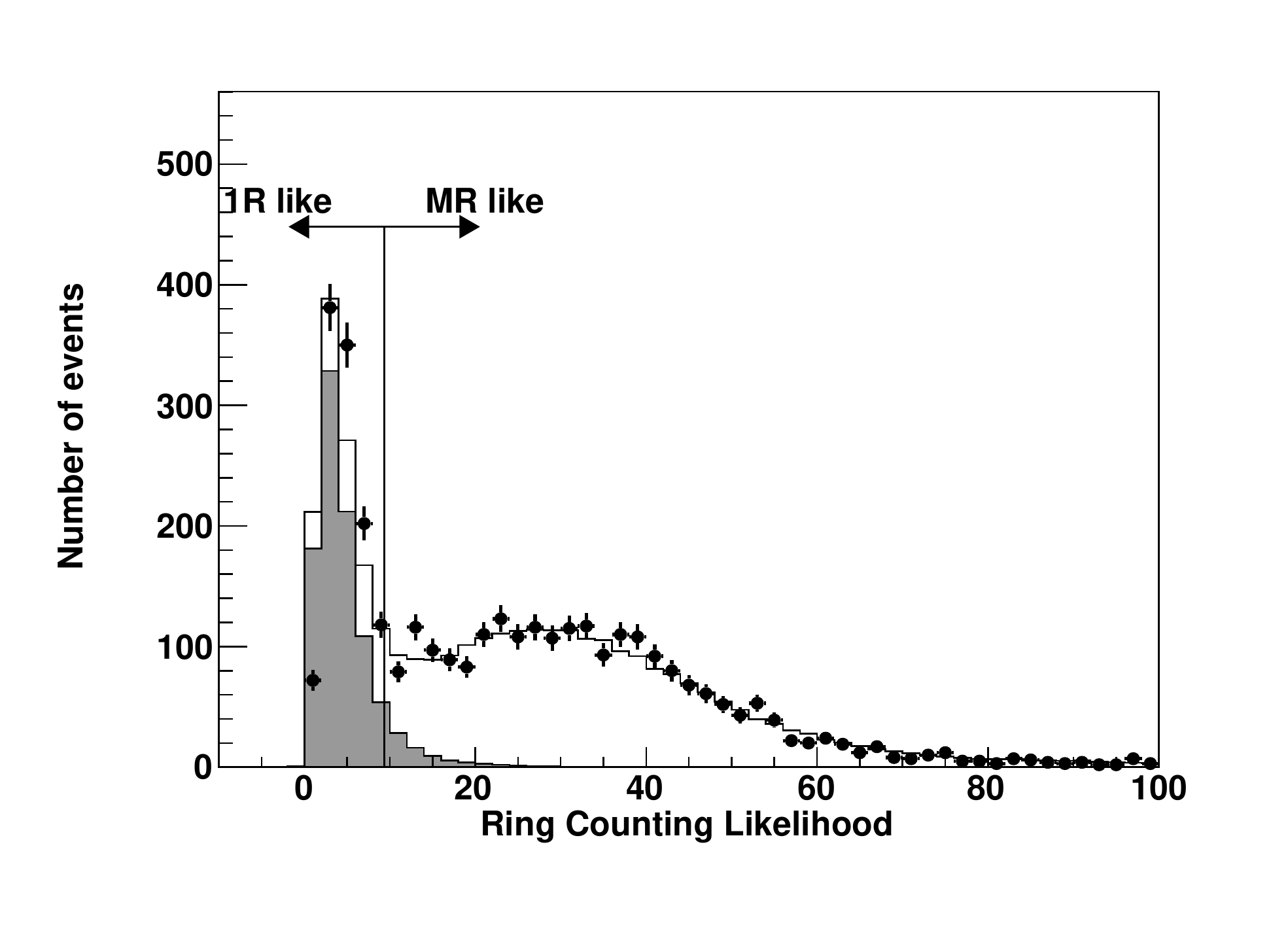}}
  \caption{ The distribution of the likelihood ratio between the best-fit single-ring hypothesis and multi-ring hypothesis 
for sub-GeV (left) and multi-GeV (right) $e$-like events. 
The points show fully contained atmospheric neutrino data and the histograms show the MC prediction including 
neutrino oscillations. 
The shaded histograms show events with one a single-ring in the final state.  
Error bars show the statistical error. 
 The terms 1R and MR represent single-ring and multi-ring, respectively. The reconstructed event vertex is required to be at least 200~cm away from the ID wall.}
\label{fig:rc}
\end{figure*}

In atmospheric neutrino analyses it is essential to reconstruct events with multiple light-producing particles 
since a large fraction of events with multi-GeV energies have multi-particle final states. 
To save computing time, the fiTQun multi-ring fitter is applied only to the time window around the primary event trigger 
and not to any additional subevents.
The process of reconstructing a multi-ring event starts by performing an iterative search for an additional ring on 
top of any existing rings from a previous fit result.
The hypothesis in the likelihood from Equation~\ref{eq:lklhdmu} is updated 
to include a new ring 
 and minimized again allowing the kinematic parameters for the rings to vary.
 Three hypotheses, including $e$-like, $\mu$-like and $\pi^+$-like rings are tested.
The result of the updated fit for each of the particle hypotheses is 
compared to the original result to determine the validity of the added ring. 
If the likelihood improves with the new ring it is accepted and the processes to search for further rings is repeated.
The cycle of adding, fitting and examining new rings iterates until either a newly added ring fails the likelihood criterion 
or six rings are found in the event. 
Figure \ref{fig:rc} shows the difference in the best-fit likelihoods from the 2-ring and a 1-ring hypotheses. 
Based on an optimization performed in MC, a cut at 9.35 (11.83) on this likelihood ratio is used to separate single- and multi-ring event when the first ring hypothesis is $e$-like ($\mu$-like).
However, ring candidates that have an angular separation of less than $20^\circ$ from the most energetic ring are discarded as spurious since MC studies indicate such candidates are typically due to the particle scattering rather than a new particle.
When this occurs the two rings are merged and refit as one for all particle hypotheses while keeping all other rings fixed. 
This procedure is repeated for all rings in the event in descending order of their energy. 

Table \ref{tbl:rcperf} shows the performance of the ring counting in both APFit and fiTQun using atmospheric neutrino MC events. 
In this comparison only charged particles with energy more than 30 MeV above Cherenkov threshold in the final states without 
any requirement on their angular separation with other particles are considered as true ring candidates.
Compared to APFit, fiTQun shows a greater ability to reconstruct multi-ring events, while the ability to correctly identify 
single-ring topologies is the same in both algorithms.
Although fiTQun tends to have more fake rings than that of APFit, the improvements largely outweigh this slight downside. 
\begin{table}
\begin{center}
\caption{ Summary table of ring counting performance of both APFit and fiTQun on fully contained atmospheric neutrino events.
 Columns denote the number of reconstructed rings and rows the number of true rings. 
 The true number of rings is counted using only particles with energy more than 30 MeV above Cherenkov threshold in the final states.
 The terms 1R, 2R, and 3R correspond to one ring, two rings, and three or more rings.}
\label{tbl:rcperf}
\begin{tabular}{l|ccc|ccc}
\hline                                                                                                        
\hline                                                                                                        
 True Number  	&  \multicolumn{3}{c|}{fiTQun Reconstruction}  &   \multicolumn{3}{c}{APFit Reconstruction} 	 	\\
 of Rings 	& 1R 	& 2R 	& $\geq$ 3R                 & 1R 	& 2R 	& $\geq$ 3R 			\\
\hline                                                                                                        
True 1R 	&95.0\%     	&4.64\%        	&0.41\%       &95.9\%        	&3.85\%        	&0.29\%    	  \\
True 2R 	&27.8\%     	&66.7\%        	&5.56\%       &42.5\%        	&52.8\%        	&4.63\%    	  \\
True $\geq$3R 	&7.04\%     	&25.5\%        	&67.5\%	      &20.2\%        	&33.0\%        	&46.8\%   	  \\
\hline 
\hline 
\end{tabular}
\end{center}

\end{table}

\begin{figure*}[htpb]
  \subfigure[ Sub-GeV events]{
  \includegraphics[width=0.45\textwidth]{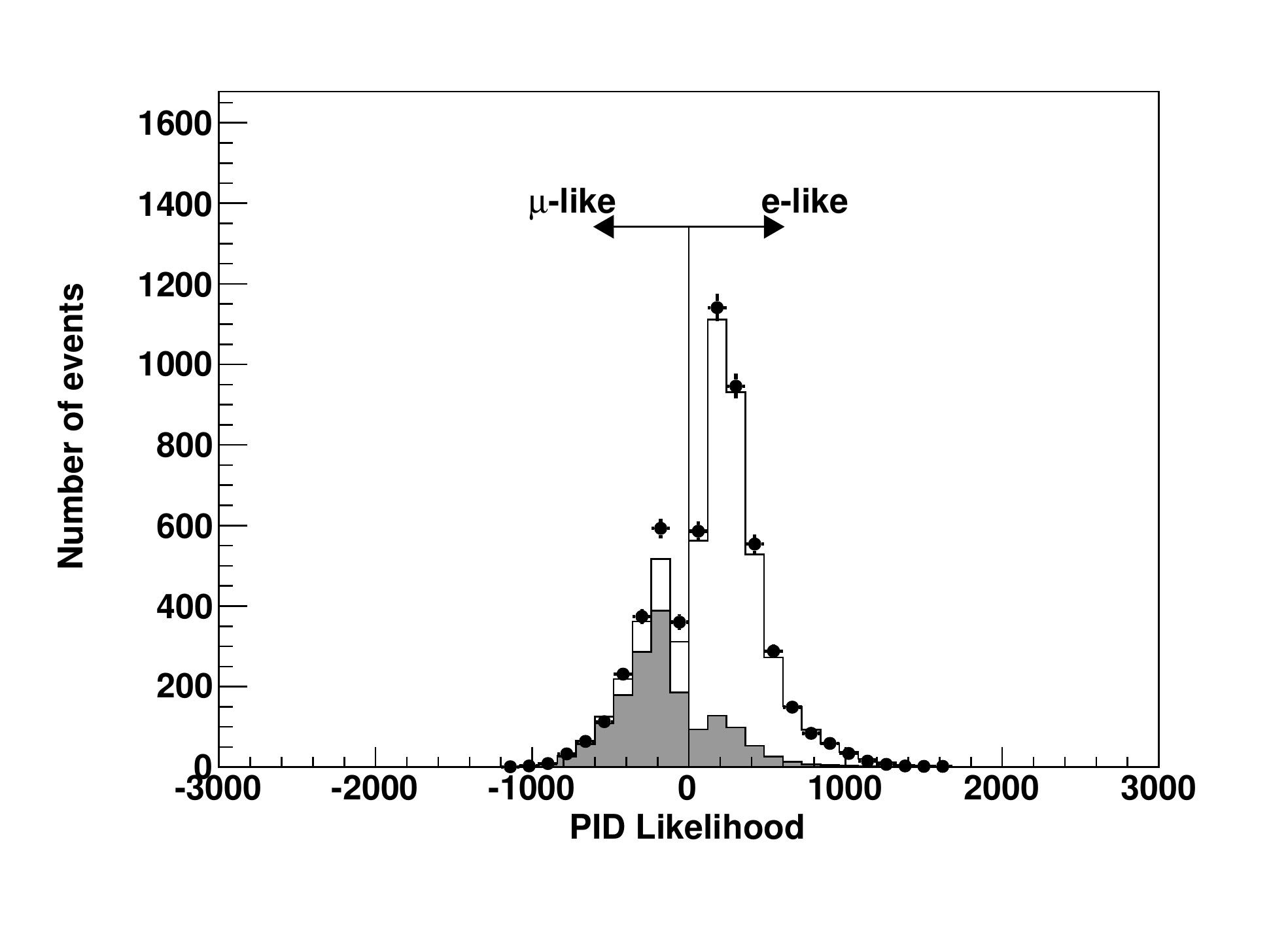}}
    \subfigure[ Multi-GeV events]{
  \includegraphics[width=0.45\textwidth]{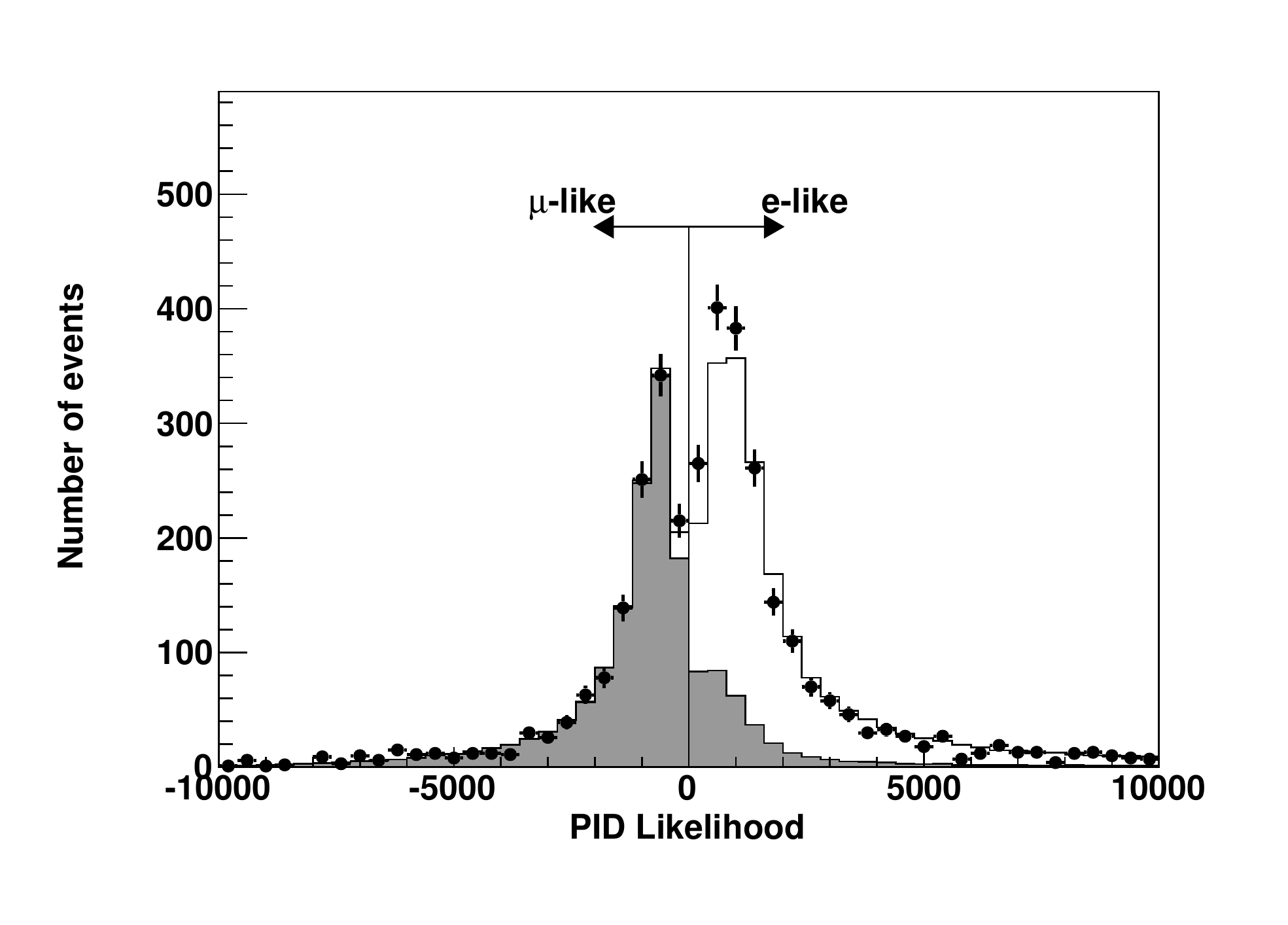}}
  \caption{ PID likelihood distributions for the most energetic ring in sub-GeV (left) and multi-GeV (right) fully contained multi-ring events. 
Points show the data and the histograms show the atmospheric neutrino MC including neutrino oscillations. 
The shaded histograms show $\nu_\mu$ charged-current interactions. 
Error bars are statistical. The reconstructed event vertex is required to be at least 200~cm away from the ID wall.}
\label{fig:mrpid}
\end{figure*}

Figure \ref{fig:mrpid} shows the PID likelihood variable distribution of the most energetic ring in fully contained multi-ring events.
 Due to the overlap of Cherenkov photons from multiple particles, the separation between $e$-like and $\mu$-like is 
not as good as that for single-ring events.

\begin{figure*}[htpb]
    \subfigure[ Event selected by fiTQun]{
  \includegraphics[width=0.45\textwidth]{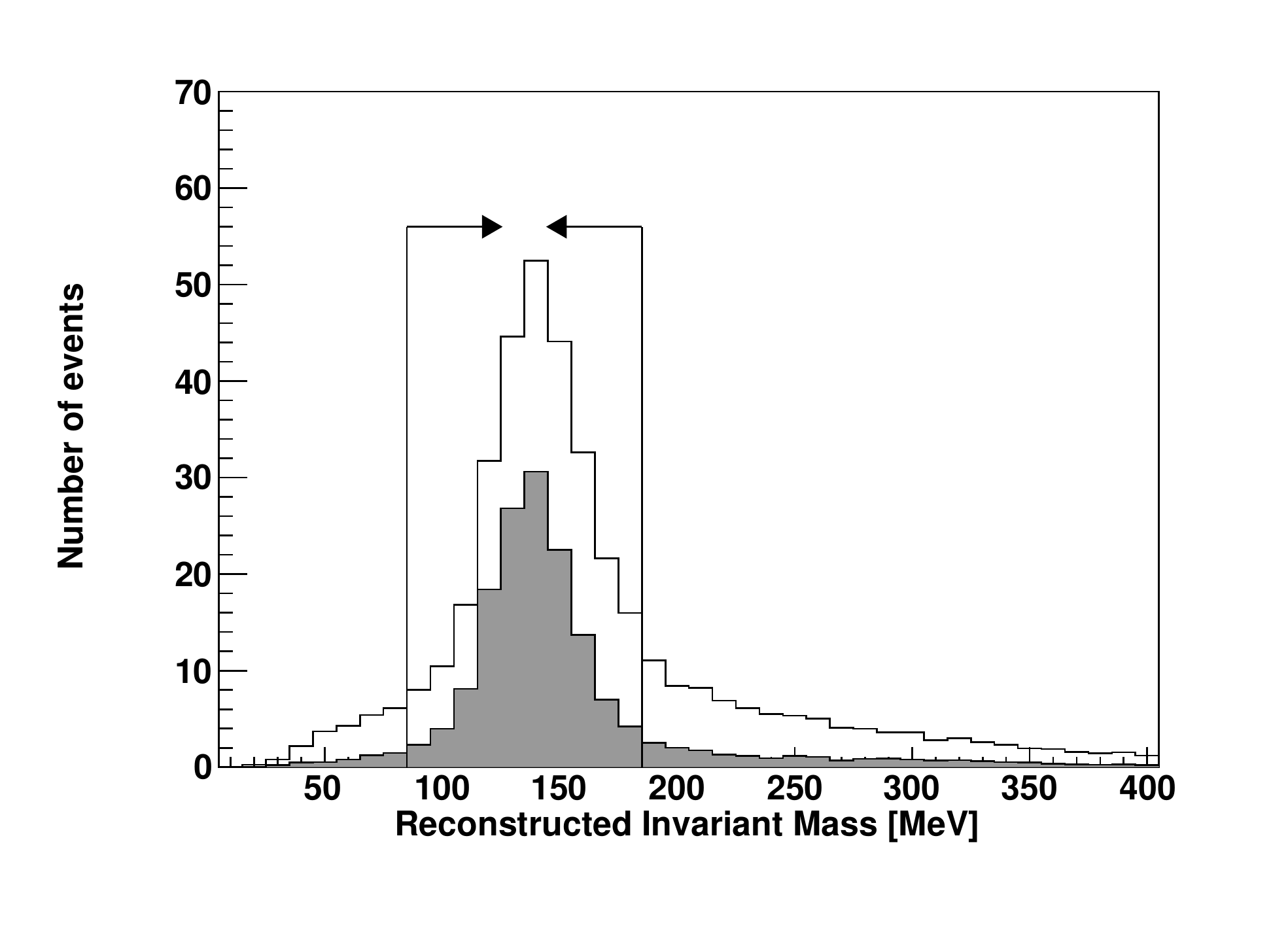}}
    \subfigure[ Event selected by APFit]{
  \includegraphics[width=0.45\textwidth]{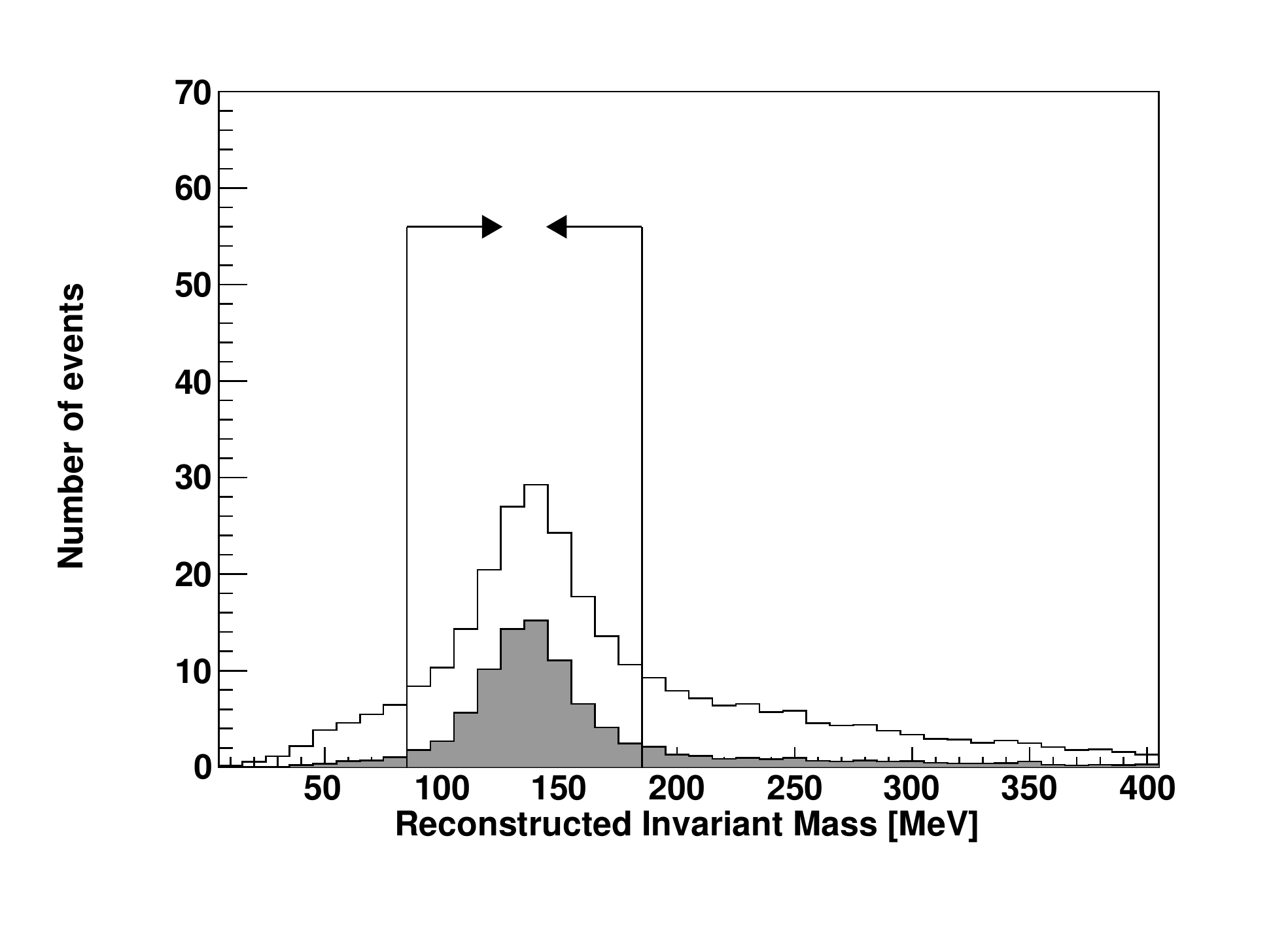}}
    \caption{ Reconstructed invariant mass calculated using the second and the third most energetic rings for events with three e-like rings by fiTQun (left) and APFit (right) from the sub-GeV atmospheric neutrino MC.
Shaded histograms show true CC$\nue 1\pi^{0}$ events and arrows indicate the invariant mass range used 
to select events in which a $\pi^{0}$ is reconstructed properly. The reconstructed event vertex is required to be at least 200~cm away from the ID wall.}
  \label{fig:pi0mass_eee}
\end{figure*}

The performance of the fitter's kinematic reconstruction on multi-ring events is checked by studying the 
invariant mass of $\pi^{0}$s in charged-current single pion (CC1$\pi^0$) events. 
Figure \ref{fig:pi0mass_eee} shows the reconstructed invariant mass 
calculated using the second and the third most energetic rings for events with three $e$-like rings ($eee$ events) for both 
reconstruction algorithms.
Shaded histograms show true CC$\nue 1\pi^{0}$ events.
Both fitters have a peak in the invariant mass distribution near the $\pi^{0}$ mass, but the 
fiTQun peak is larger, sharper, and has fewer backgrounds indicating improved reconstruction efficiency over APFit. 
The FWHM of the $\pi^0$ mass peak selected by fiTQun is 38 MeV, while this value is 42 MeV for APFit. 
The event rate and the purity of the CC$\nue 1\pi^{0}$ events with $85 < m < 185$~MeV (indicated by the arrows in the figure) are 278.5 and 52.1\% for fiTQun, which are much higher than those values of APFit: 175.9 and $42.9\%$.
Similarly, CC$\numu 1\pi^{0}$ events from the three ring sample whose leading ring is $\mu$-like and passing this selection 
are also studied and summarized in Table \ref{tbl:pi0mass_eee}.
The fiTQun algorithm demonstrates a higher efficiency and purity for multi-ring event reconstruction.

\begin{table}
\begin{center}
\caption{ Summary table of performance of CC1$\pi^0$ events selection in three ring events.
The terms $eee$ and $\mu ee$ represent the reconstructed PID of the three rings in order of momentum.
 The event rate column shows the number of events passing the invariant mass cut to select $\pi^0$. 
The purity column shows the purity of target events in the passed invariant mass region. }
\label{tbl:pi0mass_eee}
\begin{tabular}{lcccc}
\hline 
\hline 
\multirow{2}{*}{Topology} 	&  \multicolumn{2}{c}{fiTQun selection} & \multicolumn{2}{c}{APFit selection} 	  \\
 	& Event rate 	& Purity 	& Event rate 	& Purity \\
\hline 
\multicolumn{5}{l}{\textbf{$\bm{eee}$ events} (target: $\nue$CC1$\pi^0$)} \\
\hspace*{4pt} Sub-GeV 		 	& 278.5 	& 52.1\% & 175.9 	& 42.9\%\\
\hspace*{4pt} Multi-GeV 	 	& 112.1 	& 47.0\% & 32.8 		& 36.2\%\\
\multicolumn{5}{l}{\textbf{$\bm{\mu ee}$ events} (target: $\numu$CC1$\pi^0$)} \\
\hspace*{4pt} Sub-GeV 		& 384.5 	& 54.6\%& 201.2 	& 38.0\% 	 \\
\hspace*{4pt} Multi-GeV 	& 143.5 	& 64.6\%& 51.6 		& 32.0\% 	 \\

\hline 
\hline 
\end{tabular}
\end{center}

\end{table}

\begin{figure}
\centering
    \includegraphics[width=0.5\textwidth]{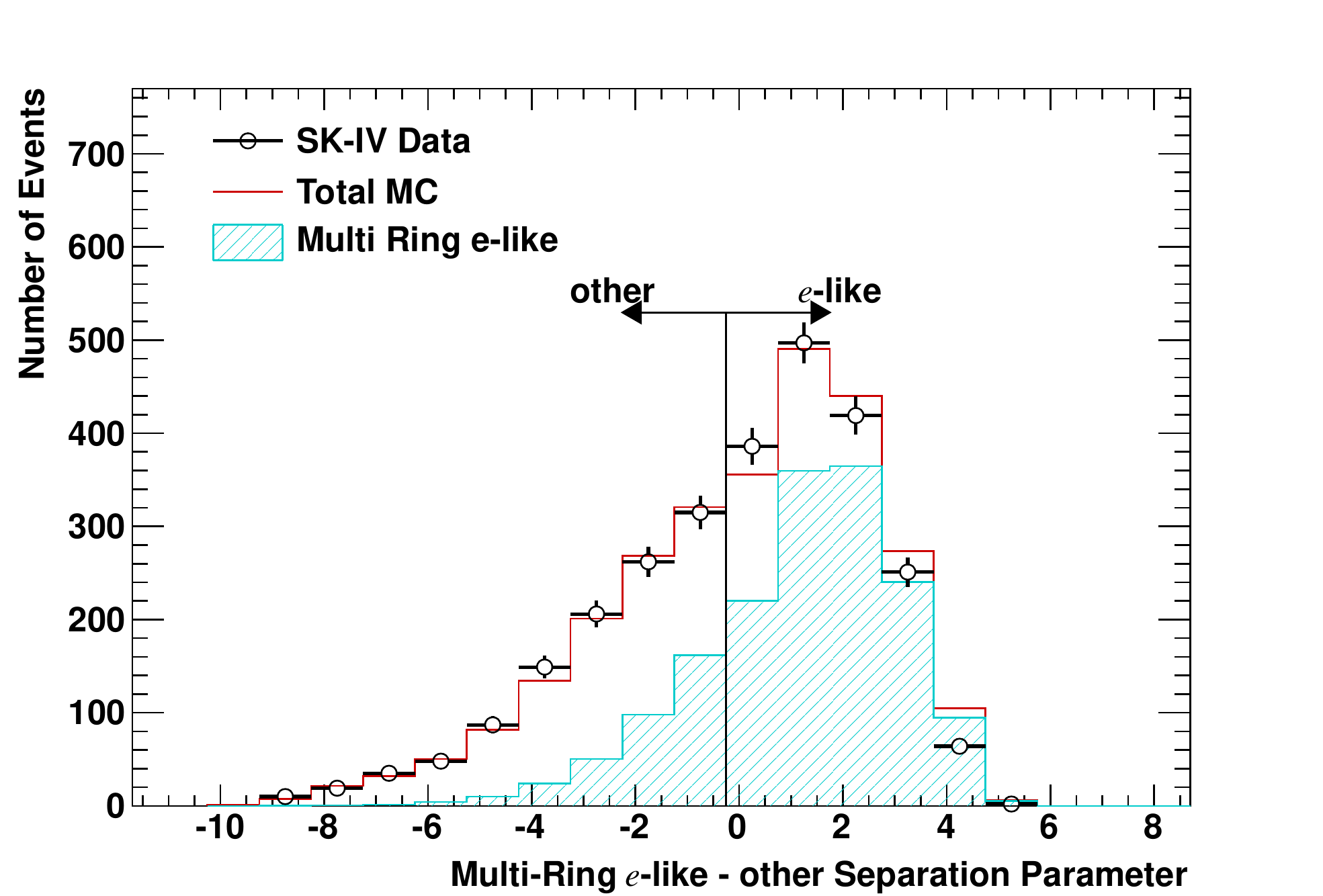}
    \caption{ Likelihood distribution used to separate SK-IV multi-ring events whose most energetic ring 
              is e-like.
              Error bars represent the statistical uncertainty of the data.
              Events with likelihood values larger than -0.25 are designated multi-ring e-like, while 
              those with lower values are termed multi-ring other. }
  \label{fig:mmeother}
\end{figure}

Another important performance indicator for multi-ring event reconstruction algorithm 
is the separation of neutrino and antineutrino components of the atmospheric neutrino sample, 
since at multi-GeV energies they have the most sensitivity to the mass hierarchy. 
A two-stage likelihood method has been developed to purify the neutrino and antineutrino components for multi-ring events.
 The first stage of the separation is designed to extract CC $\nu_{e}+\bar{\nu_{e}}$ interactions based on a likelihood selection as in~\cite{Wendell:2010md} using the APFit algorithm.
 However, in the present study the inputs have been replaced by the equivalent variables from the fiTQun reconstruction.
 Events that pass this selection are classified as ``multi-ring $e$-like'' while those that fail are termed ``multi-ring other'' as shown in Figure \ref{fig:mmeother}. 
Both are used in the oscillation analysis discussed below.

\begin{table}
\begin{center}
\caption{ Summary of the Multi-ring e-like event selection.
The efficiency represents the fraction of events passing the selection among all true charged-current $\nue$ events.
Similarly, the purity is the fraction true CC $\nue$ out of all events passing the selection. }
\label{tbl:mme}
\begin{tabular}{lcc}
\hline 
\hline 
 & fiTQun & APFit \\
 \hline 
\multicolumn{3}{l}{\textbf{First stage}} \\
\hspace*{4pt}Multi-ring $e$-like events \\
\hspace*{8pt}Efficiency &  75.7\% & 69.7\%  \\
\hspace*{8pt}Purity &  77.8\% & 69.5\% \\
\multicolumn{3}{l}{\textbf{Second stage}} \\
\hspace*{4pt}Multi-ring True CC$\nue$ events \\
\hspace*{8pt}Efficiency     & 56.8\% &  53.6\%  \\  
\hspace*{8pt}Purity     & 58.8\% & 52.6\%  \\  
\hspace*{4pt}Multi-ring True CC$\nuebar$ events \\
\hspace*{8pt}Efficiency     & 68.4\%  & 70.9\%  \\  
\hspace*{8pt}Purity     &  30.0\% &  25.9\% \\  
\hline 
\hline 
\end{tabular}
\end{center}

\end{table}

\begin{figure}
\centering
    \includegraphics[width=0.5\textwidth]{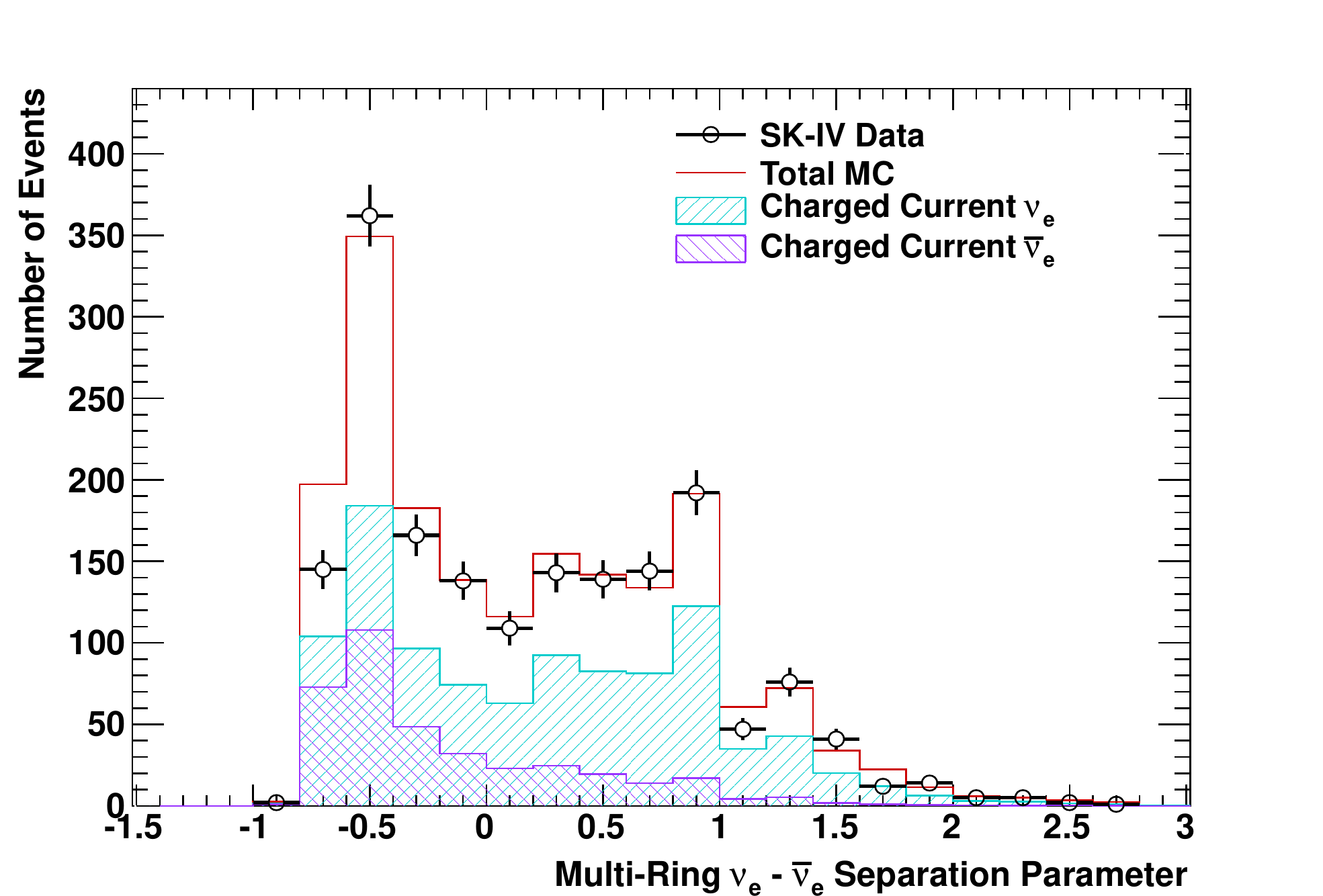}
    \caption{ Likelihood distribution used to separate multi-ring $e$-like events 
              into neutrino-like and antineutrino-like subsamples. 
              Error bars represent the statistical uncertainty of the data.
              Events with negative (positive) likelihood values are designated \nuebar-like (\nue-like). }
  \label{fig:mmebar}
\end{figure}

Neutrino and antineutrino interactions are separated from the multi-ring $e$-like sample during the second stage of the selection.
 The method is similar to the one discussed in~\cite{Abe:2017osc} and uses the number of decay electrons, the number of reconstructed rings, 
electrons, and the event's transverse momentum as reconstructed by fiTQun. 
Figure~\ref{fig:mmebar} shows the final likelihood distribution used in SK-IV. 
The efficiency and purity for selecting $e$-like events and identifying true CC $\bar{\nu}_{e}$ ($\nu_e$) events as $\bar{\nu}_{e}$-like ($\nu_{e}$-like) is summarized in Table \ref{tbl:mme}.
FiTQun shows improved performance in each of these metrics which will translate to better sensitivity 
to sin$^2\theta_{13}$ and the mass hierarchy.

 \section{Fiducial Volume expansion}
\label{sec:fvexp}
In previous analyses using the APFit algorithm the fiducial volume (FV) was defined as the region located 
200~cm from the wall of the inner detector.
With this definition approximately $30\%$ of the Super-K inner detector mass is not used for atmospheric neutrino analysis. 
On the other hand, the sensitivity to several open questions is currently limited by a lack of statistics.
The analysis sensitivity can be increased by taking advantage of fiTQun's improved reconstruction 
performance to expand the fiducial volume while keeping signal purities high and backgrounds low.

The event selection and categorization in this study is similar to the one used in \cite{Abe:2017osc}, but all cut variables used to define
the FC samples are reconstructed by fiTQun. Partially contained and Up-$\mu$ events on the other hand, are defined using APFit in the same manner as in the previous publications.

\subsection{MC study for fiducial volume expansion}

Though a larger fiducial volume provides higher statistics, the purity of signal interactions may decrease if the reconstruction performance 
deteriorates for events close to the ID wall. 
Events that are too close to the wall and producing particles that travel toward it will only illuminate a few PMTs and 
 as a result will have poorly imaged Cherenkov rings.
Further, as the FV cut is moved closer to the wall, the possibility of events with interaction vertices outside of the ID wall being reconstructed within the FV increases.
Although the OD is designed to veto such events, as introduced in Section \ref{sec:reduc}, 
events which originate in the 55~cm dead region between the ID and OD optical barriers 
or interact very near the OD wall may not produce enough light to trigger the OD cut and may remain in the analysis sample.
Such ``entering events'' are a type of mis-reconstructed background in the oscillation analysis with potentially large systematic errors. 

\begin{figure}
\centering
    \includegraphics[width=0.50\textwidth]{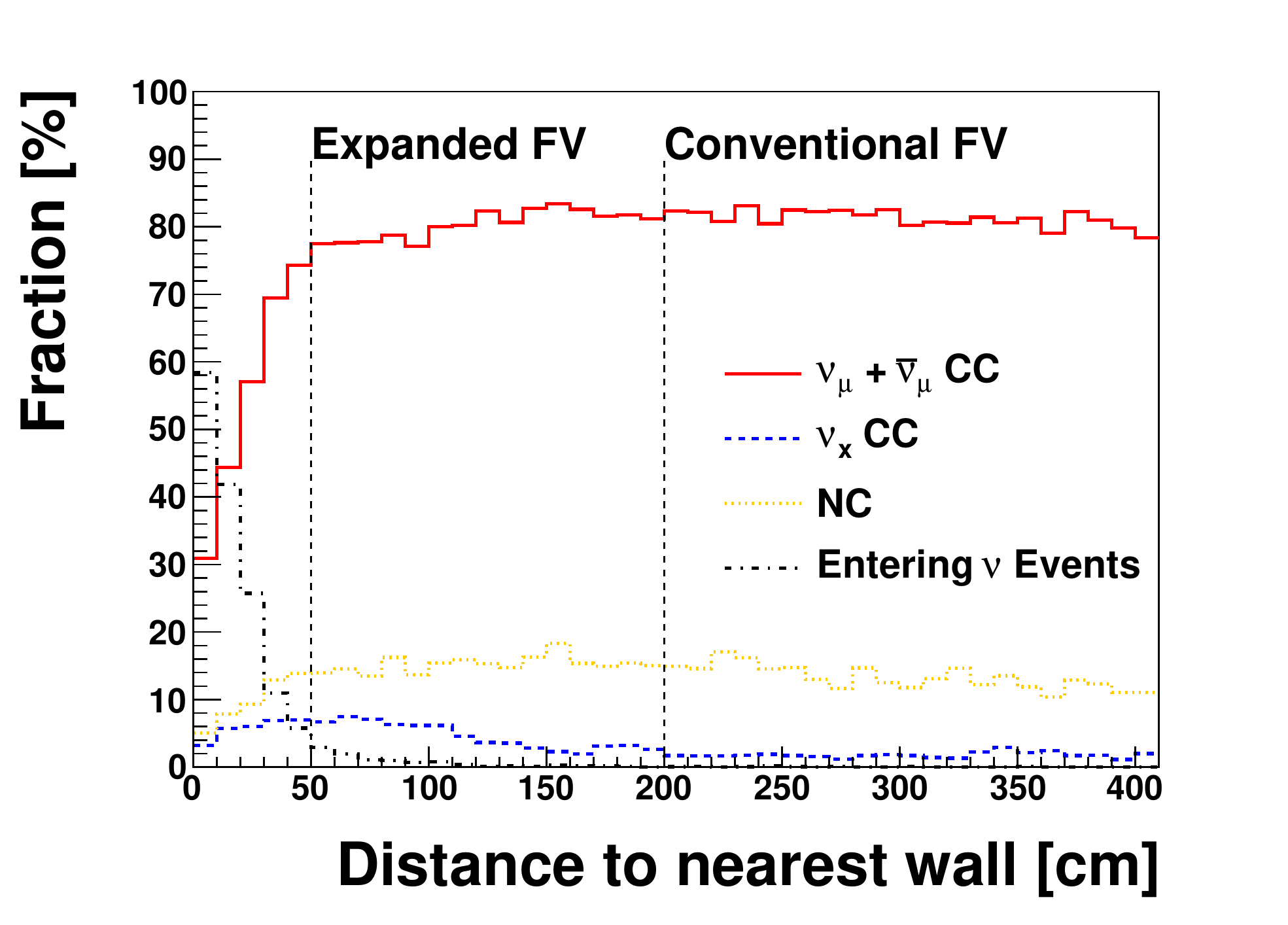}
    \caption{ The fraction of each component of the FC sub-GeV $\mu$-like 0 decay-e event sample in the atmospheric neutrino MC simulation
               as a function of the reconstructed distance from the interaction vertex to the nearest wall. }
  \label{fig:purityDwall}
\end{figure}

Figure~\ref{fig:purityDwall} shows the purity of several components of the FC sub-GeV $\mu$-like 0 decay-e sample 
as a function of the cut on the reconstructed distance from the event vertex to the nearest ID wall (dwall) as an example. 
The fraction of signal events, $\nu_{\mu}$CC and $\bar{\nu}_{\mu}$ CC interactions (red) for this sample, remains 
stable as the cut position is decreased to smaller values until it reaches 50~cm, 
where the background from  entering $\nu$ events increases dramatically. 
All of 13 FC event categories exhibit similar behavior, indicating that a 
FV cut at 50~cm is an acceptable limit due to entering $\nu$ events. 
The purity of the 13 FC samples defined using this cut and the conventional FV definition 
are summarized in Table~\ref{tbl:full_samples50vs}
(c.f. Table~II of Ref~\cite{Abe:2017osc}).

\begin{table*}
\begin{center}
\caption{Sample purity broken down by neutrino flavor assuming neutrino oscillations with $\Delta m^{2}_{32} = 2.52\times 10^{-3}\mbox{eV}^{2}$ and $\sin^{2}\theta_{23} = 0.51$ with a FV cut of 200~cm. 
The number in the parentheses is the result within the new region, where the distance to nearest wall is between 50~cm and 200~cm. }
\label{tbl:full_samples50vs}
\begin{tabular}{l@{\hspace{1em}}ccccc}
\hline
\hline
Sample  & CC $\nu_{e}$  &  CC $\bar{\nu}_{e}$ & CC $\nu_{\mu} + \bar{\nu}_{\mu} $ & CC $\nu_{\tau}$ & NC \\ 
\hline
\multicolumn{6}{l}{\textbf{Fully Contained (FC) Sub-GeV}} \\
\hspace*{4pt} e-like, Single-ring  \\
\hspace*{8pt} 0 decay-e    & 0.728   & 0.242   & 0.001   & 0.000   & 0.028    \\
  &   (0.702) &   (0.227) &   (0.025) &   (0.001) &   (0.045)  \\
\hspace*{8pt} 1 decay-e    & 0.907   & 0.020   & 0.033   & 0.001   & 0.040     \\
  &   (0.712) &   (0.015) &   (0.208) &   (0.001) &   (0.063)   \\
\hspace*{4pt} $\mu$-like, Single-ring  \\
\hspace*{8pt} 0 decay-e     & 0.010   & 0.004   & 0.795   & 0.001   & 0.191     \\
    &   (0.034) &   (0.011) &   (0.805) &   (0.001) &   (0.150)   \\
\hspace*{8pt} 1 decay-e     & 0.000   & 0.000   & 0.974   & 0.000   & 0.026     \\
 &   (0.001) &   (0.000) &   (0.968) &   (0.000) &   (0.031)   \\
\hspace*{8pt} 2 decay-e     & 0.000   & 0.000   & 0.984   & 0.000   & 0.016     \\
 &   (0.000) &   (0.000) &   (0.980) &   (0.000) &   (0.019)   \\
\hspace*{4pt} $\pi^{0}$-like  \\
\hspace*{8pt} Two-ring       & 0.051   & 0.016   & 0.011   & 0.000   & 0.922     \\
  &   (0.109) &   (0.036) &   (0.018) &   (0.000) &   (0.837)   \\
\\
\multicolumn{6}{l}{\textbf{Fully Contained (FC) Multi-GeV}} \\
\hspace*{4pt} Single-ring  \\
\hspace*{8pt} $\nu_{e}$-like   & 0.726   & 0.077   & 0.058   & 0.027   & 0.113   \\
	&   (0.748) &   (0.066) &   (0.064) &   (0.016) &   (0.105) \\
\hspace*{8pt} \nuebar-like     & 0.553   & 0.379   & 0.003   & 0.008   & 0.056    \\
   &   (0.566) &   (0.371) &   (0.003) &   (0.007) &   (0.053)  \\
\hspace*{8pt} $\mu$-like       &0.000   & 0.000   & 0.996   & 0.003   & 0.001     \\
    &  (0.000) &   (0.000) &   (0.995) &   (0.004) &   (0.001)   \\
\hspace*{4pt} Multi-ring  \\
\hspace*{8pt} \nue-like        &0.588   & 0.117   & 0.054   & 0.036   & 0.204     \\
   &  (0.609) &   (0.112) &   (0.059) &   (0.032) &   (0.188)   \\
\hspace*{8pt} \nuebar-like     &0.526   & 0.300   & 0.021   & 0.020   & 0.134     \\
   &  (0.541) &   (0.301) &   (0.023) &   (0.016) &   (0.118)   \\
\hspace*{8pt} $\mu$-like       &0.010   & 0.001   & 0.959   & 0.004   & 0.026     \\
      &  (0.016) &   (0.002) &   (0.946) &   (0.005) &   (0.031)   \\
\hspace*{8pt} Other             & 0.283   & 0.026   & 0.342   & 0.053   & 0.295    \\
       &   (0.302) &   (0.032) &   (0.342) &   (0.051) &   (0.274)  \\
\hline
\hline
\end{tabular}

\end{center}
\end{table*}

\begin{figure}
\centering
    \includegraphics[width=0.50\textwidth]{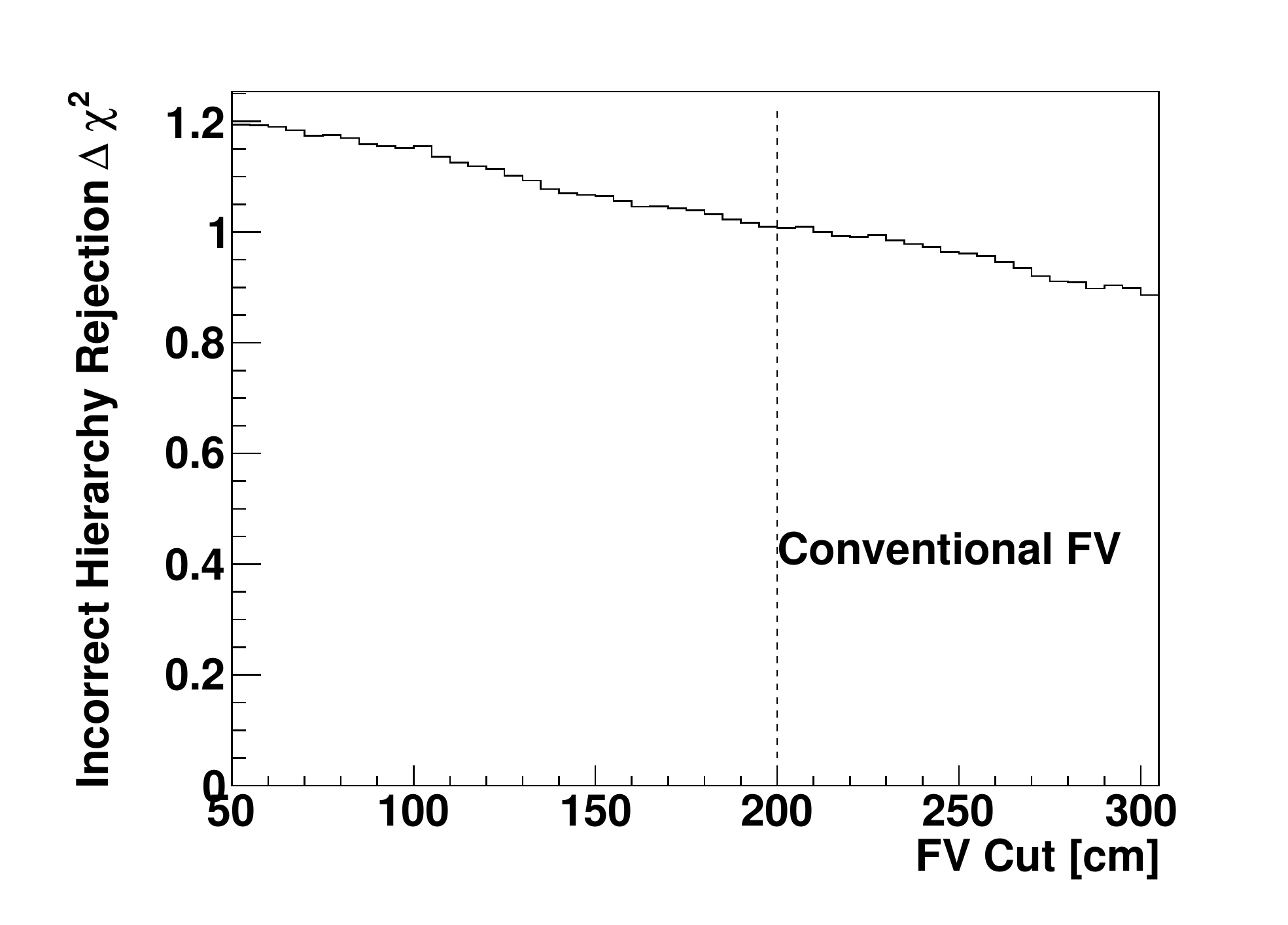}
    \caption{ Sensitivity to reject the wrong mass hierarchy as a 
              function of the FV cut for a 3118.5 day exposure. 
              The normal hierarchy is assumed to be the true hierarchy with oscillation parameters taken to be 
            $\Delta m_{23}^{2} = 2.4 \times 10^{-3} \mbox{eV}^{2}$, 
            $\mbox{sin}^{2} \theta_{23} = 0.5$, 
            $\mbox{sin}^{2} \theta_{13} = 0.0210$, and
            $\delta_{CP} = 0 $. 
            Systematic error are assumed to be the same as those for the analysis with a FV cut at 200~cm. }
  \label{fig:sensdwall}
\end{figure}

An MC study was performed to investigate the effect of decreases in purity against improved statistical power of an expanded FV.
Figure~\ref{fig:sensdwall} shows the sensitivity to the normal mass hierarchy assuming different FV cut values.
Here systematic errors are assumed to be the same as those from the standard SK analysis, which uses a cut at 200~cm.
The rise in sensitivity as the FV is expanded indicates that a FV cut at 50~cm can be expected to provide 
improved hierarchy sensitivity as long as systematic errors in the detector region outside of the conventional cut boundary are stable.
A more precise sensitivity study with updated systematic errors for this region is presented below.

\subsection{Background estimation} 

In addition to entering backgrounds and events with incorrectly assigned PID, 
non-neutrino (non-$\nu$) backgrounds from cosmic ray muons and flashing PMTs are potential background sources when 
the FV is expanded. 

Cosmic ray muons must traverse the OD before reaching the ID and are therefore 
mostly rejected by cuts on activity in the OD. 
Any remaining events are rejected by the FV cut, since their true vertex position, that is the entrance point of the muon, 
is expected to be on the wall of tank. 
However, when the FV cut is near the wall cosmic ray background events might be introduced into the 
FC sample due to the vertex resolution of the reconstruction algorithm.

Flasher events are caused by the internal electrical discharge of a PMT and represent another kind of non-neutrino background. 
The reconstructed vertex for such events is also expected to be on the wall of the detector, so as was the case 
for cosmic ray muons, more flasher backgrounds might exist in the FC sample when the FV is enlarged. 

\begin{figure}[h]
\centering
    \includegraphics[width=0.50\textwidth]{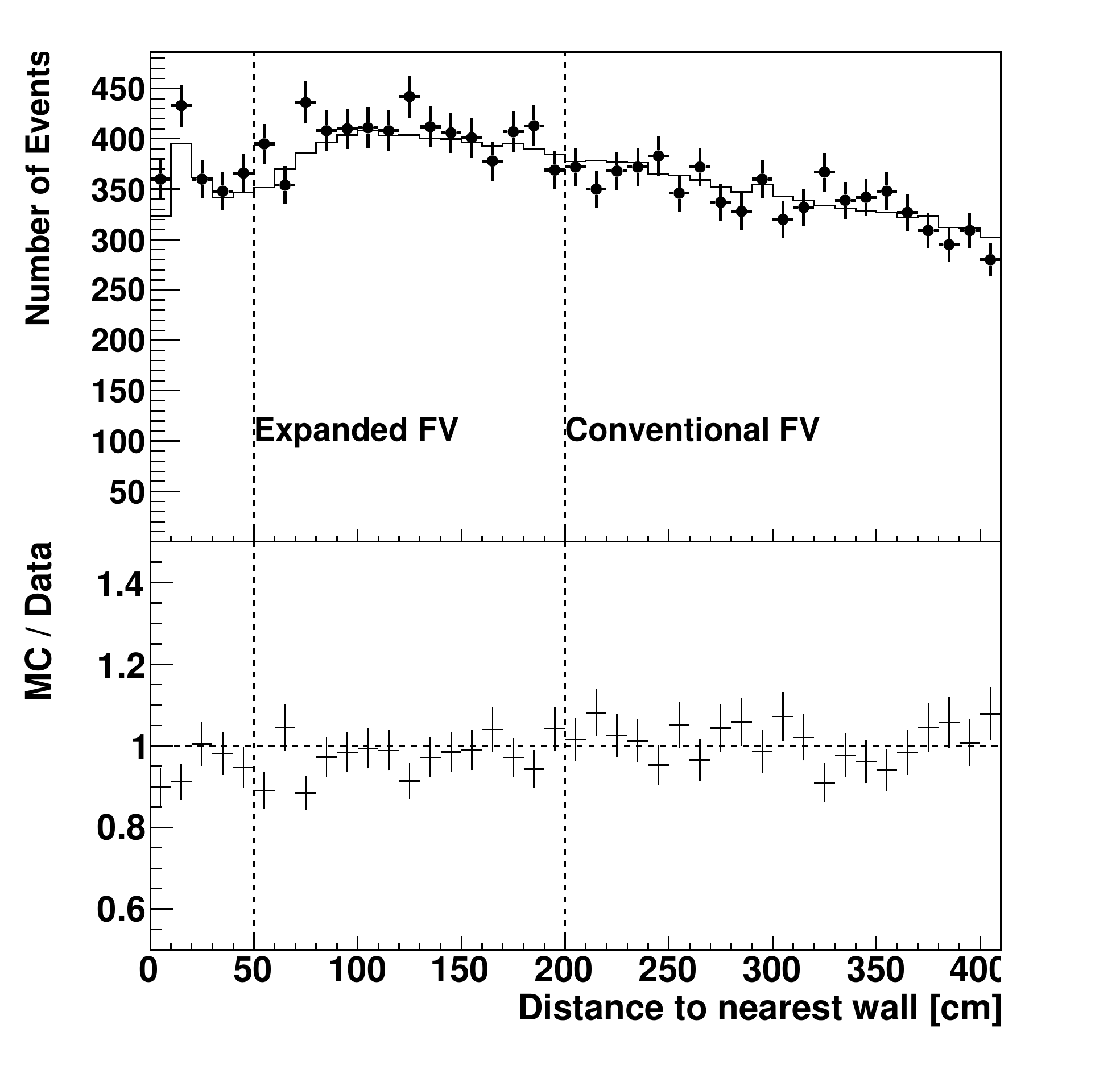}
    \caption{ The distribution of the distance from the reconstructed event vertex to the nearest ID wall of FC data (points) and MC (solid line). The lower figure shows the ratio of the MC to the data.}
  \label{fig:evtwall}
\end{figure}

In order to observe possible excesses in the data due to such backgrounds, which are not modeled in the atmospheric neutrino 
MC, the distribution of the distance from the reconstructed vertex to the nearest wall of 
FC events is compared to MC in Figure~\ref{fig:evtwall}. 
Though not shown here, the distributions of the events' momentum, direction, and particle ID have been checked 
for both the conventional FV region (dwall $>$ 200~cm) and new region (50~cm $<$ dwall $<$ 200~cm), respectively.
No evidence for the presence of a large unmodeled component to the data is observed.

\begin{figure}[h]
\centering
    \includegraphics[width=0.50\textwidth]{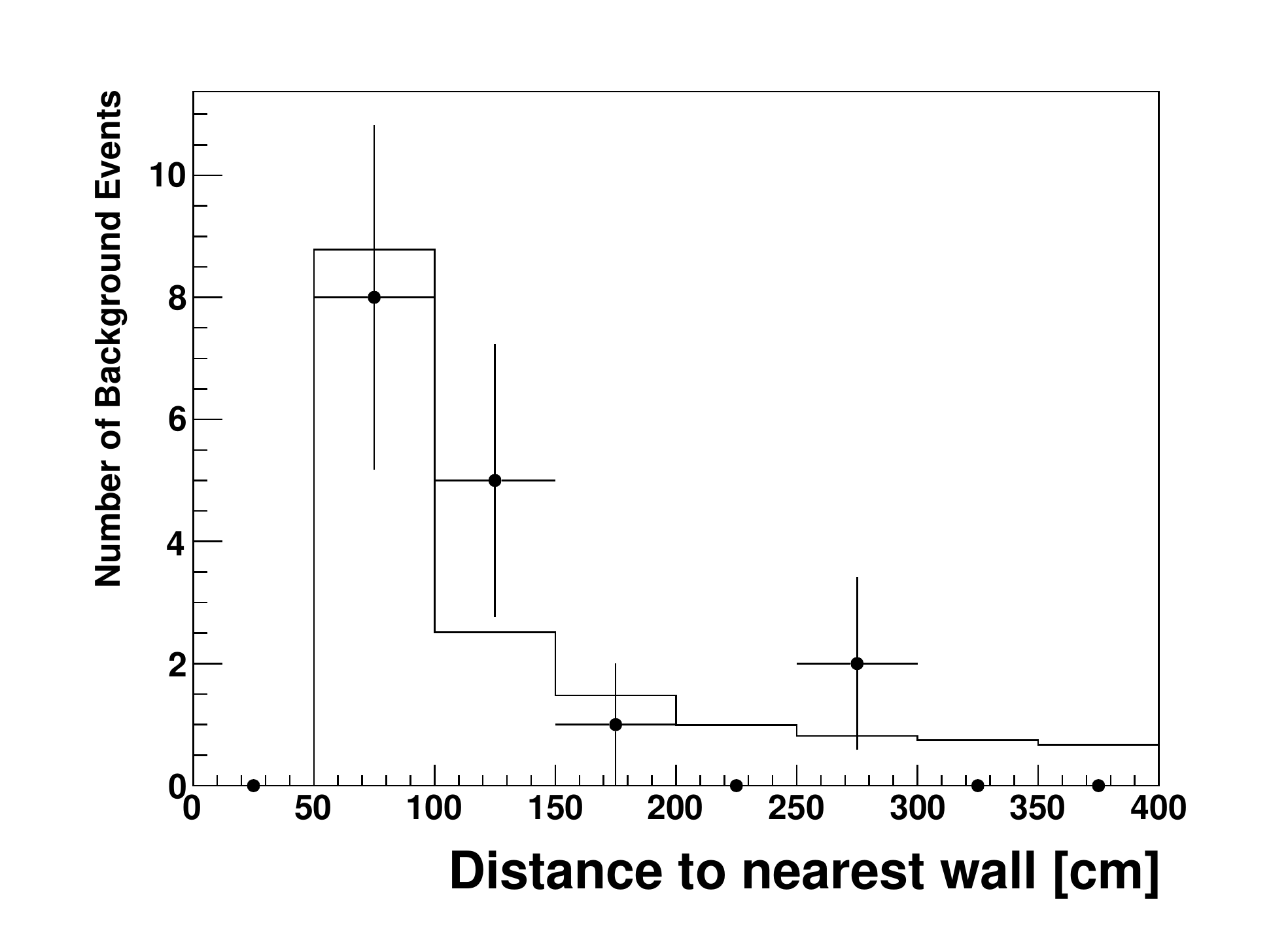}
    \caption{ Distribution of the distance to the nearest ID wall from the reconstructed vertex 
of events passing the multi-ring $\mu$-like sample selection. 
The solid line is from a cosmic ray muon control sample and the points show the distribution of events 
identified as backgrounds in the FC data.
The histograms have been normalized by the total number of events seen in the FC sample. }
  \label{fig:stmu_wall}
\end{figure}

To evaluate the background rate more precisely all FC events within the fiducial volume were scanned by eye using 
a graphical event display program. 
In total, one flasher event and 24 cosmic ray muon events were found in the observed in the FC sample with dwall $>$ 50~cm.
 Most of the cosmic ray muon backgrounds were reconstructed as multi-ring downward-going $\mu$-like events with more than 5~GeV/c of momentum. 
Eye scans of these events indicated that their vertices were incorrectly reconstructed within the new fiducial volume 
due to the presence of an erroneous second ring found by the fitter. 
Both the rate and type of such mis-reconstructions is consistent with that found in independent cosmic ray muon samples. 
Using the dwall distribution of the cosmic ray muon events, the expected cosmic ray background in the final analysis 
sample is statistically removed. 
This is done under the assumption that the FC event reduction process does not depend strongly on the muon kinematics.
By comparison of the event classification of this control sample with that observed in the 
eye-scanned background, this assumption has been validated for all particle directions and momenta.
As an example, Figure \ref{fig:stmu_wall} shows the dwall distribution for multi-ring $\mu$-like events. More than 20, 000 cosmic ray muon events (solid line) are shown here and their dwall distribution has been normalized to the eye-scanned background events (points) in the FC sample. The shape of these two samples are in good agreement.
Due to their limited contamination in the final FC sample, 
the effect of flasher PMTs is treated with a systematic error on their rate in the oscillation analysis presented below.

\begin{figure}
 \centering
    \includegraphics[width=0.49\textwidth]{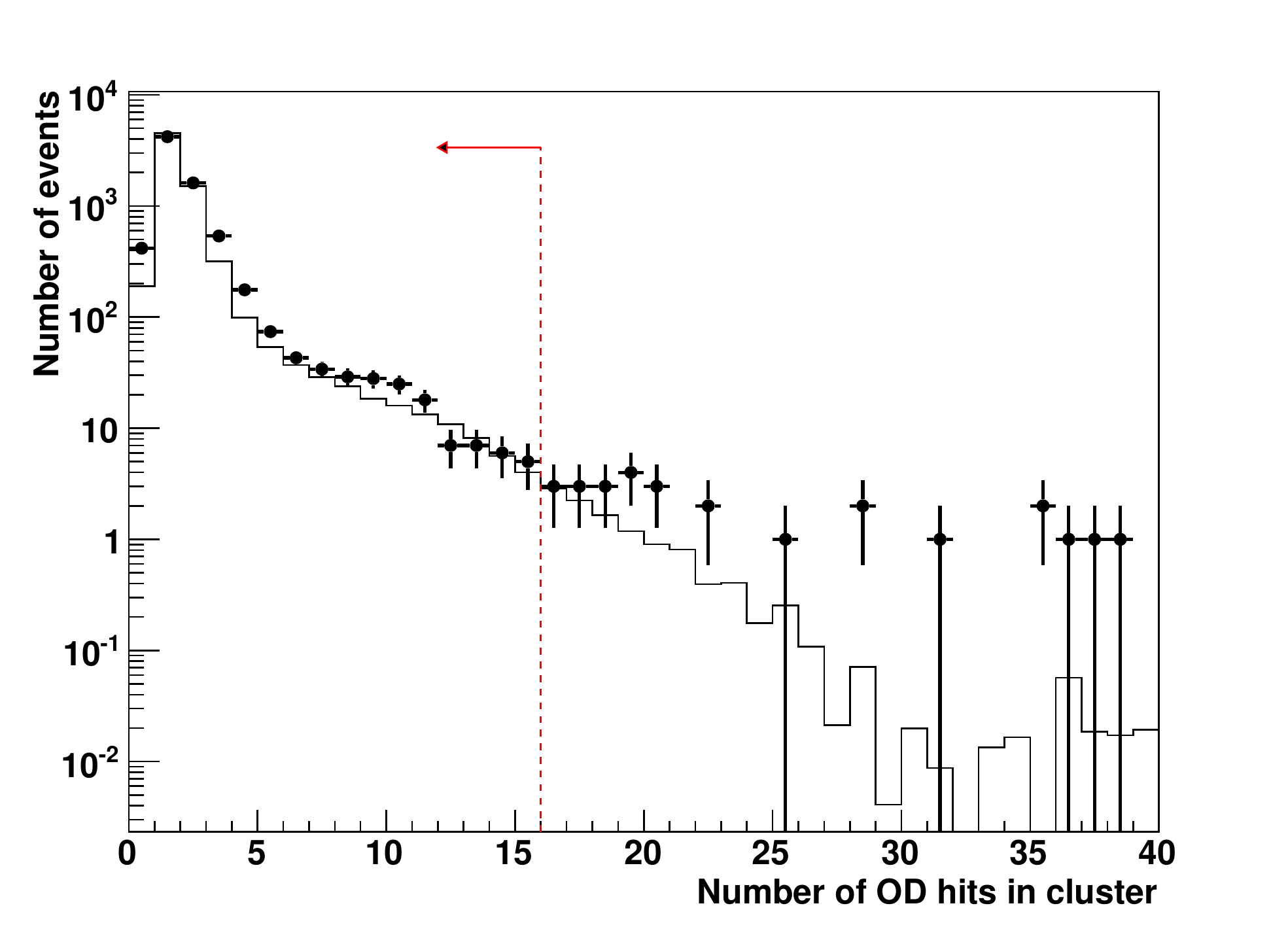}
    \caption{ Distribution number of OD hits in cluster for events within new region (50~cm $<$ dwall $<$ 200~cm) of data (points) and atmospheric neutrino MC (solid line). The red dashed line shows the threshold to select FC events.}
  \label{fig:nhitac_50to200}
\end{figure}

Expanding the FV opens the possibility that events interacting within the ID but close to its wall may produce particles which escape into the OD. 
Poor modeling of the response of the OD can thereby potentially introduce biases and relative inefficiencies in the FC sample. 
Figure~\ref{fig:nhitac_50to200} shows the distribution of OD hits used to define the FC sample after the reduction processes for data with dwall $> 50$~cm. The detection inefficiency due to the cut on OD hits for both data and MC are confirmed to be consistent, around 0.2\%.
Based on this result in conjunction with the  stability of the reconstruction algorithm, 
its systematic errors, and sample purities across the detector, the fiducial volume definition is expanded from 
its conventional value to dwall $> 50$~cm (expanded FV) in the analysis presented below.
  
\begin{figure*}[htbp]
\includegraphics[width=1.0\textwidth]{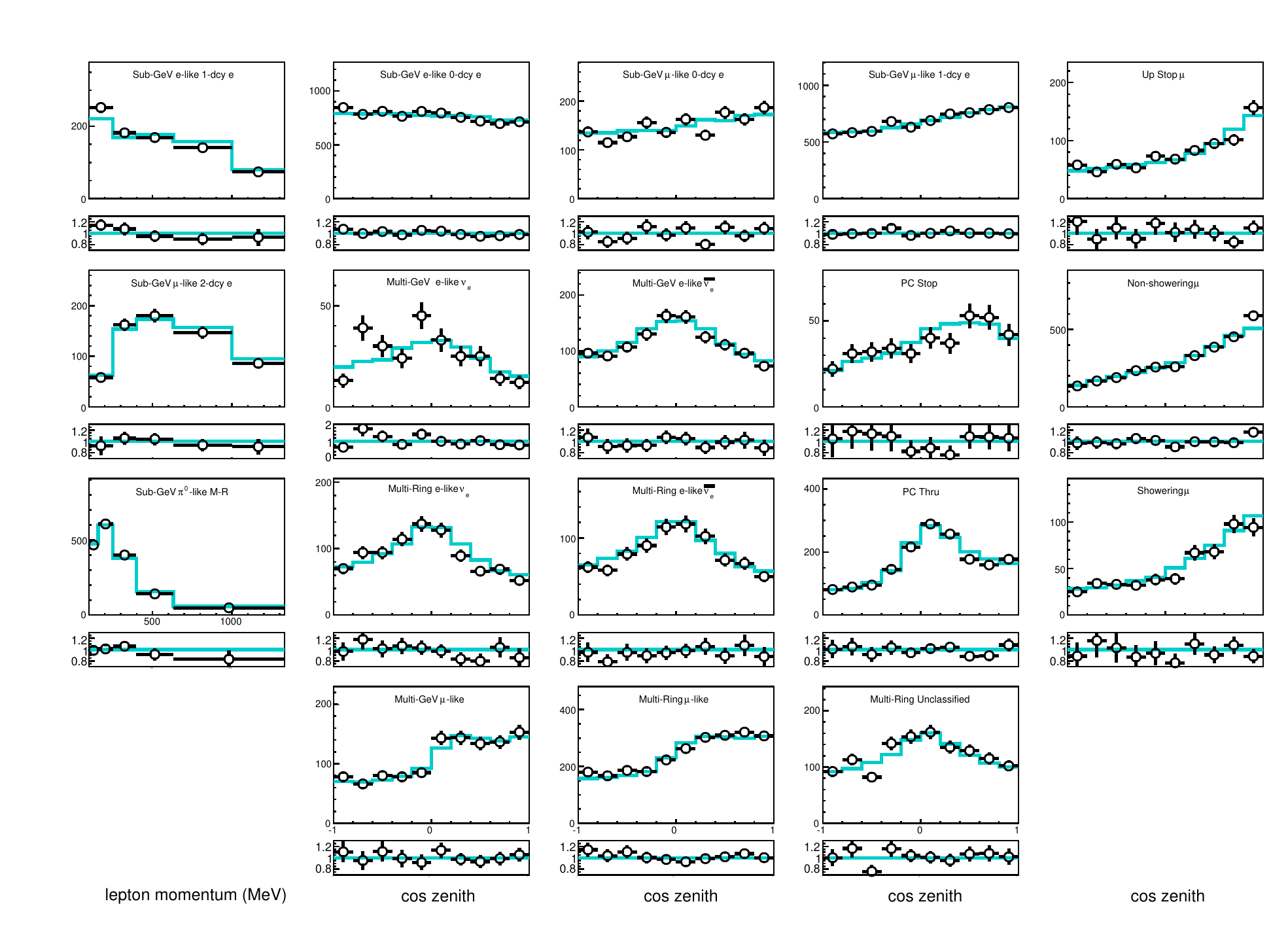}
\caption{ Data and MC comparisons for the SK-IV data divided into 18 analysis samples. The expanded FV, where dwall $>$ 50 cm, is shown here.
          Samples with more than one zenith angle bin are shown as zenith angle distributions
          (second through fifth column) and other samples are shown as reconstructed momentum
          distributions (first column). Cyan lines denote the best-fit MC assuming the normal  hierarchy.
          Narrow panels below each distribution show the ratio relative to the normal hierarchy MC.
          In all panels the error bars represent the statistical uncertainty.    }
\label{fig:sk_zenith}
\end{figure*}

\begin{figure}
\centering
    \includegraphics[width=0.50\textwidth]{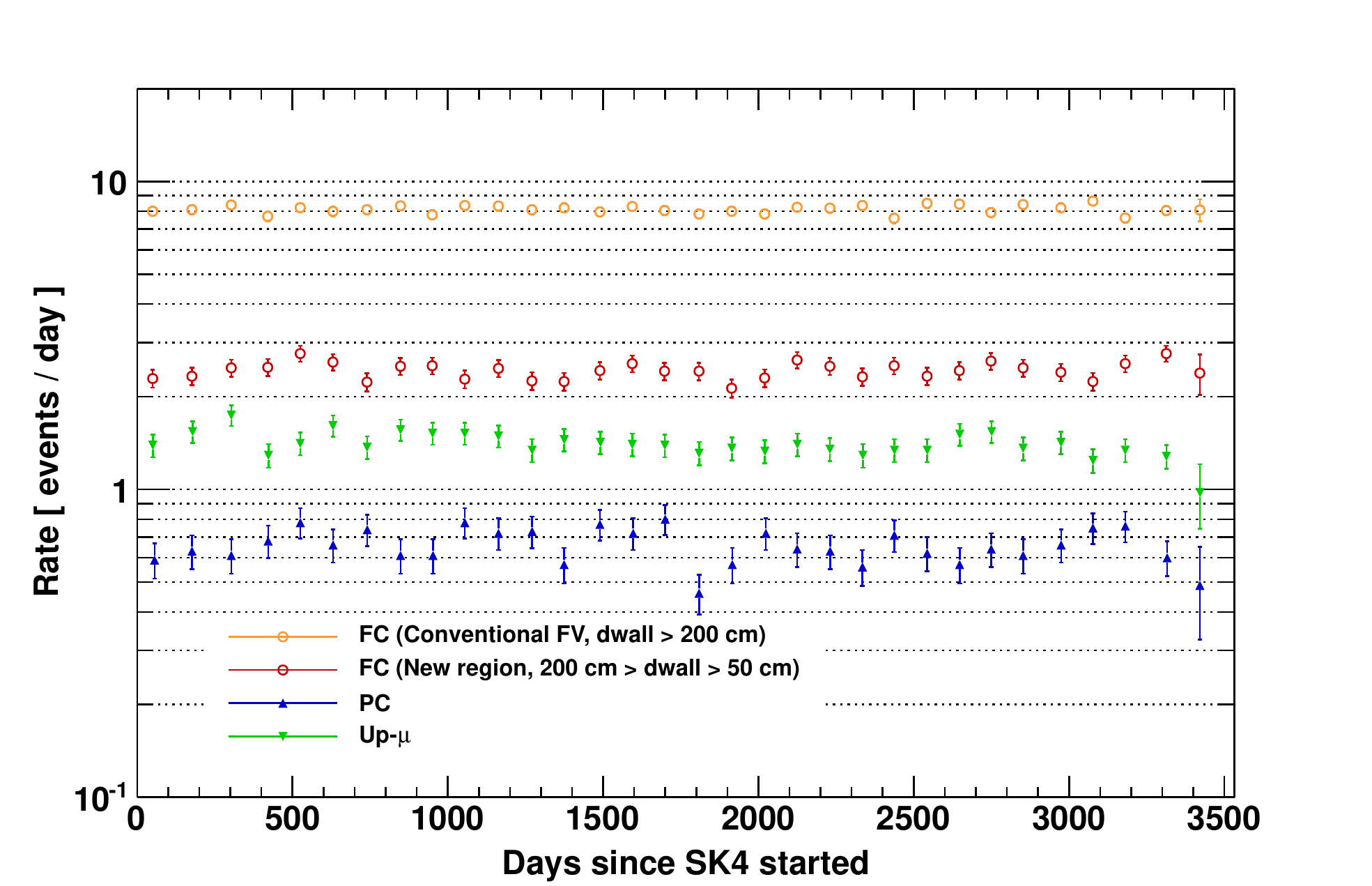}
    \caption{ Final event rates as a function of time since the start of SK-IV operations. 
              The error bars are statistical.
              Circles denote the fully contained event rate and upward-facing (downward-facing) triangles 
              show the partially contained (upward-going muon) event rates. 
	      The number in the parentheses for fully contained event shows the fiducial volume cut value.}
  \label{fig:reduction}
\end{figure}

Zenith angle distributions for each analysis sample using the expanded FV are shown in Figure~\ref{fig:sk_zenith}. 
Their event rates since the start of SK-IV of the experiment have been stable at 8.3 (2.2) 
FC events per day in the conventional FV (new region), 0.73 PC events per day, 
and 1.49 Up-$\mu$ events per day, as shown in Figure~\ref{fig:reduction}.

 \section{Systematic error}
\label{sec:syst}

In order to use the fiTQun reconstruction algorithm for physics analysis the corresponding systematic errors must be evaluated.
While systematic errors related to the atmospheric neutrino flux and cross section model are the same as those used in \cite{Abe:2017osc},
systematic errors on the event selection have been updated for FC events using the expanded fiducial volume outlined above.
Since fiTQun is not used to reconstruct PC or Up-$\mu$ events in the present work, the errors for those samples are the same 
as in the previous publications.

\begin{figure}[htbp]
\centering
  \includegraphics[width=0.5\textwidth]{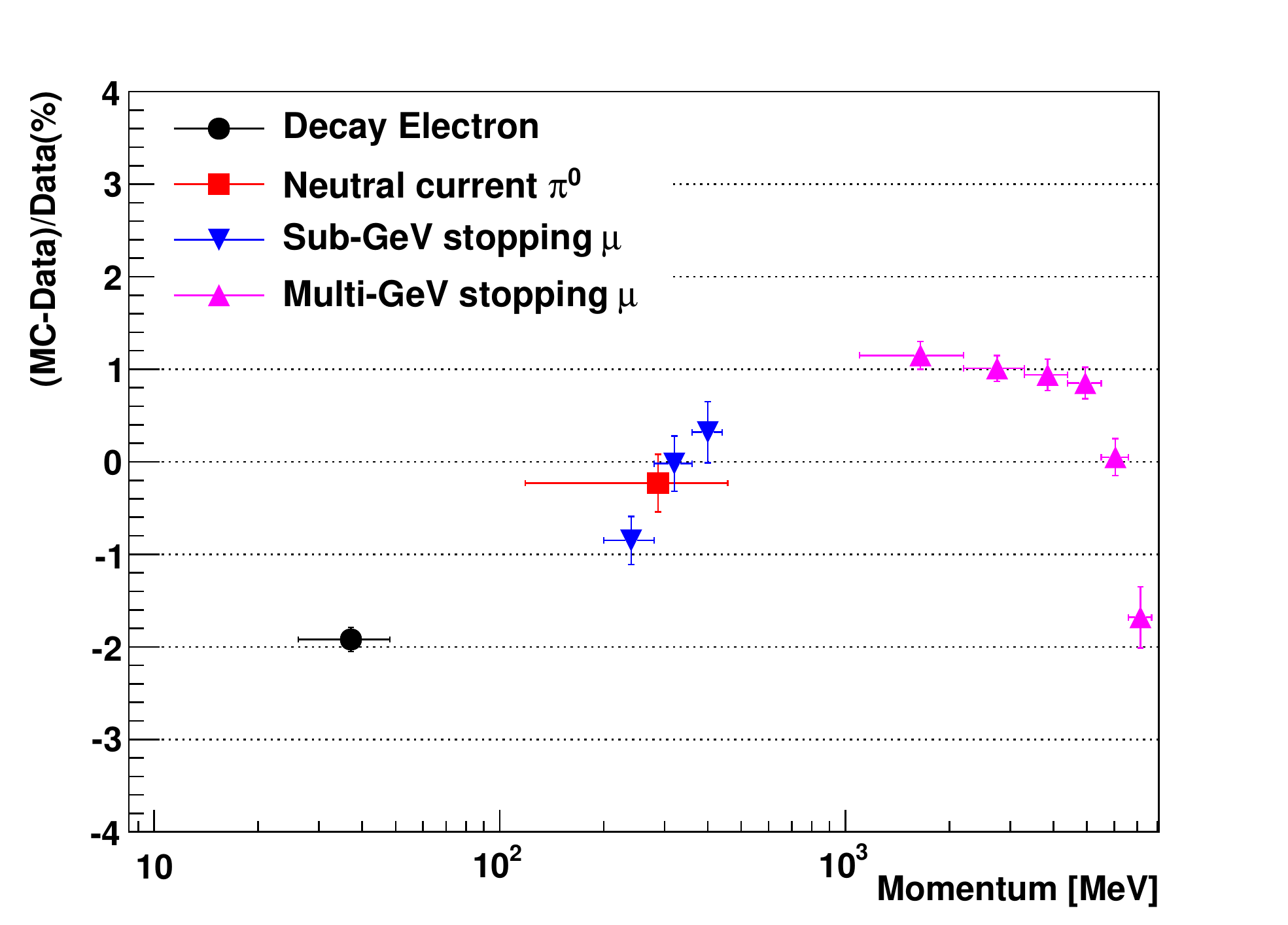}
  \caption{ Absolute energy scale measurement for SK-IV with fiTQun and the expanded FV (dwall $>$ 50~cm). Vertical error bars denote the statistical uncertainty and horizontal error bars shows the momentum range spanned by each control sample. }
\label{fig:enescale}
\end{figure}

The uncertainty on the absolute energy scale is estimated by comparison of data and MC across three control samples spanning momenta up to 10 GeV/c: electrons from cosmic ray muon decay, 
atmospheric neutrino events producing single $\pi^0$s from neutral current interactions, and stopping cosmic ray muons.
The difference between data and MC for the calibration samples reconstructed within the expanded FV is shown in Figure~\ref{fig:enescale}. 
The total systematic error follows that of \cite{Abe:2017osc} and is the sum in quadrature of the absolute energy scale, which is defined as the largest data-MC difference across all samples, and momenta with 
the average time variation of these samples throughout SK-IV.
The uncertainty from the up/down asymmetry of the detector is measured with Michel electrons from cosmic ray muons by comparison of their momenta for data and MC as a function of zenith angle. The difference between data and MC at the most deviated direction is taken as the uncertainty from the detector asymmetry.
As summarized in Table~\ref{tbl:escale}, the uncertainty of energy scale on the conventional (expanded) FV is $2.17\%$ (2.02\%) using fiTQun reconstruction, which is similar to the  value with APFit: 2.1\% \cite{Abe:2017osc}.
\begin{table}
\begin{center}
\caption{Summary of uncertainty from the energy scale. The energy scale error is the quadratic-sum of absolute energy scale and the time variation. The absolute energy scale value is obtained from the most discrepant sample, i.e. Michel electrons from cosmic ray muons in this study. } 
\label{tbl:escale}
\begin{tabular}{lccc}
\hline
\hline
Reconstruction  	& fiTQun     		& fiTQun  	& APFit\cite{Abe:2017osc} 	\\ 
Dwall range 		& $>$50~cm  	& $>$ 200~cm 	&  $>$ 200~cm 	\\
\hline
\textbf{Energy calibration error}
&2.02\%& 2.17\% & 2.1\% \\
\hspace*{4pt} Absolute energy scale 
 & 1.92\% & 2.09\% \\
\hspace*{4pt} Time variation	       
\hspace*{8pt} &  0.62\%   & 0.59\%
\\
\\
\textbf{Up/down asymmetry energy calibration error}
&0.67\% & 0.58\% & 0.4\%  \\
\hline
\hline
\end{tabular}

\end{center}
\end{table}

Lacking other control samples that span the same energies and event topologies as the atmospheric neutrino 
data, systematic errors on the event selection including the particle identification (PID), estimated number of rings (ring counting), and the two-stage 
separation of the multi-GeV multi-ring e-like event samples described above are evaluated using the atmospheric 
neutrino data.

Systematic errors are estimated for each of the FC analysis samples for both the conventional fiducial volume 
and the additional region formed between the conventional and expanded FV boundaries. 
The results are summarized in Table~\ref{tbl:syst}. 
Errors for the individual samples subject to a particular uncertainty, such as the particle ID for single ring-events, 
are assumed to be fully correlated (or fully anti-correlated). 
Anti-correlations are indicated by negative numbers in the table.
No correlation is assumed between the systematic error categories. 
The sizes of the systematic errors evaluated using fiTQun and APFit are consistent within the conventional 
fiducial volume, with improvements seen in fiTQun for many samples.
For many error categories the fiTQun reconstruction shows consistent errors 
between the conventional fiducial volume and the new region (50~cm $<$ dwall $<$ 200~cm).
In some cases however, such as the ring counting uncertainty for multi-GeV single-ring events, 
larger systematic errors are observed.
At present these are not large enough to offset the benefit to the oscillation sensitivity from 
including events in the new region in the analyses. 
A quantitative estimation of the sensitivity is presented in next the section.  
The numbers for the different FV regions are merged 
to form the final response function used in the analysis, which will be described in next section. Systematic errors and their
sizes at the best fit point of the analysis are presented in
Tables~\ref{tab:sysa}.
\begin{table}
\begin{center}
\caption{Summary of systematic errors related to the ring counting, particle identification and multi-GeV multi-ring $e$-like separation. 
The sign of the number represents the correlation between these systematic errors, with negative numbers denoting fully anti-correlated samples. 
} 
\label{tbl:syst}
\begin{tabular}{lccc}
\hline
\hline
Reconstruction  	& fiTQun     		& fiTQun  	& APFit 	\\ 
Dwall range 		& 50~cm$\sim$200~cm  	& $>$ 200~cm 	&  $>$ 200~cm 	\\
\hline
\multicolumn{4}{l}{\textbf{Ring counting}} 	\\
\hspace*{4pt} Sub-GeV , Single-ring  	\\
\hspace*{8pt} $e$-like, p $<$ 400~MeV 	& 1.94\%		& 1.20\% 	& 1.6\%	  	\\
\hspace*{8pt} $e$-like, p $>$ 400~MeV 	& 0.59\%		& 0.48\% 	& 1.0\%	  	\\
\hspace*{8pt} $\mu$-like, p $<$ 400~MeV & 1.08\%		& 0.42\% 	& -3.0\%  	\\
\hspace*{8pt} $\mu$-like, p $>$ 400~MeV & 1.25\%		& 1.21\% 	& 0.6\%	  	\\
\hspace*{4pt} Sub-GeV, Multi-ring  	\\
\hspace*{8pt} $e$-like 		& -3.58\%		& -2.39\% 	& -1.9\%   	\\
\hspace*{8pt} $\mu$-like, $E_\text{vis} >$ 600MeV & -2.32\%		& -1.66\% 	& 2.3\%  	\\ 
\hspace*{4pt} Multi-GeV Single-ring  	\\
\hspace*{8pt} $e$-like 				& 8.61\%  	& 1.21\% & 1.0\%			\\
\hspace*{8pt} $\mu$-like 			& 2.65\%  & -2.33\% 	 & -1.2\%	       \\
\hspace*{4pt} Multi-GeV Multi-ring  	\\                               
\hspace*{8pt} $e$-like 				& -4.21\% & -0.62\% 	 & -0.9\%		\\
\hspace*{8pt} $\mu$-like 			& -3.07\% 	& 0.72\% & 2.4\%		 	\\
	\\
\multicolumn{4}{l}{\textbf{Particle ID of single-ring events}} 	\\
\hspace*{4pt} Sub-GeV\\%, Single-ring  	\\
\hspace*{8pt} $e$-like   	& 0.99\% 	& 0.36\% & -0.28\%		 	\\
\hspace*{8pt} $\mu$-like& -0.89\%  	&-0.37\% & 0.22\% 		   	\\
\hspace*{4pt} Multi-GeV\\%, Single-ring  \\
\hspace*{8pt} $e$-like   	& 0.23\% 	& 0.06\% & -0.35\%		 	\\
\hspace*{8pt} $\mu$-like& -0.21\%  &-0.06\% & 0.35\%		  	\\
\\                    
\multicolumn{4}{l}{\textbf{Particle ID of the brightest ring in multi-ring events}} 	\\       
\hspace*{4pt} Sub-GeV\\%, Multi-ring  	\\
\hspace*{8pt} $e$-like   	& -3.19\%	& -0.72\%	& 4.19\%	 	  	\\
\hspace*{8pt} $\mu$-like& 1.31\%   	&0.31\%  	& -1.56\%  	        	\\
\hspace*{4pt} Multi-GeV\\%, Multi-ring  \\
\hspace*{8pt} $e$-like  	& 1.94\% 	& 1.10\% & 3.33\% 		 	\\
\hspace*{8pt} $\mu$-like& -1.06\%   	&-0.66\% & -1.56\%  		  	\\
	\\
\multicolumn{4}{l}{\textbf{Multi-GeV Multi-ring $e$-like - other separation}} 	\\
\hspace*{8pt} $e$-like  	& -0.88\% 	 	& -0.67\% 	& 3.0\% \\
\hspace*{8pt} other  	& 0.50\%   		&0.53\% 	& -3.4\%\\
\\
\multicolumn{4}{l}{\textbf{Multi-GeV Multi-ring $\nue$-like - $\nuebar$-like separation}} 	\\
\hspace*{8pt} $\nu_e$-like    	& -3.64\% 	     & -2.33\% 	& 6.82\%   	\\
\hspace*{8pt} $\bar{\nu_e}$-like	& 4.51\%	&2.10\%   & -6.04\%   	\\
\hline
\hline
\end{tabular}

\end{center}
\end{table}

\section{Atmospheric Neutrino Oscillation Analysis}

The Super-K atmospheric neutrino data are separately fit against both the normal and inverted hierarchy hypotheses. 
Since $\theta_{13}$ has been measured and is well constrained by reactor experiments, two types of analyses are done: 
one with $\theta_{13}$ as a free fitting parameter and the other with it constrained to $0.0210 \pm 0.0011$~\cite{Olive:2016xmw}.
Though the constraint on this parameter is much stronger than can be expected from Super-K alone, 
the appearance of upward-going electron events characteristic of the mass hierarchy signal is driven by $\theta_{13}$ 
and therefore measurements with atmospheric neutrinos represent an important test of the analysis' hierarchy preference. 
The solar mixing parameters, $\Delta m^{2}_{12}$ and $\sin^{2} \theta_{12}$, on the other hand, have little impact on 
the oscillations of neutrinos in the energy range considered here and will be fixed in the analysis as described below.

Data are fit to the MC expectation with a binned $\chi^{2}$ method assuming
Poisson statistics. 
The effect of systematic errors are regarded as scaling factors on the MC in each bin~\cite{Fogli:2002pt},
with different error sources assumed to be independent. 
The definition of $\chi^2$ used in the fits is
\begin{eqnarray}
\chi^{2}  = 2 \displaystyle \sum_{n} \left(  E_{n} -\mathcal{O}_{n}  + \mathcal{O}_{n} \ln \frac{ \mathcal{O}_{n} }{ E_{n} } \right)
             + \displaystyle \sum_{i} \left( \frac{ \epsilon_{i} }{ \sigma_{i} } \right)^{2}, 
\label{eq:fullchi}
\end{eqnarray}
\noindent where, 
\begin{eqnarray}
E_{n} &=& \displaystyle E_{n,0}(1 + \displaystyle \sum_{i}
f^{i}_{n} \epsilon_{i} ) 
\end{eqnarray}
\noindent In this equation $\mathcal{O}_{n}$ is the observed number of events in
the $n^{th}$ analysis bin. 
Similarly, $E_{n,0}$ represents the nominal MC expectation in that bin and $E_{n}$ is the expectation 
after incorporating the effect of the systematic errors. 
The coefficient $f^{i}_{n}$ describes the fractional change in the
bin's MC under a $1{\sigma_{i}}$ variation of the $i^{th}$
systematic error source. 
The sum over $(\epsilon_{i}/\sigma_{i} )^2$ penalizes the $\chi^{2}$ for adjusting the systematic errors when 
bringing the MC into agreement with the data. 
For each set of oscillation parameters tested, the MC fit to the data by minimizing $\chi^{2}$ over 
the error parameters $\epsilon_{i}$. 
All fits are performed over 515 analysis bins in SK-IV.  

\subsection{Analysis with unconstrained $\theta_{13}$}
\label{sec:atm_only}

\begin{figure}[htbp]
\centering
  \includegraphics[width=0.49\textwidth]{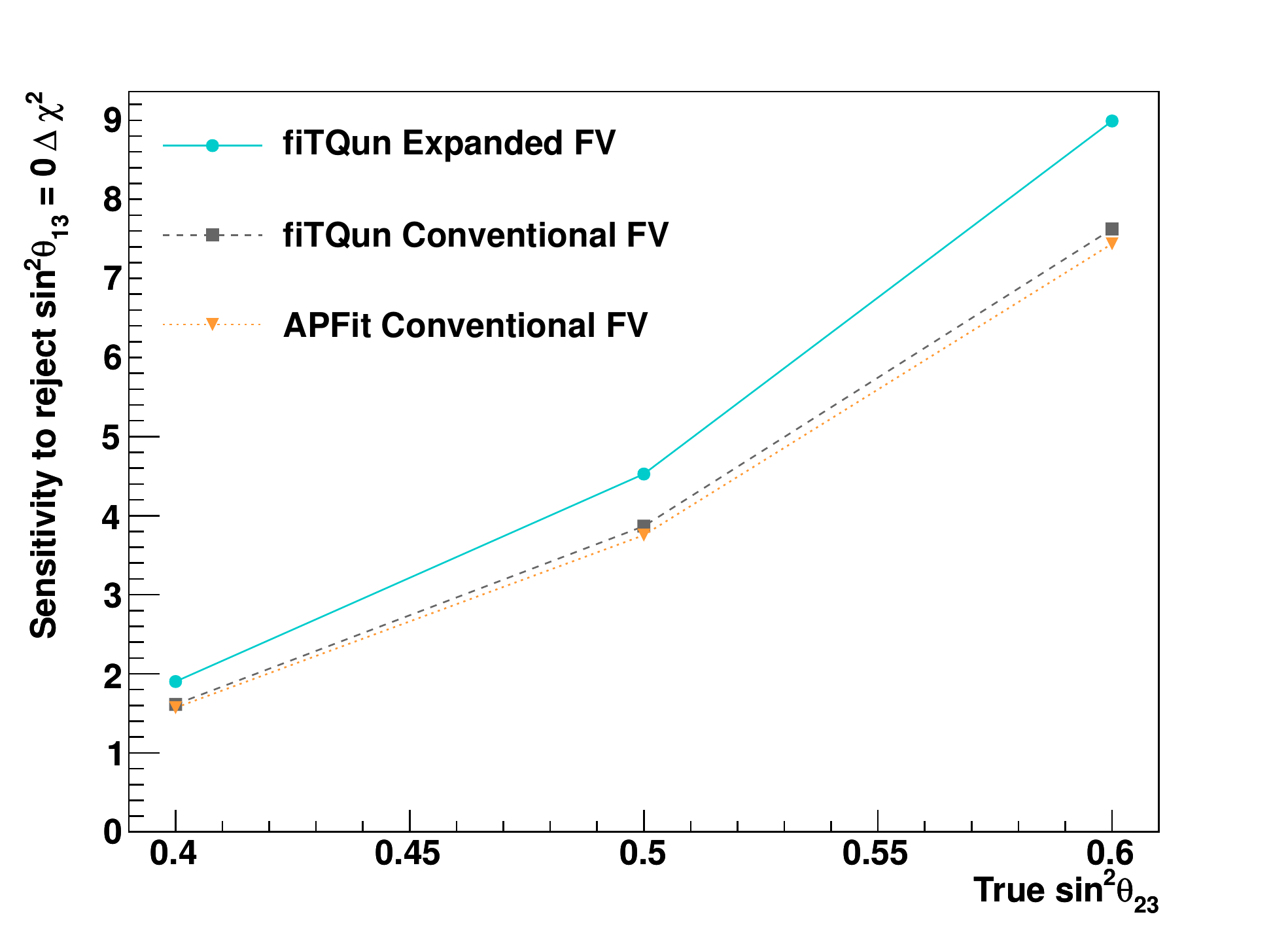}
  \caption{ Expected sensitivity to reject $\sin^2 \theta_{13}$ = 0 assuming its true value is $0.0210$
as a function of the true value of $\sin^2 \theta_{23}$ for a livetime of 3118.5 days. 
Grey and blue lines show the sensitivity for samples reconstructed with fiTQun and using the conventional FV and the expanded FV, respectively. 
The orange line denotes the sensitivity when events are reconstructed with APFit in the conventional FV. 
}
  \label{fig:sens_sk_q13_free}
\end{figure}

\begin{figure*}[htbp]
  \includegraphics[width=0.24\textwidth]{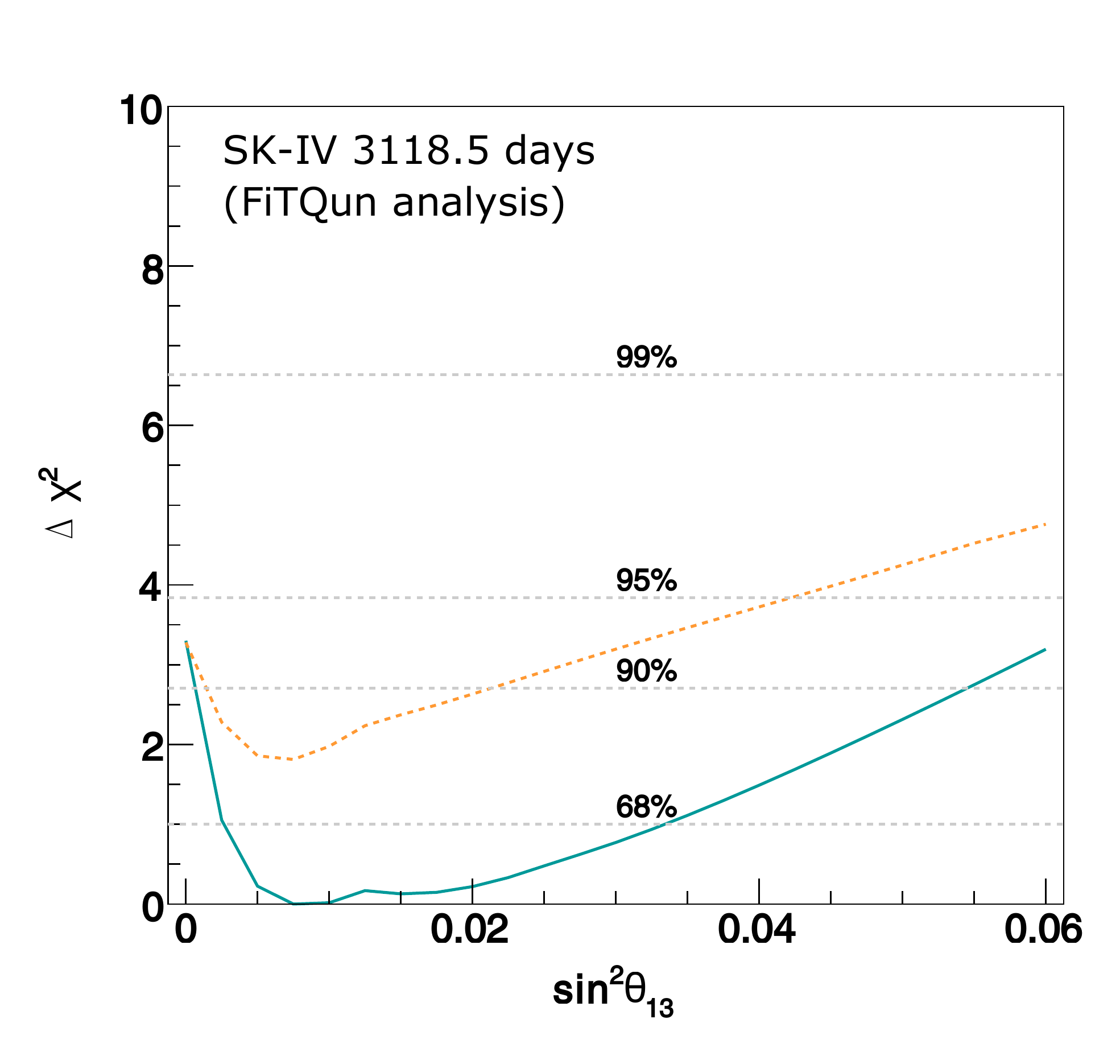}
  \includegraphics[width=0.24\textwidth]{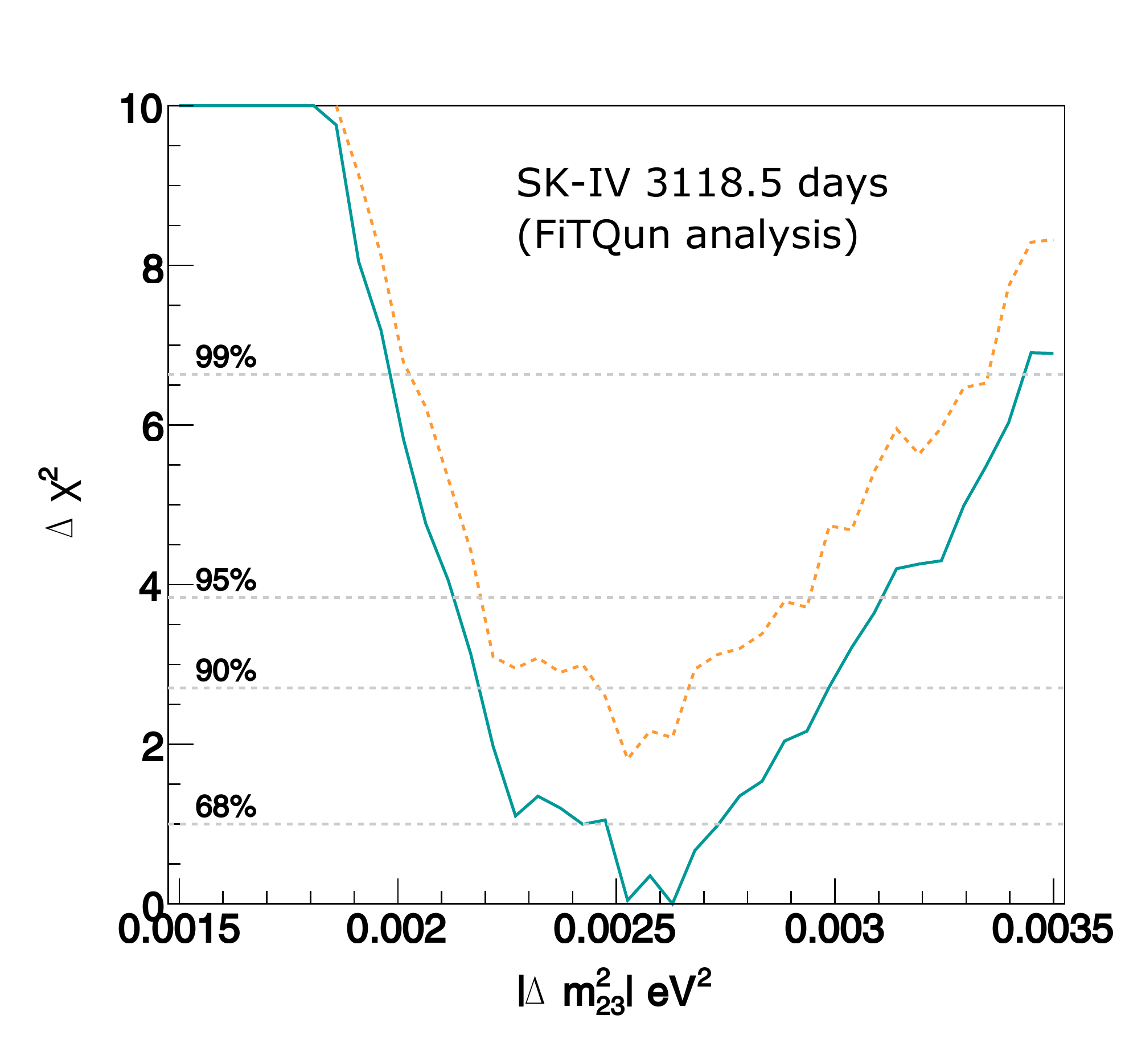}
  \includegraphics[width=0.24\textwidth]{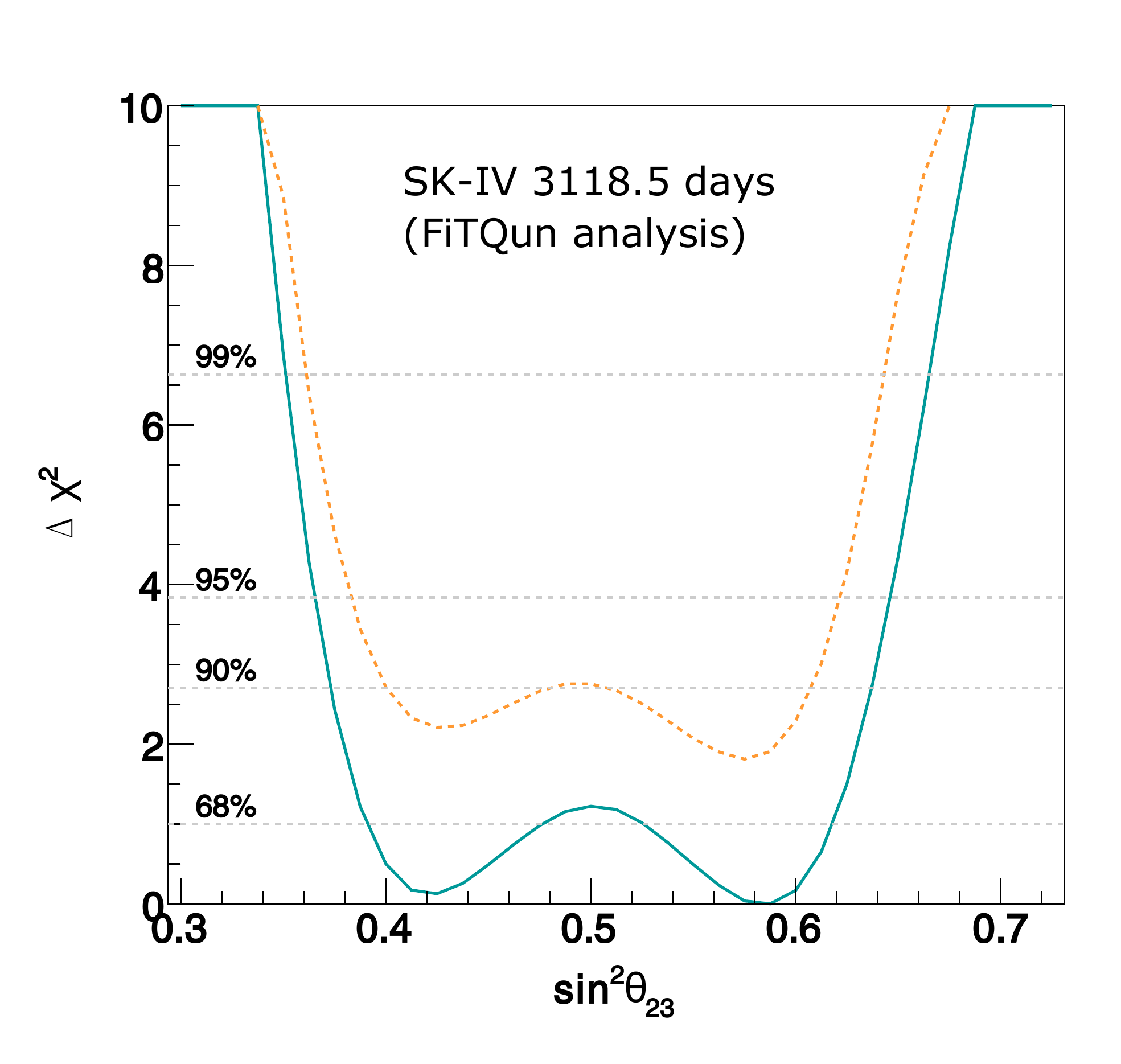}
  \includegraphics[width=0.24\textwidth]{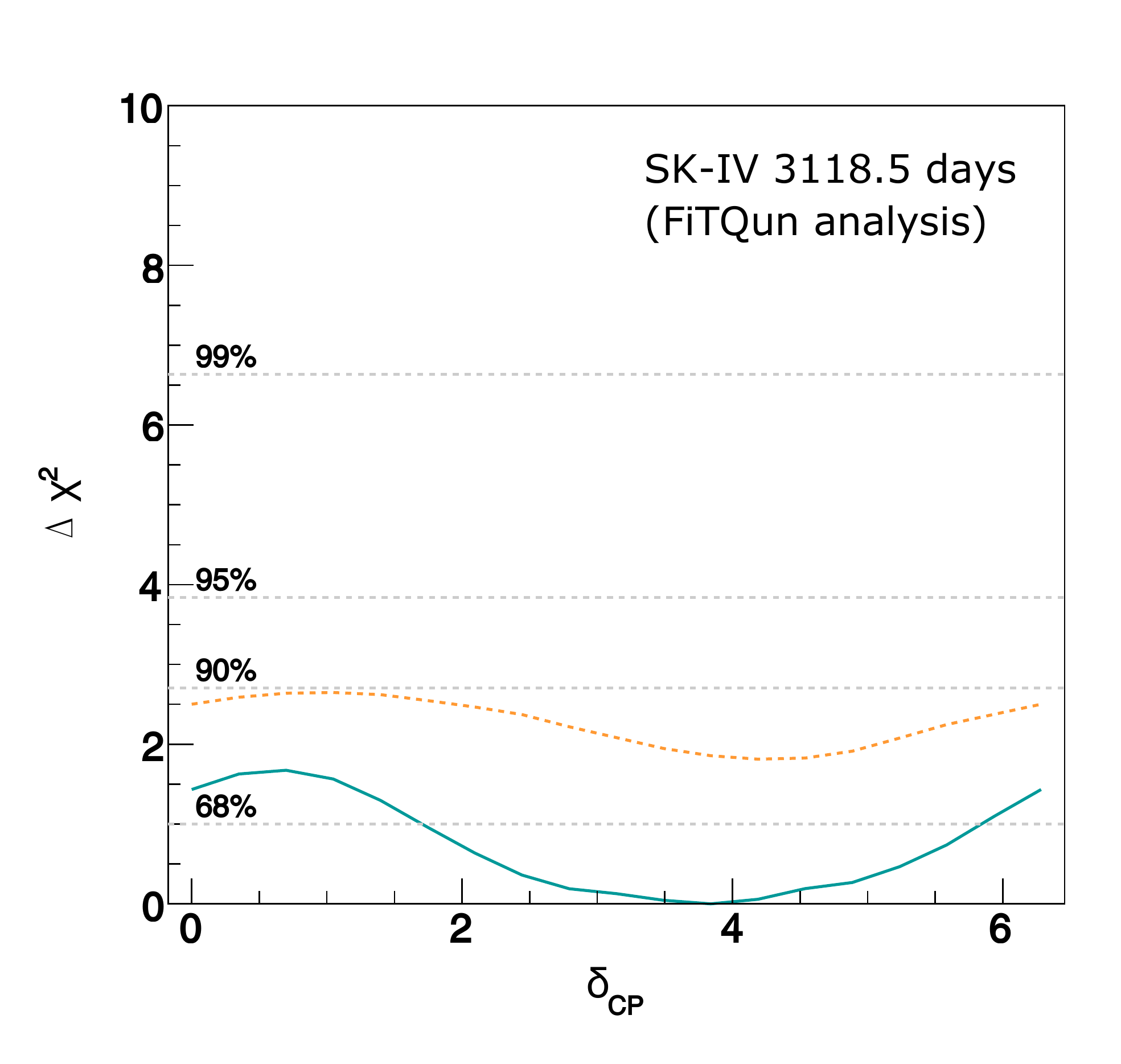}
  \caption{ Constraints on neutrino oscillation parameters from
            the Super-K atmospheric neutrino data with no assumed constraint on $\theta_{13}$. 
            The solid blue and dashed orange lines denote the normal and inverted hierarchy fit results, respectively.
            The latter has been offset from the former 
            by the difference in their minimum $\chi^{2}$ values.}
  \label{fig:q13free_sk_cont}
\end{figure*}

\begin{table}
\setlength\tabcolsep{10pt}
\begin{center}
\caption{Parameter grid used in the fit.}
\label{tbl:grid}
\begin{tabular}{lcc}
\hline 
\hline 
Parameter & Range & Scan points  \\
\hline 

$|\Delta m^{2}_{32,31}|$             & $1.5 \times 10^{-3} \sim 3.5 \times 10^{-3}$& 40 points  \\   
$\sin^{2}\theta_{13}$ &  $0.0 \sim 0.06$& 25 points   \\  
$\sin^{2}\theta_{23}$ &  $0.3 \sim 0.725$& 35 points \\  
$\delta_{CP}$ &  $0.0 \sim 2 \pi $& 19 points \\  
\hline 
\hline 
\end{tabular}
\end{center}

\end{table}

\begin{table}
\begin{center}
\caption{Values of oscillation parameters fixed in the analysis and
  their systematic errors.  Note that $\sin^{2} \theta_{13}$ is
  only fixed in the ``$\theta_{13}$ constrained''analyses described in
  Section~\ref{sec:gaibu}.}
\label{tbl:oscparm}
\begin{tabular}{lc}
\hline 
\hline 
Parameter & Value  \\
\hline 

$\Delta m^{2}_{21}$             & $(7.53 \pm 0.18) \times 10^{-5} \mbox{eV}^{2}$  \\  
$\sin^{2}    \theta_{12}$ & $0.307  \pm 0.013$  \\  
$\sin^{2}    \theta_{13}$ & $0.0210 \pm 0.0011$  \\  
\hline 
\hline 
\end{tabular}
\end{center}

\end{table}

The fit with $\theta_{13}$ unconstrained by reactor measurements 
is performed over four parameters, $|\Delta m^{2}_{32,31}|$, $\sin^{2} \theta_{13}$,$\sin^{2} \theta_{23}$ and $\delta_{CP}$ with a total of 71 systematic errors. 
Fits to the normal hierarchy use $\Delta m^{2}_{32}$ while those for the inverted hierarchy use $\Delta m^{2}_{31}$.
The agreement between the data and MC is evaluated using Equation~\ref{eq:fullchi} at each point in
the grid of points shown in Table~\ref{tbl:grid}, while keeping the solar mixing parameters, $\Delta m^{2}_{12}$ and $\sin^{2} \theta_{12}$, 
fixed to the values in Table~\ref{tbl:oscparm}. 
Systematic errors representing the uncertainty in these parameters are included in the analysis. 
The best-fit parameters for each hierarchy are defined as the parameter set returning the smallest $\chi^{2}$ value. 
Between the two hierarchy fits the one with smallest minimum $\chi^{2}$ value is taken as the global best-fit and 
is taken to indicate the mass hierarchy preference.

\subsection*{Results and Discussion }
Figure~\ref{fig:sens_sk_q13_free} shows the expected sensitivity to reject a zero value of $\sin^{2} \theta_{13}$ 
as a function of the true value of sin$^{2}\theta_{23}$ for 
the fiTQun-based analysis samples with both the conventional and expanded FV as well as an analysis with APFit-reconstructed samples with the 
conventional FV. 
The sensitivities for the two algorithms using the same conventional FV are similar since the fraction of events common between them is larger than 97\%. The expected number of atmospheric neutrino events selected by APFit and fiTQun are 24188.7 and 24184.3, respectively, while the number of the events passed both the cut of the two algorithms is 23520.2.
The 3\% difference is mainly due to the vertex resolution near  FV boundary.

However, since the contamination of $\nu_{\mu}$ events in the e-like samples most sensitive to $\theta_{13}$ is lower 
for the fiTQun-selected sample as shown in Table \ref{tbl:mubkg}, there is a slight boost in its sensitivity.
Expanding the FV naturally incorporates more signal events in the analysis and the sensitivity improvement is clear.

\begin{table}
\begin{center}
\caption{ $\nu_{\mu} + \bar\nu_{\mu}$ CC backgrounds in multi-GeV e-like sample. }
\label{tbl:mubkg}
\begin{tabular}{lccc}
\hline
\hline
Reconstruction & fiTQun&  fiTQun &  APFit  \\
Dwall range & 50 $\sim$ 200~cm& $> 200$~cm & $>$ 200~cm  \\
\hline  
Single-ring  \\
\hspace*{8pt} $\nu_{e}$-like & 0.064  &  0.058&  0.109 \\
\hspace*{8pt} \nuebar-like   & 0.003  &  0.003&  0.009 \\
Multi-ring  \\                                
\hspace*{8pt} \nue-like      &  0.059&  0.054 & 0.121   \\
\hspace*{8pt} \nuebar-like   & 0.023 &  0.021 & 0.040   \\
\hline 
\hline 
\end{tabular}
\end{center}
\end{table}

One-dimensional allowed regions
for $\theta_{13}$, $|\Delta m^{2}_{32,31} |$, $\sin^{2} \theta_{23}$ 
 and $\delta_{CP}$ from the fit to the data are shown in Figure~\ref{fig:q13free_sk_cont}. 
The normal hierarchy hypothesis yields better data-MC agreement than the inverted hierarchy hypothesis with 
$\chi^{2}_{\text{NH,min}} - \chi^{2}_{\text{IH,min}} = -1.81$.
The 1$\sigma$ allowed region for $\sin^{2} \theta_{13}$ is from 0.003 to 0.033 (from 0.001 to 0.023) for 
the normal (inverted) hierarchy fit, which is consistent with the globally preferred value. 
The point at $\sin^{2} \theta_{13} = 0.0$ is disfavored at approximately $1.8{\sigma}$ (1.2 $\sigma$) for normal (inverted) fit. 
A summary of the best-fit information and parameter constraints is presented in Table~\ref{tbl:bestfits}.

\begin{figure*}[htpb]
\includegraphics[width=0.90\textwidth]{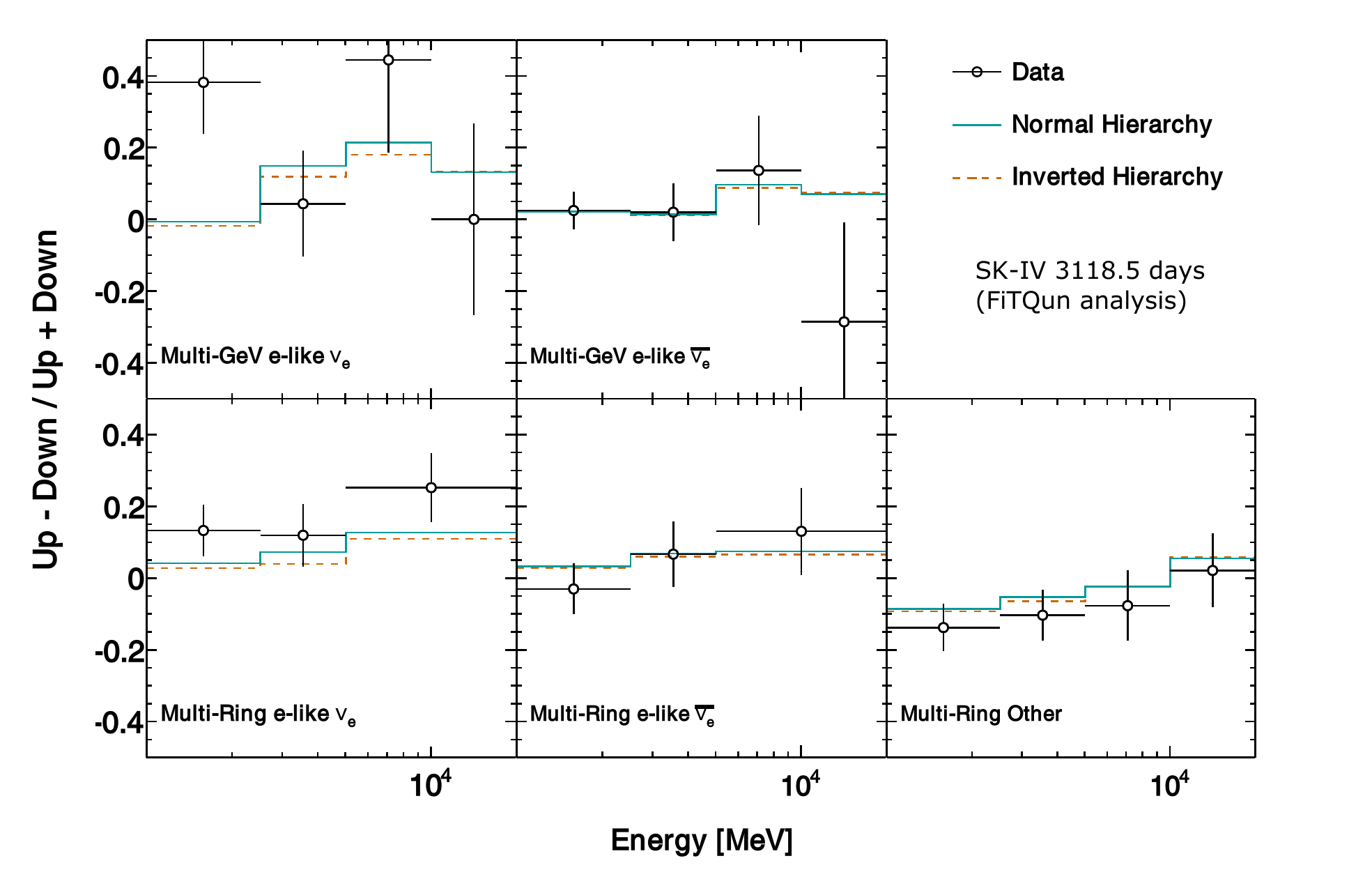}
\caption{ Ratio of upward- ($\mbox{cos}\theta < -0.4$) to downward-going
  ($\mbox{cos}\theta > 0.4$) events as a function of energy for 
  the mass hierarchy-sensitive analysis samples. 
  energy. The error bars are statistical. 
For the single-ring samples the energy is taken to be the visible energy 
  assuming the observed event was an electron. 
  The energy estimator for multi-ring samples is the total 
  observed energy summed over each reconstructed ring  
  after adjusting for each ring's PID.
  The orange line shows the best-fit result assuming the inverted hierarchy hypothesis 
  and the cyan that from the normal hierarchy hypothesis. }
\label{fig:ud_ratio}
\end{figure*}

As discussed in Section~\ref{sec:oscillations}, the determination of $\sin^{2} \theta_{13}$ and 
the mass hierarchy using atmospheric neutrinos is achieved using upward-going electron neutrino appearance at multi-GeV energies.
Figure~\ref{fig:ud_ratio} shows the up-down asymmetry of the multi-GeV single- and multi-ring electron-like samples,
where the asymmetry is defined as the ratio of the difference of the number of downward-going and downward-going events relative to their sum. 
Here downward-going (upward-going) events are events whose zenith angle satisfy $\mbox{cos}\theta_{z} < -0.4$ ($\mbox{cos}\theta_{z} > 0.4$). 
Excesses between a few and ten GeV in the multi-GeV e-like $\nu_{e}$ and the Multi-Ring e-like $\nu_{e}$ samples, 
drive the normal hierarchy preference.

The best-fit atmospheric mixing parameters from the normal hierarchy fit are   
$\Delta m^{2}_{32} = 2.63^{+0.10}_{-0.21}\times 10^{-3} \mbox{eV}^{2}$ and $\sin^{2}
\theta_{23} = 0.588^{+0.030}_{-0.062}$ for second octant (best-fit) and $0.425^{+0.051}_{-0.034}$ for first octant. 
Maximal mixing ($\sin^{2} \theta_{23} = 0.5$) is weakly disfavored at around $1{\sigma}$ significance. 
Besides the data excesses in the upward-going regions of the single-ring e-like $\nu_{e}$ sample and multi-ring other sample, 
the data deficits in the multi-GeV $\mu$-like sample also contribute to this preference. 
It should be noted that the preference for $\sin^{2}\theta_{23}$ is coupled to the $\theta_{13}$ measurement,
since both parameters feature in the $\nu_{\mu}\rightarrow \nu_{e}$ oscillation probability.

The best-fit value of $\delta_{CP}$ is 3.84 radians
for both the normal and inverted hierarchy fits.  
Comparing with the least preferred parameter value of 0.8 radians, 
more electron neutrino appearance in the sub-GeV e-like samples are expected to be observed due to 
$\nu_{\mu} \rightarrow \nu_{e}$ oscillations. 
In the multi-GeV region, $\delta_{CP}$ similarly modulates the amount of electron neutrino appearance 
but it is subdominant to the effects of $\theta_{13}$.
In the inverted hierarchy fit, the value of $\theta_{13}$ can be more easily 
adjusted to bring the MC prediction into better agreement with data since only 
antineutrino events experience matter effects in the earth.
Since there are more neutrino events than antineutrino events in the atmospheric sample, 
the freedom to move $\theta_{13}$ ultimately leads to a weaker constraint on $\delta_{CP}$.

 \subsection{Analysis with constrained $\theta_{13}$}
\label{sec:gaibu}

Since atmospheric neutrinos come from all directions, carry a wide range of energies, and often produce particles that are invisible 
to Super-K it is not possible in general to fully reconstruct the neutrino kinematics. 
As a result atmospheric neutrinos themselves do not typically have oscillation parameter sensitivity at the same level 
as that in long-baseline or reactor experiments, such that the introduction of constraints from 
those measurements can improve sensitivity to the mass hierarchy and and $\delta_{CP}$.
Indeed, Equation~\ref{eqn:oscprob_inmatter} shows that the size of MSW resonance relies on the value of $\mbox{sin}^{2}\theta_{13}$ directly.
Since reactor neutrino experiments constrain this parameter more precisely than the analysis in the 
previous section, the analysis presented here constrains the value of $\sin^{2} \theta_{13}$ to $0.0210 \pm 0.0011$~\cite{Olive:2016xmw}.
During the fit the parameter is held at its central value and a systematic error that imparts the impact of 
its uncertainty is added to the analysis.
The fit uses the same data samples, binning, and parameter grid for all other oscillation parameters as the analysis presented above. 

\subsection*{Results and Discussion }

\begin{figure*}[htpb]
\includegraphics[width=0.47\textwidth]{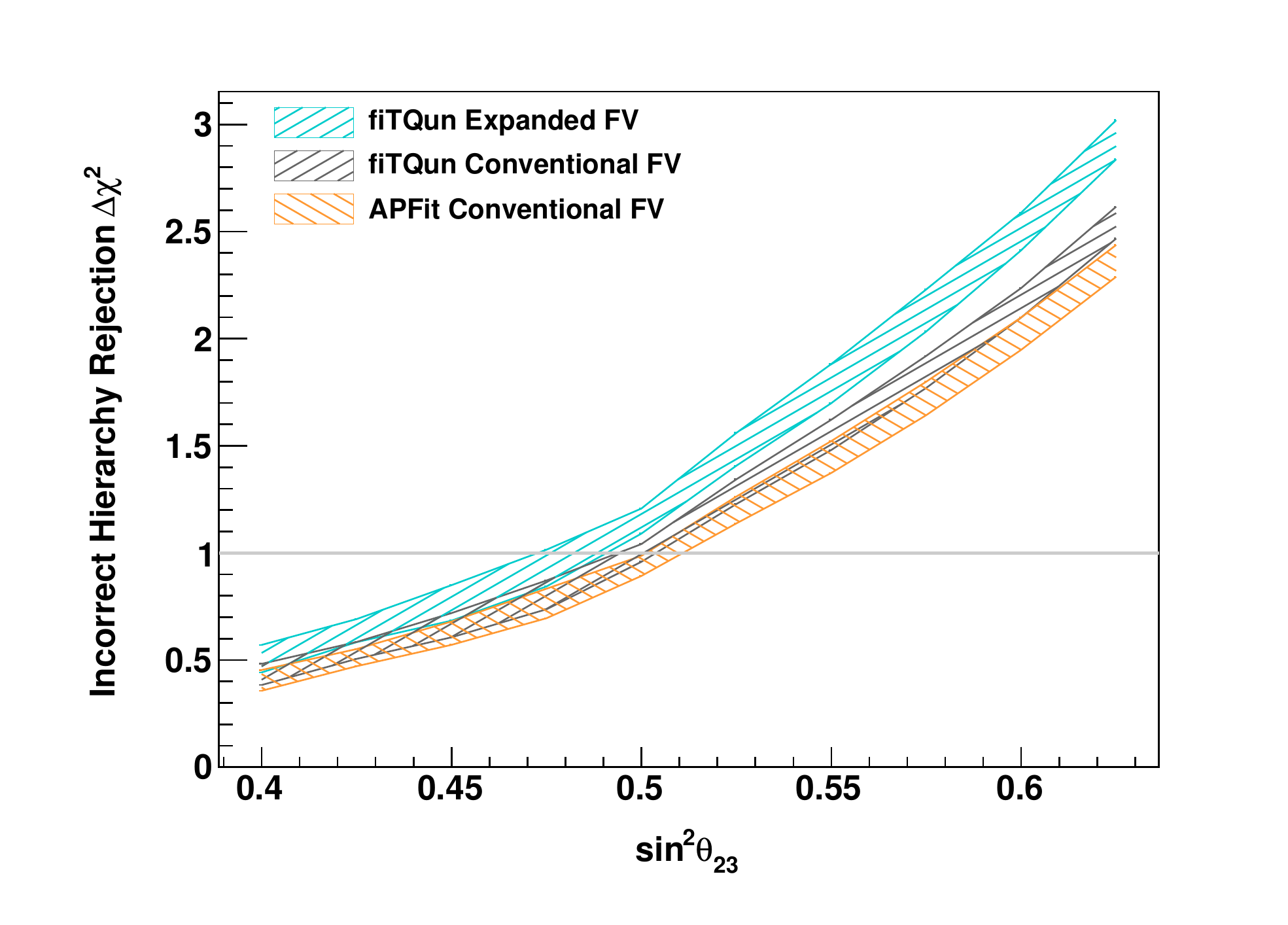}
\includegraphics[width=0.47\textwidth]{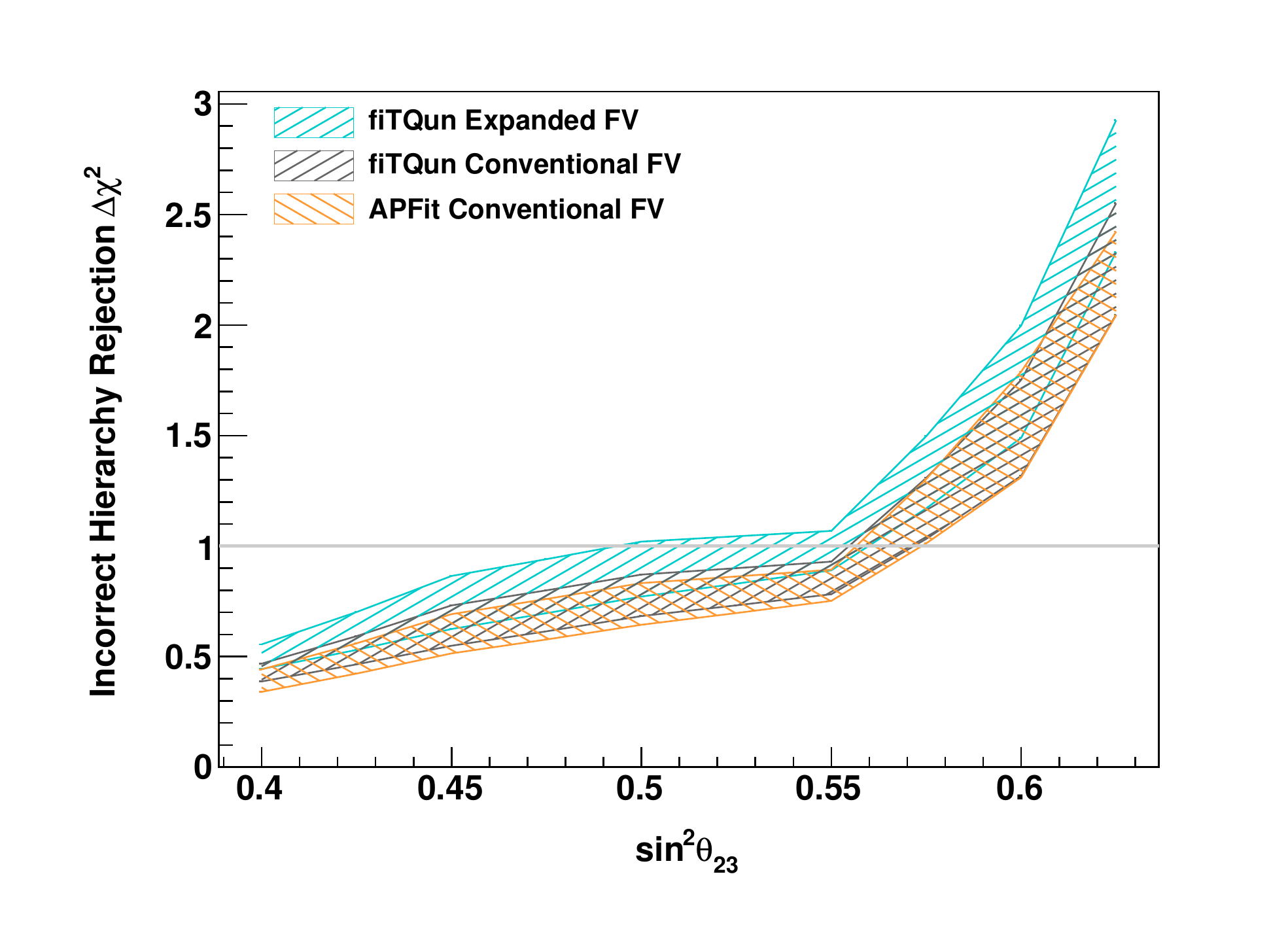}
\caption{ Expected sensitivity to the normal mass hierarchy (left) and inverted hierarchy (right) as a function 
of the true value of $\sin^2 \theta_{23}$. 
Here $\sin^{2} \theta_{13} = 0.0210 \pm 0.0011$ and  
the assumed livetime is 3118.5 days. Grey and blue bands
show the sensitivity of the analysis with event samples reconstructed with fiTQun in the conventional FV and expanded FV, respectively. 
Orange lines denote the sensitivity when events are reconstructed using the APFit algorithm with the conventional FV. 
The width of the bands corresponds to the uncertainty from $\delta_{CP}$. 
}
\label{fig:hier_sens}
\end{figure*}

\begin{figure}[htpb]
\centering
\includegraphics[width=0.49\textwidth]{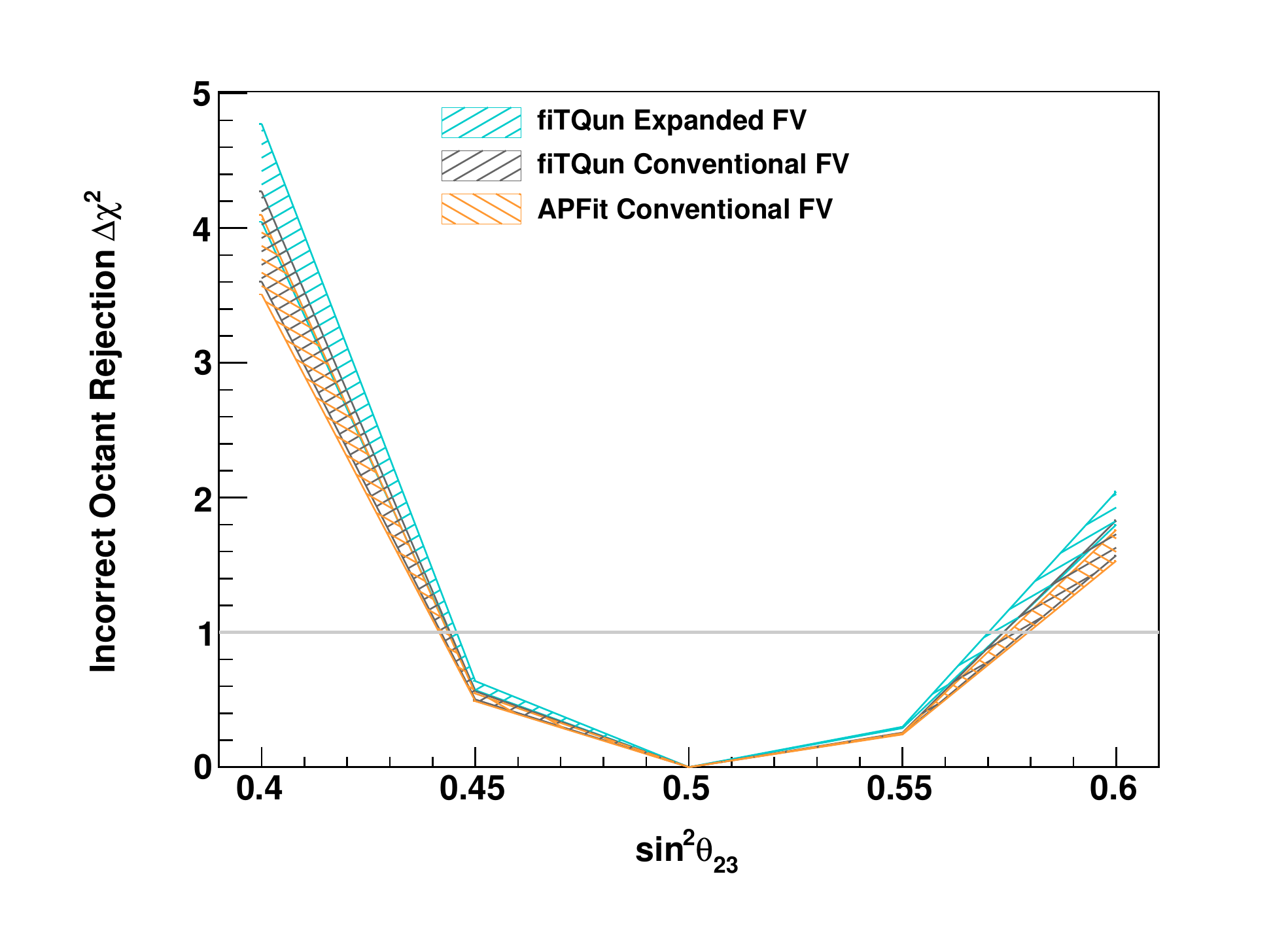}
\caption{ Expected sensitivity to reject wrong octant as a function of the true value of $\sin^2 \theta_{23}$ for true normal mass hierarchy. The other conditions are the same as Figure~\ref{fig:hier_sens}.
}
\label{fig:oct_sens}
\end{figure}

The expected sensitivity to the mass hierarchy and $\theta_{23}$ octant for true normal hierarchy are illustrated in Figure~\ref{fig:hier_sens} and Figure~\ref{fig:oct_sens} for 
the fiTQun-reconstructed analysis samples using both the conventional and expanded FV and 
for the APFit-based analysis with the conventional FV. 
Improved sensitivity is found with the fiTQun-based analyses regardless of the fiducial volume used 
for the same reasons as produced the improved sensitivity to $\theta_{13}$ in the previous section.

\begin{figure*}[htpb]
\centering
  \includegraphics[width=0.25\textwidth]{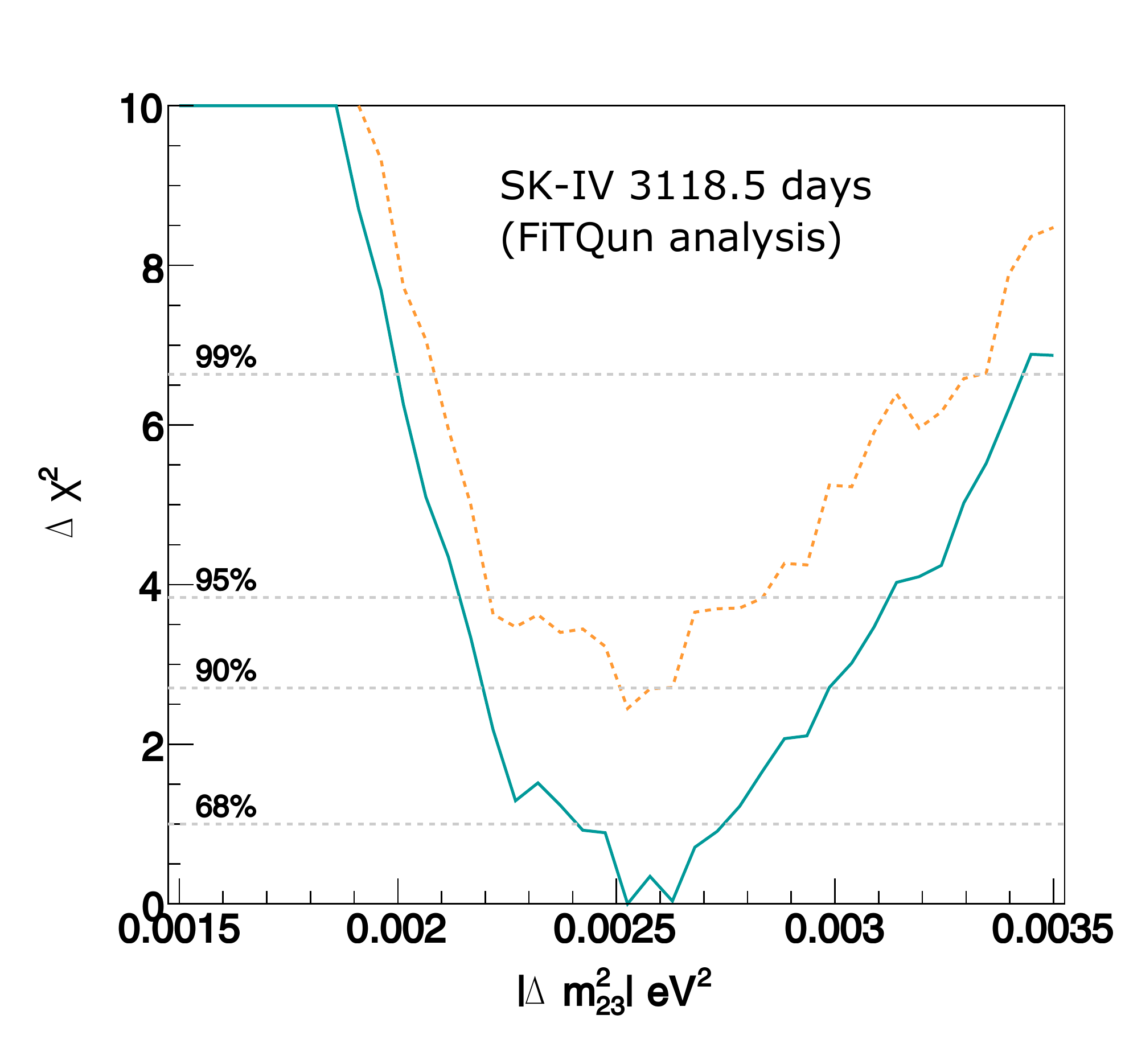}
  \includegraphics[width=0.25\textwidth]{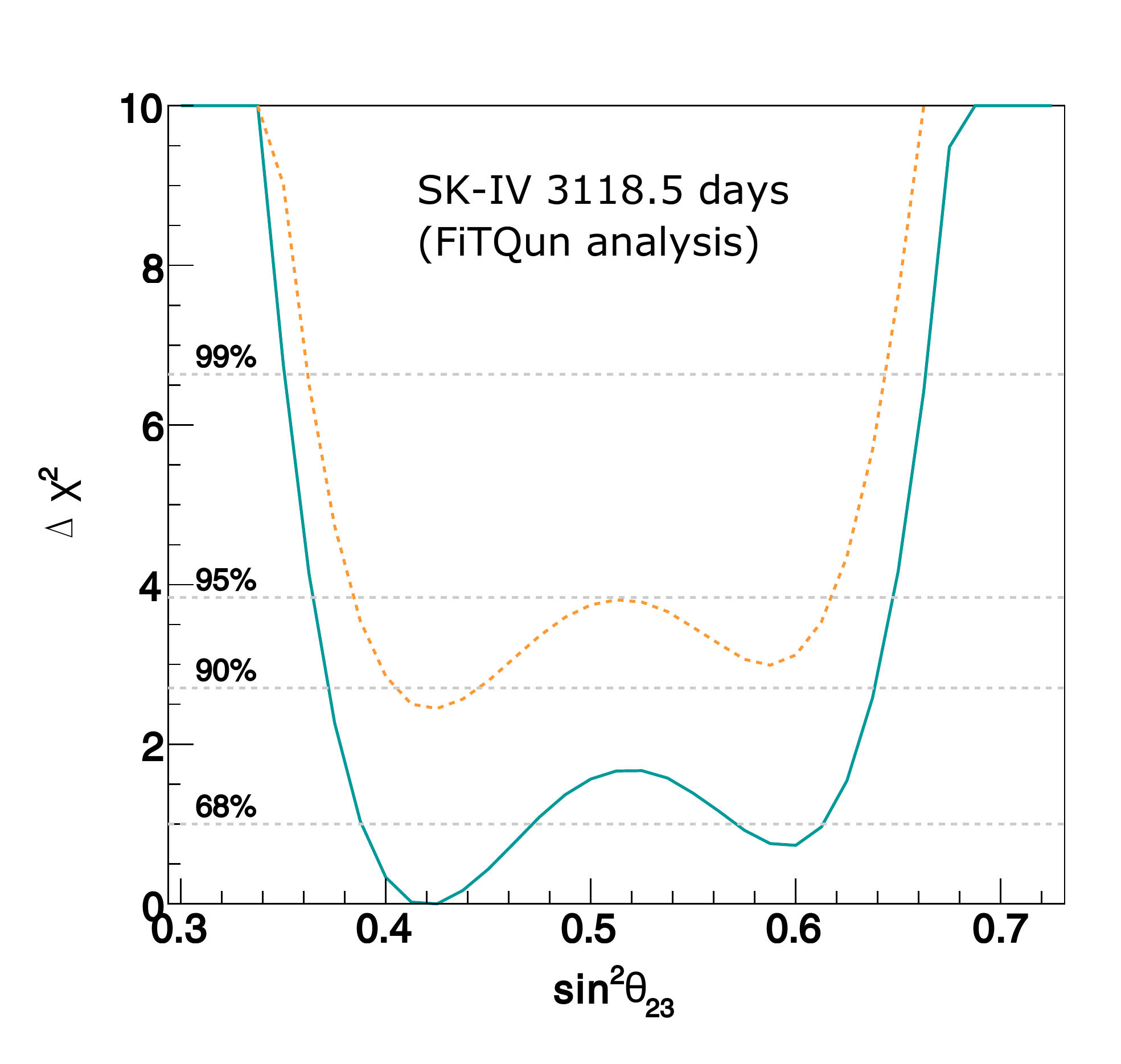}
  \includegraphics[width=0.25\textwidth]{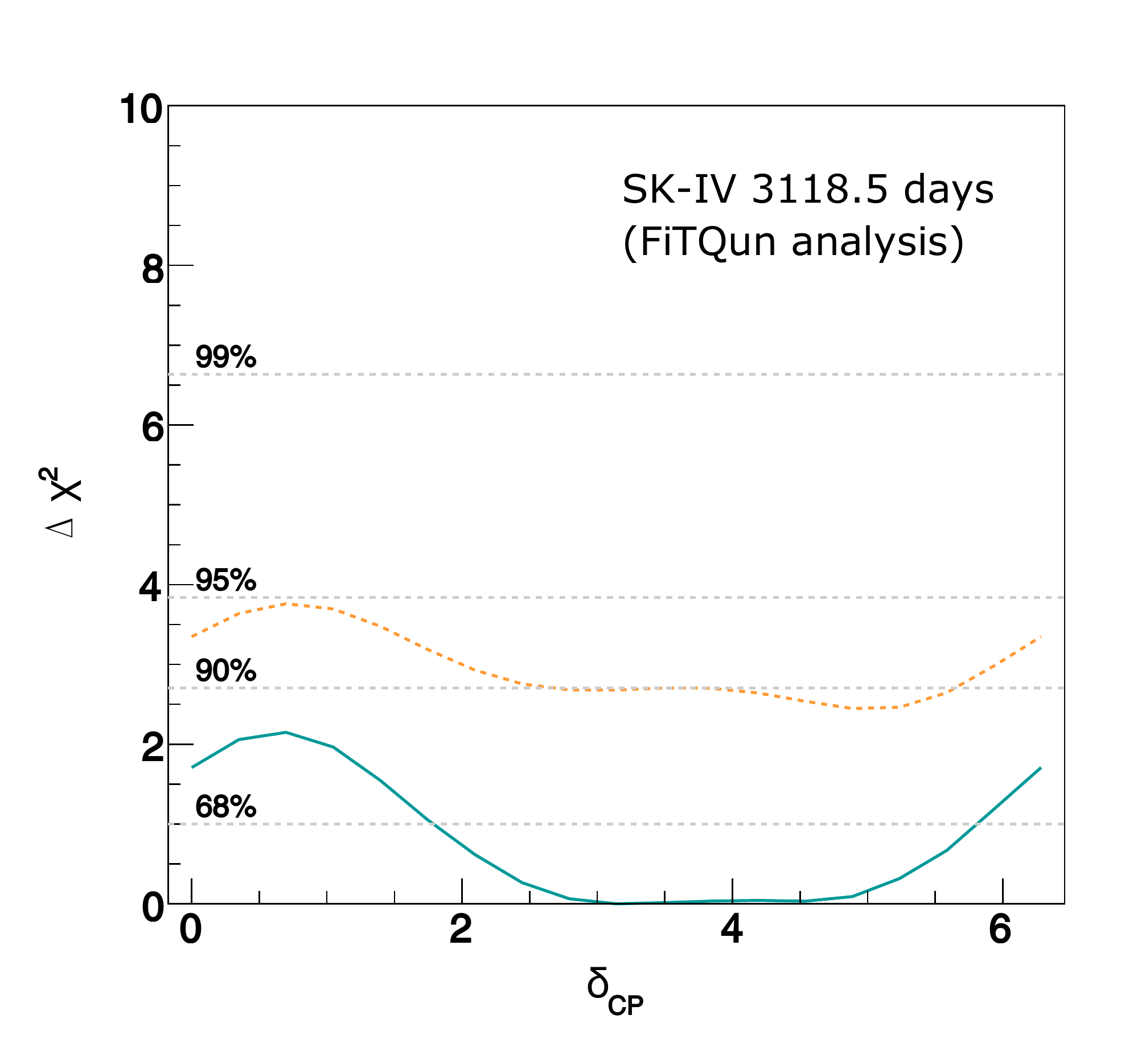}
  \caption{ Constraints on neutrino oscillation parameters from SK-IV atmospheric neutrino data using the expanded FV 
            and assuming $\mbox{sin}^{2} \theta_{13} = 0.0210 \pm 0.0011 $.
            The solid blue and dashed orange lines denote the normal and inverted hierarchy fit results, respectively.
            The latter has been offset from the former 
            by the difference in their minimum $\chi^{2}$ values.}
  \label{fig:skefv_cont}
\end{figure*}

\begin{figure}[htpb]
\centering
  \includegraphics[width=0.45\textwidth]{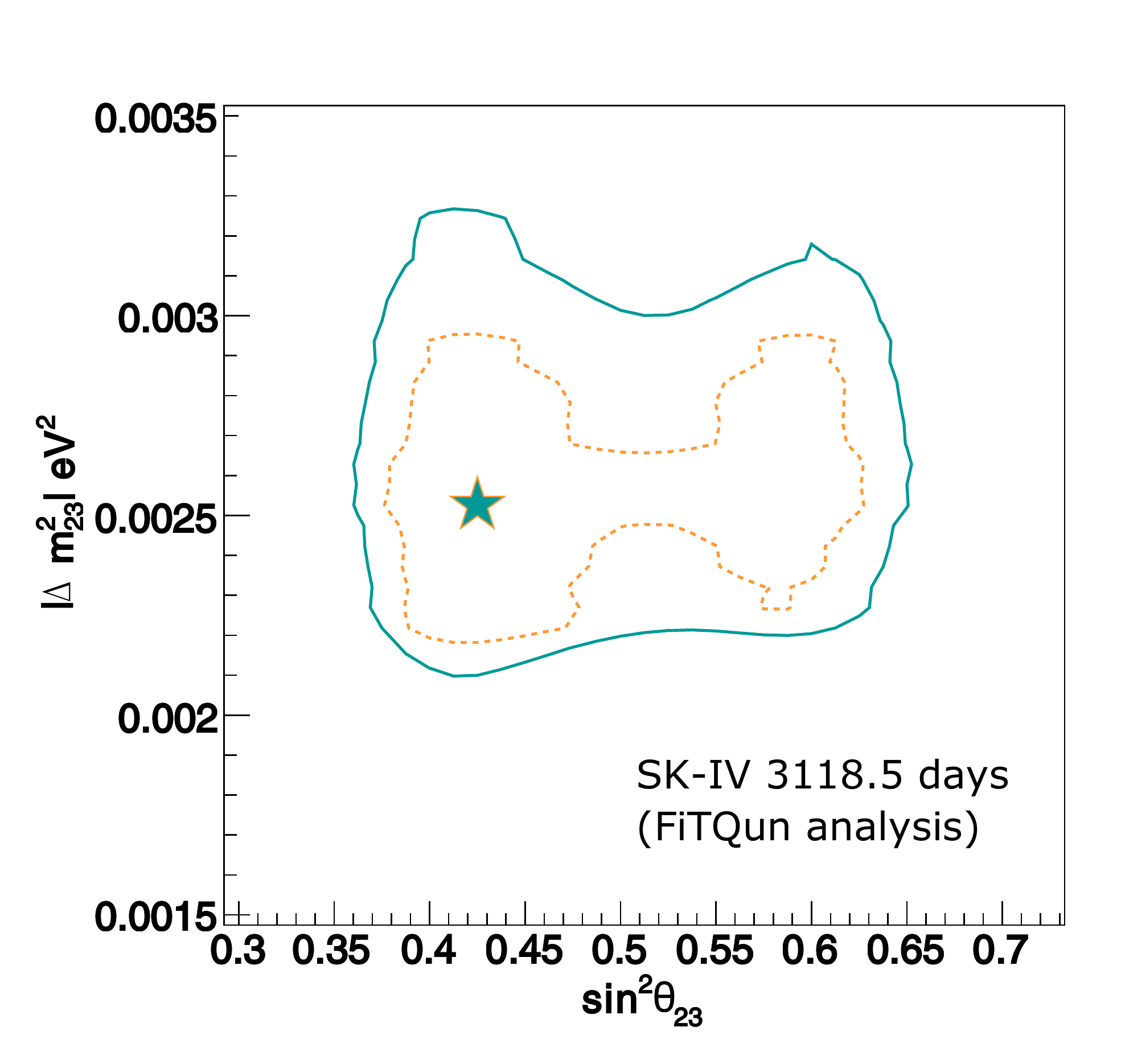}
  \caption{ Constraints on the atmospheric mixing parameters using SK-IV atmospheric neutrino and the expanded FV. 
            The solid blue (dashed orange)  line shows 90\% C.L. for the normal (inverted) hierarchy. The star denotes the best-fit value, which is at the same point for normal and inverted hierarchy, as shown in Table~\ref{tbl:bestfits}. 
            In each contour $\mbox{sin}^{2} \theta_{13}$ is constrained to be $0.0210 \pm 0.0011$.
            The contours have both been drawn relative to the global best-fit.}
  \label{fig:twod}
\end{figure}

Figure~\ref{fig:skefv_cont} shows the $\chi^2$ value as a 
function of the atmospheric neutrino mixing parameters and $\delta_{CP}$ in the $\theta_{13}$-constrained fit. 
As in the unconstrained fit the data prefer the normal mass hierarchy with 
$\Delta \chi^{2} = -2.45$.
The 1$\sigma$ allowed region for $|\Delta m^{2}_{32}|$ is from 2.41 to 2.75 
$\times 10^{-3} \mbox{eV}^2$ (2.36 to 2.67 $\times 10^{-3}\mbox{eV}^2$) for the normal (inverted) hierarchy fit, which is consistent with the result of unconstrained fit.
The contour for the allowed region with 90\% CL is shown in Figure~\ref{fig:twod}.
The preference for the second octant of $\theta_{23}$ in the unconstrained fit is changed. 
Adding the constraint on $\sin^{2} \theta_{13}$ the preferred value of $\theta_{23}$ shifts 
from the second to first octant, though both are allowed at 1$\sigma$. 
Indeed the difference in $\chi^{2}$ between the minimum value of each octant is $\Delta \chi^{2}_{1^{st}-2^{nd}}$ = -0.73 (0.13) 
for the constrained (unconstrained) fit.
This result can be explained by degeneracies in the appearance probability that arise for certain combinations of $\sin^22\theta_{13}$ and $\sin^2\theta_{23}$ as
shown in Equation~\ref{eqn:oscprob_atm_vac}; an increase in one can be compensated by a decrease in the other to describe the data.
The same upward-going event excesses (deficits) in the $e$-like ($\mu$-like) samples 
that drive the hierarchy,  $\theta_{13}$, and octant preferences in the unconstrained analysis are 
not strong enough to support the increased 
excess (deficits) expected for higher values of $\theta_{13}$ and the second octant. 
Since the $\theta_{13}$ is now fixed, the data can be accommodated by reducing the expected number of $e$-like events 
and increasing the number of $\mu$-like events by moving $\theta_{23}$ to the first octant.

The best-fit value of $\delta_{CP}$ is 3.14 (4.89) for the normal (inverted) hierarchy hypothesis, with a tighter constraint compared with the unconstrained fit. 
Parameter values and their $1{\sigma}$ errors are summarized in Table~\ref{tbl:bestfits}.

\begin{table*}
\caption{Parameter estimates for each analysis and hierarchy hypothesis considered. 
         Here NH (IH) refers to the normal (inverted) hierarchy fit.
         The terms ``Unconstrained'' and ``Constrained'' refer to fits without and with a constraint on $\mbox{sin}^{2} \theta_{13}$, respectively.
         For $\sin^{2} \theta_{23}$ parameter ranges are shown for both octants, with the best-fit octant enclosed in a box. 
         The expected absolute $\chi^{2}$ value for the constrained fit is 523.6 
         and the probability for obtaining larger value is 0.244 for the NH. } 
\label{tbl:bestfits}
\begin{center}
\begin{tabular}{lcccc}
\hline 
\hline 
& \multicolumn{2}{c}{$\theta_{13}$ Free}  & \multicolumn{2}{c}{$\theta_{13}$ Constrained}  \\
Hierarchy & NH & IH & NH & IH \\
\hline 
$\chi^{2}$ & 576.3 & 578.1 &576.5 & 579.0 \\
$\mbox{sin}^{2} \theta_{13}$ &  $0.008^{+0.025}_{-0.005}$  &  $0.008^{+0.015}_{-0.007}$  & -- & -- \\
$\mbox{sin}^{2} \theta_{23}$ ($1^{st}$ oct.) & $0.425^{+0.051}_{-0.034}$ &$0.425^{+0.075}_{-0.027 }$ & \mybox{$0.425^{+0.046}_{-0.037}$}  &  \mybox{$0.425^{+0.055}_{-0.036}$} \\
$\mbox{sin}^{2} \theta_{23}$ ($2^{nd}$ oct.) &  \mybox{$0.588^{+0.030}_{-0.062}$}   & \mybox{$0.575^{+0.034}_{-0.075 }$}  & $0.600^{+0.013}_{-0.030}$  & $0.588^{+0.022}_{-0.037}$ \\
$|\Delta m^{2}_{32,31}|$ [$\times 10^{-3}$ eV$^2$] & $2.63^{+0.10}_{-0.21} $ &  $2.53^{+0.14}_{-0.08} $  &  $2.53^{+0.22}_{-0.12} $ &  $2.53^{+0.14}_{-0.31} $  \\
$\delta_{CP}$ &$3.84^{+2.00}_{-2.14}$ & $4.19^{+2.09}_{-4.19}$    &$3.14^{+2.67}_{-1.35}$ &  $4.89^{+1.51}_{-3.46}$    \\
\hline
\hline
\end{tabular}
\end{center}

\end{table*}

\begin{figure}[htpb]
\centering
  \includegraphics[width=0.49\textwidth]{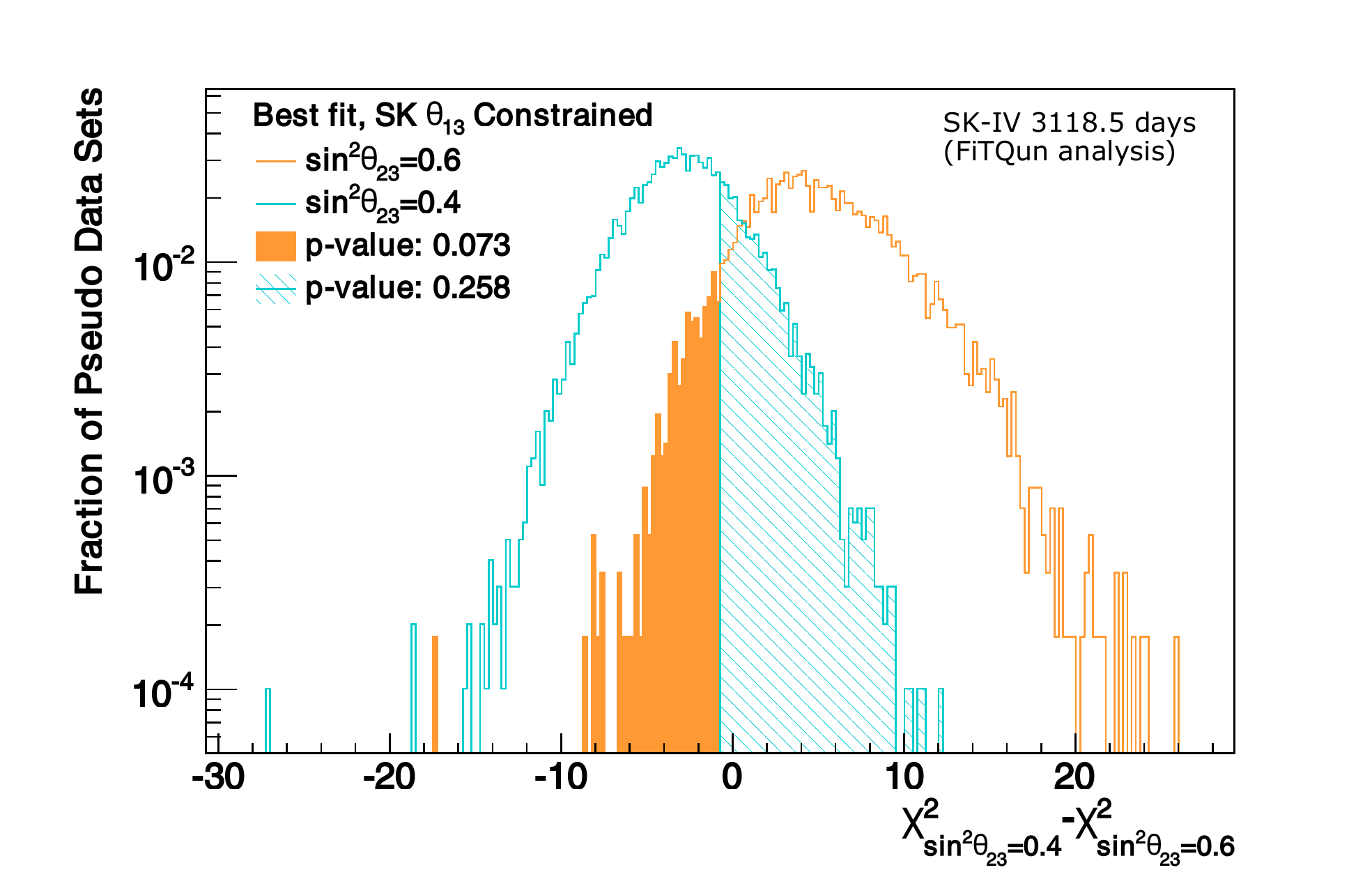}
  \caption{ Distributions of the difference in best-fit $\chi^{2}$ 
            values between first octant and second octant fits 
            to pseudo data sets used in the generation of the $\mbox{CL}^{\mbox{O}}_{s}$ value 
            for the SK $\theta_{13}$ constrained analysis. 
            In the cyan (orange) histogram the pseudo data have been generated 
            assuming $\sin^2\theta_{23}=0.4$ ($\sin^2\theta_{23}=0.6$).
            Shaded portions of the histograms denote the fraction of 
            pseudo data sets with more extreme values than that observed in the data, 
             $\Delta \chi^{2}_{data} = -0.73$.}
  \label{fig:oct_pvalue}
\end{figure}

The $\mbox{CL}_{s}$ method \cite{Read:2002hq} is used to address the significance of the octant and mass hierarchy preferences observed in the data.
As in the previous Super-K study~\cite{Abe:2017osc} 
MC ensembles assuming different parameter combinations of the octant and hierarchy were generated with statistical and systematic error variations
applied to each pseudo data set.
Statistical variations have been applied assuming the current detector exposure and systematic errors have been varied as Gaussian parameters 
with widths specified by their uncertainties.
The $\mbox{CL}_{s}$ parameter for the octant study, 
$\mbox{CL}^{\mbox{O}}_{s}$, is defined as:
\begin{equation}
\mbox{CL}^{\mbox{O}}_{s} = \frac{ p_{0}( O_2 ) }{ 1 - p_{0}(O_1) },
\end{equation}
\noindent where $p_{0}(O_{i})$ represents the $p$-value with respect to obtaining a difference 
in the best-fit $\chi^2$ values between both octant hypotheses smaller (larger) than the value of data assuming 
the true value of $\mbox{sin}^2\theta_{23}$ is $O_2 = 0.6$ ($O_1 = 0.4$).
These pseudo MC distributions of these hypotheses are shown as the orange and cyan shaded histograms in Figure~\ref{fig:oct_pvalue}, respectively.
The $\mbox{CL}^{\mbox{O}}_{s}$ value is found to be just 0.098 and accordingly, the data show no strong preference for the octant.

Similarly the parameter for the mass hierarchy is defined as 
\begin{equation}
\mbox{CL}^{\mbox{H}}_{s} = \frac{ p_{0}(\mbox{IH}) }{ 1 - p_{0}(\mbox{NH}) },
\end{equation}
where $p_{0}(\mbox{IH})$ and $p_{0}(\mbox{NH})$ are p-values for obtaining a difference 
in the $\chi^{2}$ of the best-fit mass hierarchies more extreme than that of the data 
assuming a true IH and NH, respectively.

\begin{figure}[htpb]
\centering
  \includegraphics[width=0.49\textwidth]{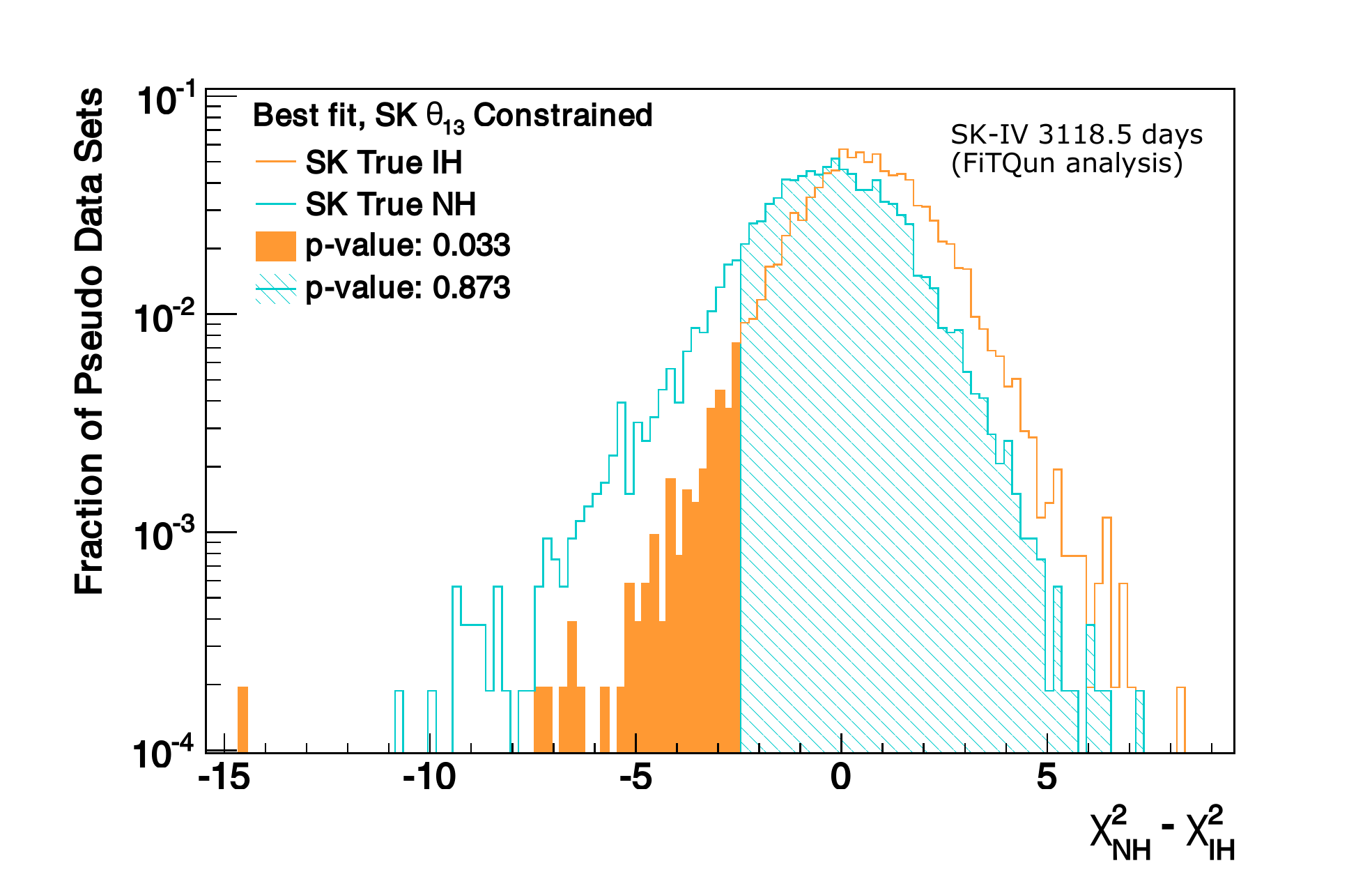}
  \caption{ Distributions of the difference in best-fit $\chi^{2}$ 
            values between normal- and inverted-hierarchy fits 
            to pseudo data sets used in the generation of the $\mbox{CL}^{\mbox{H}}_{s}$ value 
            for the SK $\theta_{13}$ constrained analysis. 
            In the cyan (orange) histogram the pseudo data have been generated 
            assuming the normal (inverted) hierarchy at the analysis best-fit 
            shown in Table~\ref{tbl:bestfits}. 
            Shaded portions of the histograms denote the fraction of 
            pseudo data sets with more extreme values than that observed in the data, 
             $\Delta \chi^{2}_{data} = -2.45$.}
  \label{fig:cls}
\end{figure}

Figure~\ref{fig:cls} shows the distribution for the mass hierarchy determination. Due to the large uncertainty on $\theta_{23}$, MC ensembles 
have been generated with different assumed values $\theta_{23}$ and with the other oscillation parameters fixed to their best-fit values.
The $p$-values and $\mbox{CL}^{\mbox{H}}_{s}$ values for hierarchy test are summarized in Table \ref{tbl:pval}.
As sin$^2\theta_{23}$ ranges from 0.4 to 0.6 the observed $\mbox{CL}^{\mbox{H}}_{s}$ values 
decrease from a 30.8\% C.L. preference for the inverted mass hierarchy to 14.3\%.
At the analysis best-fit value the preference is 26.0\%.
A smaller value of $\mbox{sin}^{2} \theta_{23}$ predicts less electron neutrino appearance, which results in a smaller 
$p_{0}(\mbox{IH})$ values and larger $\mbox{CL}^{\mbox{H}}_{s}$ 
due to the multi-GeV $e$-like samples' event excesses seen in data.
\begin{table}
\caption{Normal hierarchy significance summarized in terms of the
  probability of observing a $\chi^{2}$ preference for the NH more
  extreme than that observed in the data assuming an IH, $p_{0}(\mbox{IH})$ 
  , and
  $\mbox{CL}^{\mbox{H}}_{s}$ values for a range of assumed parameters. 
  The first row
  shows the true $\theta_{23}$ used to generate MC ensembles used in the calculations. 
  Other oscillation parameters are taken from the analysis' best-fit. The value of $\theta_{13}$ is constrained to 0.0210 $\pm$ 0.0011.} 
\label{tbl:pval}\begin{center}
\begin{tabular}{lcccc}
\hline
\hline 
True sin$^2\theta_{23}$ &0.4 &0.425 & 0.5&0.6 \\
\hline 
$p_{0}(\mbox{IH})$ & 0.025 & 0.033 & 0.065 & 0.072  \\
$\mbox{CL}^{\mbox{H}}_{s}$ & 0.308 & 0.260 & 0.229 & 0.143 \\
\hline 
\hline 
\end{tabular}
\end{center}

\end{table}

 \section{Conclusion}
\label{sec:conclusion}

\begin{figure}[h]
\centering
\includegraphics[width=0.49\textwidth]{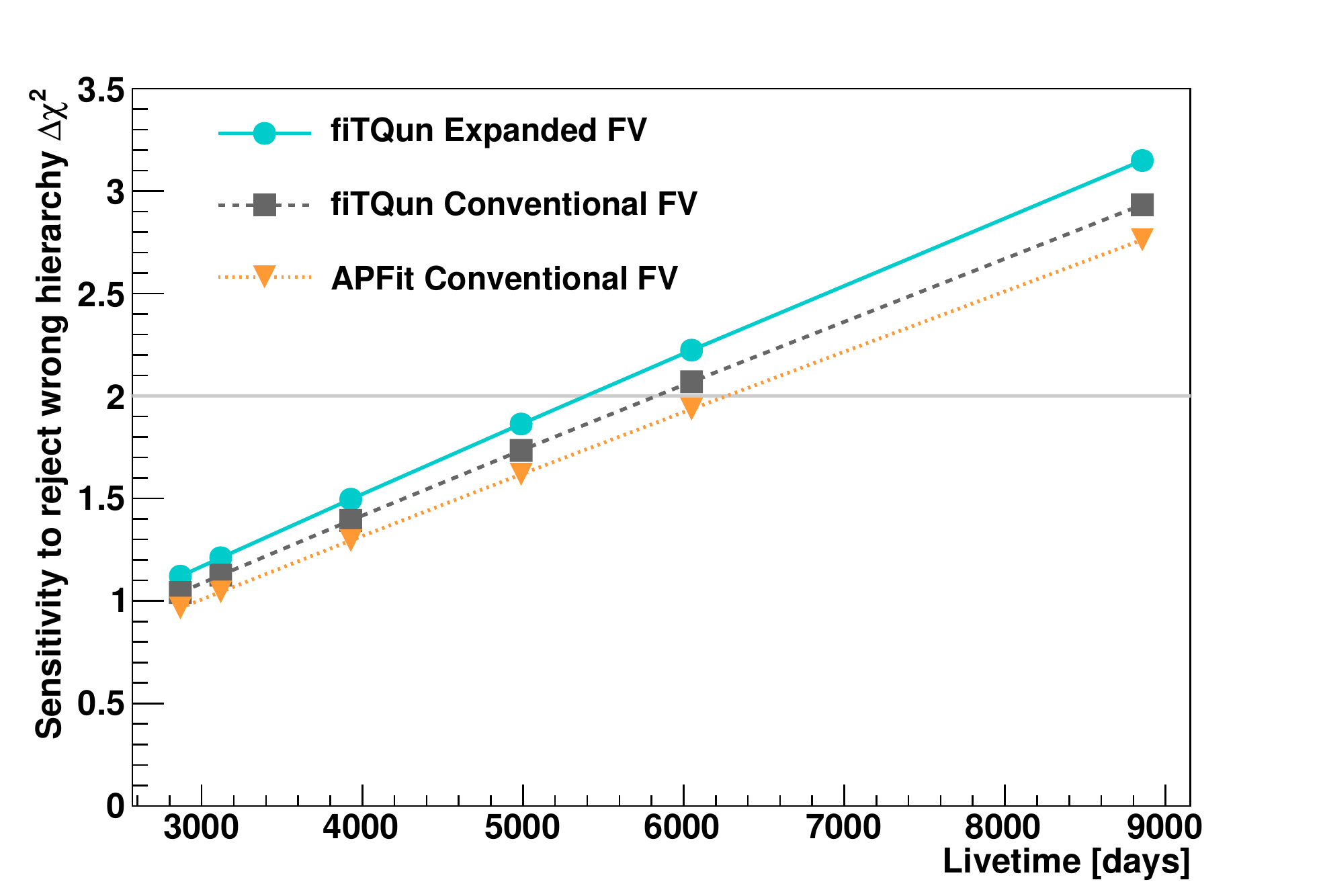}
\caption{ Expected sensitivity to the normal mass hierarchy as a function 
of livetime assuming $\sin^2 \theta_{23}$=0.5. Grey and blue bands
show the sensitivity of the atmospheric neutrino sample reconstructed with fiTQun in the conventional FV and expanded FV, respectively. 
Orange lines denote the sensitivity when events are reconstructed with APFit in the conventional FV. }
\label{fig:hier_sens_pot}
\end{figure}

A new reconstruction algorithm based on a maximum likelihood method has been developed for Super-K.
Compared to the conventional reconstruction algorithm, the new algorithm shows improved performance 
in a variety of metrics including event vertex resolution, particle momentum resolution, and particle identification.  
The new algorithm has further demonstrated reliable performance over a larger volume of the detector than the previous volume.
Accordingly, in the present analysis the fiducial volume definition has been expanded to include all events 
reconstructed more than 50~cm from any ID wall.
This represents a 32\% increase in the number of events available for analysis relative to the 200~cm threshold used in previous 
SK analyses. 

Using the new algorithm with its expanded fiducial volume definition an analysis of a 253.9~kton-year exposure of the SK-IV 
atmospheric data has yielded oscillation parameter estimates consistent with both previous Super-K measurements and 
results from other experiments. 
Assuming the normal mass hierarchy constraints on the atmospheric
mixing parameters are $\mbox{sin}^{2} \theta_{23} =
0.425^{+0.046}_{-0.037}$ ($0.600^{+0.013}_{-0.030}$) for first (second) octant and $\Delta m^{2}_{32} = 2.53^{+0.22}_{-0.12}$, with $\delta_{CP} = 3.14^{+2.67}_{-1.35}$.  
The data show a weak preference for the normal mass hierarchy, disfavoring the inverted mass hierarchy at
74.0\% assuming oscillation parameters at the analysis best-fit point.
No strong preference for the $\theta_{23}$ octant is observed.

Figure~\ref{fig:hier_sens_pot} shows the expected sensitivity to mass hierarchy as the function of livetime for both 
reconstruction algorithms and their fiducial volumes. 
The expected improvement in sensitivity with the new algorithm becomes more apparent as data is accumulated, 
even assuming the conventional FV.
While the new reconstruction has only been applied to the 3118.5 day SK-IV fully contained data set,
future efforts will expand this study to include other Super-K samples and run periods, which constitute 
an additional 2,800 days of livetime.
 
\section{Acknowledgments}
We gratefully acknowledge the cooperation of the Kamioka Mining and
Smelting Company. The Super-Kamiokande experiment has been built and
operated from funding by the Japanese Ministry of Education, Culture,
Sports, Science and Technology, the U.S. Department of Energy, and the
U.S. National Science Foundation. Some of us have been supported by
funds from the National Research Foundation of Korea NRF-2009-0083526
(KNRC) funded by the Ministry of Science, ICT, and Future Planning,
the European Union H2020 RISE-GA641540-SKPLUS, the Japan Society for
the Promotion of Science, the National Natural Science Foundation of
China under Grants No. 11235006, the National Science and Engineering
Research Council (NSERC) of Canada, the Scinet and Westgrid consortia
of Compute Canada, and the National Science Centre of Poland
(2015/17/N/ST2/04064, 2015/18/E/ST2/00758).

\bibliographystyle{ptephy}

\section{Systematic Uncertainties}\label{sec:systematics}

\renewcommand{\arraystretch}{1.0}
\newcolumntype{E}{D{.}{.}{-1}}
\renewcommand{\tabcolsep}{0pt}

\begin{center}
\begin{longtable}{lllEE}
\caption{Systematic error used in this analysis.
The second column shows the best-fit value of the systematic error parameter in percent and the third column shows the estimated $1{\sigma}$ error size in percent. The fit result within expanded FV and $\theta_{13}$ constrained to $0.0210 \pm 0.0011$. The same applies below. The systematic errors about ring counting, particle identification and Multi-ring likelihood selection within conventional fiducial volume (dwall $>$ 200 cm) and within new region (200 cm $>$ dwall $>$ 50 cm) have been merged.}\\
\label{tab:sysa}\\
\hline \hline
\multicolumn{3}{l}{Systematic Error} & \multicolumn{1}{c}{Fit Value (\%)} & \multicolumn{1}{c}{$\sigma$ (\%)} \\
\hline
\endfirsthead
\multicolumn{5}{c}{\tablename\ \thetable\ -- \textit{Continued from previous page}} \\
\hline \hline
\multicolumn{3}{l}{Systematic Error} & \multicolumn{1}{c}{Fit Value (\%)} & \multicolumn{1}{c}{$\sigma$ (\%)} \\
\hline
\endhead
\hline \multicolumn{5}{r}{\textit{Continued on next page}} \\
\endfoot
\hline
\hline
\endlastfoot
Flux normalization                              & $E_\nu < \val{1}{GeV}$\footnotemark[1] &                    
&   7.3 &    25 \\
                                                & $E_\nu > \val{1}{GeV}$\footnotemark[2] &                    
&    9.6 &    15 \\
$(\numu+\numubar)/(\nue+\nuebar)$               & \multicolumn{2}{l}{$E_\nu < \val{1}{GeV}$}            
&   1.3 &     2 \\
                                                & $1 < E_\nu < \val{10}{GeV}$      &                    
&  2.7  &     3 \\
                                                & $E_\nu > \val{10}{GeV}$\footnotemark[3] &                    
&    6.6 &     5 \\
$\nuebar/\nue$                                  & \multicolumn{2}{l}{$E_\nu < \val{1}{GeV}$}            
&    -1.1 &     5 \\
                                                & $1 < E_\nu < \val{10}{GeV}$      &                    
&    -2.7 &     5 \\
                                                & $E_\nu > \val{10}{GeV}$\footnotemark[4]  &                    
&   -0.34 &     8 \\
$\numubar/\numu$                                & \multicolumn{2}{l}{$E_\nu < \val{1}{GeV}$}            
&    0.36 &     2 \\
                                                & $1 < E_\nu < \val{10}{GeV}$      &                    
&    3.24 &     6 \\
                                                & $E_\nu > \val{10}{GeV}$\footnotemark[5]  &                    
&   9.2 &    15 \\
Up/down ratio                                   & $< \val{400}{MeV}$               & $e$-like           
& 0.079 &   0.1 \\
                                                &                                  & $\mu$-like         
& 0.24 &   0.3 \\
                                                &                                  & 0-decay $\mu$-like 
& 0.87 &   1.1 \\
                                                & $> \val{400}{MeV}$               & $e$-like           
& 0.63 &   0.8 \\
                                                &                                  & $\mu$-like         
& 0.40 &   0.5 \\
                                                &                                  & 0-decay $\mu$-like 
& 1.34 &   1.7 \\
                                                & Multi-GeV                        & $e$-like           
& 0.55 &   0.7 \\
                                                &                                  & $\mu$-like         
& 0.16 &   0.2 \\
                                                & Multi-ring Sub-GeV               & $e$-like           
& 0.32 &   0.4 \\
                                                &                                  & $\mu$-like         
& 0.16 &   0.2 \\
                                                & Multi-ring Multi-GeV             & $e$-like           
& 0.24 &   0.3 \\
                                                &                                  & $\mu$-like         
& 0.16 &   0.2 \\
                                                & PC                               &                    
& 0.16 &   0.2 \\
Horizontal/vertical ratio                       & $< \val{400}{MeV}$               & $e$-like           
&  -0.023 &   0.1 \\
                                                &                                  & $\mu$-like         
&  -0.023 &   0.1 \\
                                                &                                  & 0-decay $\mu$-like 
&  -0.069 &   0.3 \\
                                                & $> \val{400}{MeV}$               & $e$-like           
&  -0.32 &   1.4 \\
                                                &                                  & $\mu$-like         
&  -0.44 &   1.9 \\
                                                &                                  & 0-decay $\mu$-like 
&  -0.32 &   1.4 \\
                                                & Multi-GeV                        & $e$-like           
&  -0.74  &   3.2 \\
                                                &                                  & $\mu$-like         
&  -0.53 &   2.3 \\
                                                & Multi-ring Sub-GeV               & $e$-like           
&   -0.32 &   1.4 \\
                                                &                                  & $\mu$-like         
&  -0.30 &   1.3 \\
                                                & Multi-ring Multi-GeV             & $e$-like           
&   -0.64 &   2.8 \\
                                                &                                  & $\mu$-like         
&   -0.35 &   1.5 \\
                                                & PC                               &                    
&   -0.39 &   1.7 \\
\multicolumn{3}{l}{K/$\pi$ ratio in flux calculation\footnotemark[6]}	
&   -5.3 &    10 \\
\multicolumn{3}{l}{Neutrino path length}                                                                
&  -2.3 &    10 \\
Sample-by-sample                                & \multicolumn{2}{l}{FC Multi-GeV}                      
&   -0.22 &     5 \\
                                                & PC + Stopping \UP                &                    
&   2.46 &     5 \\
\multicolumn{3}{l}{Matter effects}                                                                      
&    0.06 &   6.8 \\
\multicolumn{3}{l}{Solar Activity}                                                                      
&    0.31 &    7 
\\ $M_A$ in QE                                     &                                  &                    
&   -4.79 &    10 \\
\multicolumn{3}{l}{Single $\pi$ Production, Axial Coupling}         
&   -7.03 &    10  \\
\multicolumn{3}{l}{Single $\pi$ Production, $C_{A5}$}               
&   9.33 &    10  \\
\multicolumn{3}{l}{Single $\pi$ Production, BKG}                   
&   -3.03 &    10  \\
\multicolumn{3}{l}{CCQE cross section\footnotemark[7] } 
&    7.37 &    10 \\
\multicolumn{3}{l}{CCQE $\bar \nu/\nu$ ratio\footnotemark[7]}                                           
&    4.81 &    10 \\
\multicolumn{3}{l}{CCQE $\mu/e$ ratio\footnotemark[7]}                                                  
&   -2.74 &    10 \\
\multicolumn{3}{l}{DIS cross section}                                                                   
&   5.72 &     10 \\
\multicolumn{3}{l}{DIS model comparisons\footnotemark[8]} 
&    4.84 &    10 \\
\multicolumn{3}{l}{DIS $Q^2$ distribution (high W)\footnotemark[9]} 
&    -5.70 &    10 \\
\multicolumn{3}{l}{DIS $Q^2$ distribution (low W)\footnotemark[9]}                                      
&   -9.21 &    10 \\
\multicolumn{3}{l}{Coherent $\pi$ production}                                                           
&   -9.41 &   100 \\
\multicolumn{3}{l}{NC/CC}                                                                               
&   -0.10 &    20 \\
\multicolumn{3}{l}{\nutau cross section}                                                                
&  9.08 &    25 \\
\multicolumn{3}{l}{Single $\pi$ production, $\pizero/\pi^\pm$}                                          
&   1.75 &    40 \\
\multicolumn{3}{l}{Single $\pi$ production, $\bar \nu_{i} /\nu_{i}$ (i=$e,\mu $)\footnotemark[10] } 
&    2.30 &    10 \\
\multicolumn{3}{l}{NC fraction from hadron simulation}                                                  
&     -5.07 &    10 \\
\multicolumn{3}{l}{$\pi^+$ decay uncertainty Sub-GeV 1-ring}        \\& \multicolumn{2}{l}{$e$-like 0-decay}                  
&  -0.050  &   0.6 \\
                                                & $\mu$-like 0-decay               &                    
&  -0.066 &   0.8 \\
                                                & $e$-like 1-decay                 &                    
&   -0.34  &   4.1 \\
                                                & $\mu$-like 1-decay               &                    
&   -0.025 &   0.3 \\
                                                & $\mu$-like 2-decay               &                    
&   -0.47 &   5.7 \\
\multicolumn{3}{l}{Final state and seconday interactions\footnotemark[11]} \\& \multicolumn{2}{l}{1-ring}      
&   -6.83 &    10 \\
&Multi-ring&
&   6.93 &    10 \\
\multicolumn{3}{l}{Meson exchange current\footnotemark[12] } 
&   1.08 &    10 \\
\multicolumn{3}{l}{\dmsq{21} \cite{Olive:2016xmw}}
&  0.0 &  0.0002 \\
\multicolumn{3}{l}{\sn{12} \cite{Olive:2016xmw}}                                                           
&   0.05 &  1.3 \\
\multicolumn{3}{l}{\sn{13} \cite{Olive:2016xmw}}                                                                 
&   0.00 &  0.11 \\
 FC reduction                                    &                                  &                    
&  0.27 &   1.3 \\
\multicolumn{3}{l}{PC reduction}                                                                        
&  -0.26 &     1 \\
\multicolumn{3}{l}{FC/PC separation}                                                                    
& 0.000 &  0.02 \\
\multicolumn{3}{l}{PC stopping/through-going separation (bottom)}                                       
&  -1.75 &   6.8 \\
\multicolumn{3}{l}{PC stopping/through-going separation (barrel)}                                       
&  1.13 &   8.5 \\
\multicolumn{3}{l}{PC stopping/through-going separation (top)}                                          
&   -29.0 &    40 \\
Non-$\nu$ background                            & \multicolumn{2}{l}{Sub-GeV $\mu$-like}                
& -0.001 &   0.02 \\
                                                & \multicolumn{2}{l}{Multi-GeV $\mu$-like}                                
& -0.002 &   0.07 \\
                                                & \multicolumn{2}{l}{Sub-GeV 1-ring 0-decay $\mu$-like}                     
& -0.001 &   0.02 \\
                                                & \multicolumn{2}{l}{PC}                                                 
& -0.015 &  0.49 \\
                                                & \multicolumn{2}{l}{Sub-GeV $e$-like (flasher event) }                                    
& -0.000 &   0.03 \\
                                                & \multicolumn{2}{l}{Multi-GeV $e$-like (flasher event) }                                
& -0.001 &   0.07 \\
\multicolumn{3}{l}{Fiducial Volume}                                                                     
&   -0.40 &     2.00 \\
Ring separation                                 & $< \val{400}{MeV}$               & $e$-like           
&   -0.44 &   -0.86 \\
                                                &                                  & $\mu$-like         
&   0.43 &     0.84 \\
                                                & $> \val{400}{MeV}$               & $e$-like           
&  0.19 &     0.37 \\
                                                &                                  & $\mu$-like         
& 1.03 &   2 \\
                                                & Multi-GeV                        & $e$-like           
&  1.32 &     2.57 \\
                                                &                                  & $\mu$-like         
&   -0.74 &   -1.44 \\
                                                & Multi-ring Sub-GeV               & $e$-like           
&   -0.42 &   -0.82 \\
                                                &                                  & $\mu$-like         
&   0.42 &   0.82 \\
                                                & Multi-ring Multi-GeV             & $e$-like           
&  -0.66 &   -1.28 \\
                                                &                                  & $\mu$-like         
&  0.59 &   1.14 \\
\multicolumn{3}{l}{Particle identification (1 ring)}\\                & Sub-GeV                          & $e$-like           
& -0.11 &  0.52 \\
                                                &                                  & $\mu$-like         
&  0.10&  -0.49 \\
                                                & Multi-GeV                        & $e$-like           
& -0.02 &  0.09 \\
                                                &                                  & $\mu$-like         
&  0.02 &  -0.09 \\
\multicolumn{3}{l}{Particle identification (multi-ring)} \\            & Sub-GeV                          & $e$-like           
&   0.29 &   -1.33 \\
                                                &                                  & $\mu$-like         
&   -0.12  &   0.53 \\
                                                & Multi-GeV                        & $e$-like           
&   -0.27 &   1.25 \\
                                                &                                  & $\mu$-like         
&   0.16 &   -0.73 \\
Multi-ring likelihood selection                 &  Multi-ring $e$-like             &   $\nu_{e}$,$\bar{\nu}_{e}$                    
&   -0.42  &   -0.71 \\
                                                &  Multi-ring Other                            &                    
&    0.32  &   0.52 \\
\multicolumn{3}{l}{Energy calibration}                                                                  
&  -0.24 &   2.02 \\
\multicolumn{3}{l}{Up/down asymmetry energy calibration}                                                
& -0.53 &   0.67 \\
\UP reduction                                   & \multicolumn{2}{l}{Stopping}                          
&  0.048 &   0.5 \\
                                                & Through-going                    &                    
&  0.029 &   0.3 \\
\multicolumn{3}{l}{\UP stopping/through-going separation}                                               
& 0.018 &   0.6 \\
\multicolumn{3}{l}{Energy cut for stopping \UP}                                                         
&   0.30 &   1.7 \\
\multicolumn{3}{l}{Path length cut for through-going \UP}                                               
&  -0.33 &   1.5 \\
\multicolumn{3}{l}{Through-going \UP showering separation}                                              
&  3.26 &     3 \\
\multicolumn{3}{l}{Background subtraction for \UP}\\                  & \multicolumn{2}{l}{Stopping\footnotemark[13]} 
&   -4.9 &    11 \\
                                                & Non-showering\footnotemark[13]    &                    
&    2.3 &    17 \\
                                                & Showering\footnotemark[13]        &                    
&   -6.55 &    24 \\
\multicolumn{3}{l}{$\nue/\nuebar$ Separation}                                                           
&  1.89 &   2.57 \\
\multicolumn{3}{l}{Sub-GeV 2-ring \pizero}                                                              
&   -0.039 &   1.03 \\
\multicolumn{3}{l}{Decay-e tagging}                                                                     
&    0.29 &    0.7 
 \end{longtable}
\footnotetext[1]{Uncertainty decreases linearly with $\log E_{\nu}$ from 25\,\%(0.1\,GeV) to 7\,\%(1\,GeV).} 
\footnotetext[2]{Uncertainty is 7\,\% up to 10\,GeV, linearly increases with $\log E_{\nu}$ from 7\,\%(10\,GeV) to 12\,\%(100\,GeV) and then to 20\,\%(1\,TeV)} 
\footnotetext[3]{Uncertainty linearly increases with $\log E_{\nu}$ from 5\,\%(30\,GeV) to 30\,\%(1\,TeV).} 
\footnotetext[4]{Uncertainty linearly increases with $\log E_{\nu}$ from 8\,\%(100\,GeV) to 20\,\%(1\,TeV).} 
\footnotetext[5]{Uncertainty linearly increases with $\log E_{\nu}$ from 6\,\%(50\,GeV) to 40\,\%(1\,TeV).}
\footnotetext[6]{Uncertainty increases linearly from 5$\%$ to 20$\%$ between 100GeV and 1TeV.}
\footnotetext[7]{Difference from the Nieves~\cite{Nieves:2004wx} model is set to 1.0}
\footnotetext[8]{Difference from CKMT~\cite{CKMT94} parametrization is set to 1.0}
\footnotetext[9]{Difference from GRV98~\cite{Gluck:1998xa} is set to 1.0}
\footnotetext[10]{Difference from the Hernandez\cite{Hernandez07} model is set to 1.0}
\footnotetext[11]{Ref.~\cite{Pinzon:2017fsi}}
\footnotetext[12]{Difference from NEUT without model from~\cite{Nieves:2004wx} is set to 1.0}
\footnotetext[13]{The uncertainties in BG subtraction for upward-going muons are only for the most horizontal bin,  $-0.1 < \cos\theta < 0$.} \end{center}

\end{document}